\documentstyle[12pt,epsfig]{article}

\newcommand{\beq}{\begin{equation}}
\newcommand{\eeq}{\end{equation}}
\newcommand{\bea}{\begin{eqnarray}}
\newcommand{\eea}{\end{eqnarray}}
\newcommand{\bp}{\bar p}
\newcommand{\bq}{\bar q}

\def\wh{\widehat}

\def\la{\mathrel{\mathpalette\fun <}}
\def\ga{\mathrel{\mathpalette\fun >}}
\def\fun#1#2{\lower3.6pt\vbox{\baselineskip0pt\lineskip.9pt
\ialign{$\mathsurround=0pt#1\hfil##\hfil$\crcr#2\crcr\sim\crcr}}}

\begin{document}

\title{Systematics of quark-antiquark states and scalar exotic mesons}

\date{}
\author{ V.V. Anisovich}
\maketitle

\begin{abstract}
The analysis of the experimental data of Crystal Barrel Collaboration on
the $p\bp$ annihilation in flight with the production of mesons in the
final state resulted in a discovery of a large number of mesons over
the region  1900--2400~MeV, thus allowing us to systematize
quark-antiquark states in the
$(n,M^2)$  and $(J,M^2)$ planes, where $n$ and $J$ are radial quantum
number and spin of the meson with the mass $M$. The data point to meson
trajectories in these planes being approximately linear, with a
universal slope. The sector of scalar mesons is discussed in more
detail,
where, on the basis of the recent  $K$-matrix analysis, the nonet
classification of quark--antiquark
 states was performed: in the region below 2000
MeV, two scalar nonets are fixed, that is, the basic one
and the nonet of the first radial excitation.  In the
scalar sector, there are two states with the isospin  $I=0$, which are
extra ones with respect to the quark--antiquark classification, i.e.
exotic states:  the broad resonance $f_0(1200-1600)$ and the light
$\sigma$-meson.  The ratios of coupling constants for hadronic decays
to the states $\pi\pi, K\bar K, \eta\eta,\eta\eta'$
point to the gluonium nature of the broad state $f_0(1200-1600)$.
\end{abstract}

\newpage

\begin{center}

Contents.
\end{center}

\noindent
1. Introduction.\\
2. Experimental data of the Crystal Barrel Collaboration and
systematics of mesons states.\\
2.1 Systematics of meson states.\\
2.2 Exotics: the states off quark--antiquark
trajectories.\\
3. Scalar-meson sector.\\
3.1 The $K$-matrix analysis of the $(IJ^{PC}=00^{++})$-wave.\\
3.1.1 $K$-matrix amplitude.\\
3.1.2 Partial amplitude for the $00^{++}$ wave:  unitarity,
analyticity and the problem of the left-hand cut.\\
3.1.3 Three-meson production in the reactions of the
$p\bar p$ and  $n\bar p$ annihilation.\\
3.1.4 Peripheral two-meson production in meson--nucleon
collisions at high energies.\\
3.2 Classification of scalar bare states.\\
3.3 Overlapping $f_0$-resonances in the region
            1200--1700~MeV:  the accumulation of widths of
            quark-antiquark states by the glueball.\\
3.4 Evolution of couplings of the $00^{++}$-states with
channels $\pi\pi$, $\pi\pi\pi\pi$, $K\bar K$, $\eta\eta$ $\eta\eta'$
with the onset of the decay processes.\\
3.5 Evaluation of the glueball component in the
resonances $f_0(980)$,
$f_0(1300)$, $f_0(1500)$, $f_0(1750)$ and  broad state
$f_0(1200-1600)$ based on the analysis of hadronic channels.\\
3.6 The light $\sigma$-meson: Is there a  pole of the $00^{++}$-wave
amplitude?\\
3.7 Systematics of scalar states on the $(n,M^2)$  and
$(J,M^2)$ planes and the problem of basic multiplet $1^3P_0q\bar q$.\\
3.7.1 The $K$-matrix classification of scalars and
$q\bar q$ trajectories in the  $(n,M^2)$ planes.\\
3.7.2 Systematics of kaons in the $(J,M^2)$ plane.\\
3.8 Exotic scalar states, $f_0(1200-1600)$ and $f_0(300-500)$.\\
4. Conclusion.
\newpage

\section{Introduction}

To understand the structure of strong interactions at low and
moderate energies is one of the highest-priority problems of the modern
particle physics. In the last decades, great efforts have been
paid to develop the strong-interaction theory, the strong QCD,
with a considerable progress
though
without determinate breakthrough. One may believe that, in part,
 this was due to a poor knowledge, up to recent time, of the
experimental situation in meson sector.

To enlighten experimental situation, a number of
experiments had been undertaken  during 1990--2000 directed
purposefully to the search for new meson resonances, and a detailed
investigation of mesons discovered before was carried out.

The Crystal Barrel Collaboration  accumulated one of the richest
statistics on the  $p\bp$  and $n\bar p$ annihilation reactions; the
group of physicists from PNPI took part in the analysis and
interpretation of the data  by Crystal Barrel Collaboration
(a brief review of this activity is given in \cite{pnpi}); this
survey presents the results of recent investigations.

In 1993--1994, two groups,  QMWC (London) and PNPI (St.Petersburg),
were analysing meson spectra obtained in reactions of the
$p\bp$ annihilation at rest. As a result, the scalar-isoscalar
resonances $f_0(1370)$ and  $f_0(1500)$ have
been discovered \cite{f1500,f1500cb,f1500PR,a0cb}; now these resonances are
actively discussed in connection with the glueball problem. To
establish the systematics of quark-antiquark states, the study of
$p\bp$ annihilation in flight carried out in 1999-2001 \cite{ral} was of a
particular importance, for it allowed us to
investigate the mass region 1900--2400~MeV: a large number of
resonances discovered in this region made it possible to fix the $q\bar
q$ trajectories on the
$(n,M^2)$  and  $(J,M^2)$ planes \cite{pnpi,syst}. We speak about
the systematics of the $q\bar q$ states in Section 2 of this
paper.

Scalar mesons play  crucial role for understanding  the strong
QCD.  Starting from 1995, the PNPI group worked upon the $K$-matrix
fit to the wave $(IJ^{PC}=00^{++})$ based on the simultaneous analysis
of all data available by that time. The necessity of a combined
analysis was dictated by the existence of large interference effects
"resonance--background" as well as the effects associated with the
resonance overlapping. In a situation of such a type, only a combined
fitting to a large number of reactions allows one to hope for reliable
results.  In Section 3, a current understanding of scalar-meson sector
is presented  based on recent results of the $K$-matrix fit to meson
spectra \cite{K}.

Previous analyses carried out in 1997--1998 \cite{YF,YF99,ufn} were
based on
the experimental data as follows:\\
(1) GAMS data on the $S$-wave two-meson  production in the reactions
 $\pi p\to \pi^0\pi^0 n$,
$\eta\eta n$ and $\eta\eta' n$
at small nucleon momenta transferred, $|t|<0.2$ (GeV/$c$)$^2$
\cite{GAMS,GAMS-eta};\\
(2) GAMS data on the $\pi\pi$ $S$-wave production in the reaction
 $\pi p\to \pi^0\pi^0 n$
at large momentum transfers squared,  $0.30<|t|<1.0$ (GeV/$c$)$^2$
\cite{GAMS};\\
(3) BNL data on the reaction $\pi^- p\to K\bar K n$ \cite{BNL};\\
(4) CERN-M\"unich data on $\pi^+\pi^- \to\pi^+\pi^-$ \cite{C-M};\\
(5)  Crystal Barrel data on
$p\bar p$(at rest, from liquid hydrogen)$\to \pi^0\pi^0\pi^0$,
$\pi^0\pi^0\eta$, $\pi^0\eta\eta$ \cite{f1500cb,a0cb,cbc}.

Now the experimental basis has  much broadened, and
additional samples of data are included into the analysis
\cite{K} of the
$00^{++}$ wave as follows:\\
(6) Crystal Barrel data on proton-antiproton annihilation in gas:
$p\bar p$(at rest, from gaseous hydrogen)$\to
\pi^0\pi^0\pi^0$, $\pi^0\pi^0\eta$ \cite{cbc_new},\\
(7) Crystal Barrel data on proton-antiproton annihilation
in liquid:
$p\bar p$(at rest, from liquid hydrogen)$ \to
\pi^+\pi^-\pi^0$,
$K^+K^-\pi^0$, $K_SK_S\pi^0$, $K^+K_S\pi^-$ \cite{cbc_new};\\
(8) Crystal Barrel data on neutron-antiproton annihilation in
liquid deuterium: $n\bar p$(at rest, from liquid
deuterium)$\to \pi^0\pi^0\pi^-$, $\pi^-\pi^-\pi^+$,
$K_SK^-\pi^0$, $K_SK_S\pi^-$ \cite{cbc_new};\\
(9) E852 Collaboration data on the $\pi\pi$ $S$-wave production in the
reaction $\pi^-p\to \pi^0\pi^0n$ at the nucleon momentum transfers
squared $0<|t|<1.5 \; ({\rm GeV/c})^2$ \cite{BNL-new}.

In addition to Ref. \cite{YF}, the reactions of the $p\bar p$
annihilation in gas have been also  included into
the analysis \cite{K}. One should
keep in mind that in liquid hydrogen the $p\bar p$
annihilation is going dominantly from the
$S$-wave state, while in  gas
there is a considerable admixture of the $P$-wave, thus giving
an opportunity to analyse  the three-meson Dalitz plots in more
detail.  New Crystal Barrel data allowed us to study the two-kaon
channel with a more confidence as compared to what had been done
before.  This is undoubtedly important for the conclusion about the
quark-gluon content of the scalar--isoscalar $f_0$-mesons under
investigation.

Experimental data of the E852 Collaboration on the reaction
$\pi^- p \to \pi^0\pi^0n$  at $p_{lab}=18$ GeV/c \cite{BNL-new} together
with the GAMS data on the reaction $\pi^- p \to \pi^0\pi^0n$
at $p_{lab}=38$ GeV/c \cite{GAMS} give us a solid ground for the
study of  the resonances $f_0(980)$ and $f_0(1300)$, for at large
momenta transferred to the nucleon, $|t| \sim 0.5\sim 1.5$ (GeV/c)$^2$,
the production of resonances is accompanied by a small background,
thus allowing us to fix reliably their masses and widths. This is
especially important for $f_0(1300)$: in the compilation \cite{PDG} this
resonance is quoted as $f_0(1370)$, with the mass in the
interval $1200-1500$ MeV, though
experimental data favour rather definitely the mass near 1300 MeV.

The $K$-matrix amplitude determines both the amplitude poles
(masses and widths of resonances)
and  $K$-matrix poles (masses of bare states). The
$K$-matrix poles differ from the amplitude poles in two items: \\
(i) The states corresponding to the $K$-matrix poles do not contain any
 component
associated with the decay processes, i.e. the transitions into real
mesons. The
absence of a cloud of  real mesons allows us to refer conventionally
to these states  as  bare ones \cite{YF,ufn,km}. \\
(ii) Due to the transition {\it bare  state(1) $\to$ real mesons $\to$
bare state(2)} the observed resonances are mixtures of bare states ---
in the first place this is related to the $f_0$-mesons.  So, for quark
systematics, the  bare states are primary objects rather than  the
resonances.

This can be explained with the example of the behaviour of a level
in the potential picture. Consider a potential
well and the levels which
correspond to stable states. Then gradually we switch the
decay channels on, that is, replace impenetrable wall by potential
barrier --- at the beginning this leads to a broadening of levels,
but masses of states
remain almost the same. But when the widths and
resonance positions are such that resonances overlap, a cardinal
reconstruction of the structure of levels occurs. Namely, one of
resonances accumulates the widths of its resonance-neighbours thus
making them comparatively narrow. In this way an intensive mixing of
resonance states takes place, while the positions of masses are
shifted in a value of the order of the resonance width,  in 100--200
MeV. The $K$-matrix amplitude allows us to trace the transformation of
stable levels (bare states)
into  resonances. The $K$-matrix amplitude works
with parameters related to stable levels.
 Multichannel unitarization of the amplitude,
which is inherent in the $K$-matrix representation, with the account
for analyticity, allows one to trace the transformation of stable
levels into real resonances. A characteristic feature of the $K$-matrix
fit is its ability to reconstruct the picture of stable levels as well
as re-create a real picture of complex masses and partial widths.

The latest $K$-matrix fit \cite{K} gave us rather definite
information on the resonances $f_0(980)$, $f_0(1300)$, $f_0(1500)$,
$f_0(1750)$ and broad state $f_0(1200-1600)$.  Relying on the
extracted partial widths for transitions to the channels $\pi\pi, K\bar
K, \eta\eta, \eta\eta'$, one can analyse the quark-gluonium content of
 these states. In this way the following properties of resonances are
to be formulated:

{\bf 1. $f_0(980)$}: This resonance is dominantly the $q\bar q$ state,
$q\bar q =n\bar n\sin \varphi+s\bar s \cos \varphi$,
with a large $s\bar s$ component.
Assuming the  glueball
admixture to be not greater than 20\%,
$W_{gluonium} \la 20\%$,
the hadronic decays give
us the following constraints for mixing angle:
 $-95^\circ \le \varphi \la -40^\circ$.
Rather large uncertainties in the
determination of mixing angle  are due to a high sensitivity of coupling
constants to a plausible small admixture of the gluonium. At
$W_{gluonium}=0$,
hadronic decays provide us with $\varphi = -67^\circ\pm 10^\circ$.

{\bf 2. $f_0(1300)$} (in the compilation \cite{PDG}
this resonance is denoted as $f_0(1370)$): This resonance is the
descendant of the bare $q\bar q $ state
which is close to the  flavour singlet. The resonance $f_0(1300)$
is formed due to a strong mixing with the primary gluonium
and neighbouring $q\bar q$ states.  The quark-antiquark
content of $f_0(1300)$, $q\bar q=n\bar n \cos \varphi+s\bar s\sin \varphi$
determined from
the transitions $f_0(1300)\to \pi\pi, K\bar K, \eta\eta$
 strongly depends on the admixture of the gluonium component.
At $W_{gluonium} \la 30\%$ the mixing angle changes, depending on
$W_{gluonium}$ and interference sign, in the interval
$-45^\circ \la \varphi[f_0(1300)] \la 25^\circ$;
at $W_{gluonium} = 0$ the hadronic decays provide us
$\varphi[f_0(1300)] = - 6^\circ\pm 10^\circ$.

{\bf 3. $f_0(1500)$}: This resonance is the
descendant of a bare state with large $n\bar n$ component.
Like $f_0(1300)$, the resonance
$f_0(1500)$ is formed by mixing with the gluonium
and  neighbouring $q\bar q$ states. The quark-antiquark
content, $q\bar q=n\bar n \cos \varphi+s\bar s\sin \varphi$,
depends on the admixture of the gluonium:
at $W_{gluonium} \la 30\%$ the mixing angle changes, depending on
$W_{gluonium}$, in the interval
$-20^\circ \la \varphi[f_0(1300)] \la 25^\circ$;
at $W_{gluonium} = 0$ one has
$\varphi[f_0(1500)] =  11^\circ \pm 8^\circ$.

{\bf 4. $f_0(1750)$}: This resonance is the descendant of the bare state
belonging to the radial-excitation nonet $2^3P_1 q\bar q$,
the wave function of which has large $s\bar s$ component.
The $K$-matrix analysis permits the two solutions, with
 different
values of  the $s\bar s$ component. In the first solution, the $s\bar s$
component dominates;
in the absence of  gluonium component,
$\varphi[f_0(1750)] = - 72^\circ \pm 5^\circ$, and
if the gluonium admixture does not exceed 30\%,
 then
$-110^\circ \la \varphi[f_0(1750)] \la - 35^\circ$.
 In the second solution, in the absence of  gluonium component,
$\varphi[f_0(1750)] = - 18^\circ \pm 5^\circ$, and
with the gluonium admixture  $W_{gluonium}\la 30\%$,
$-50^\circ \la \varphi[f_0(1750)] \la 10^\circ$.

{\bf 5. $f_0(1200-1600)$}: The broad state is the descendant of
the primary glueball. The
analysis of hadronic decays  of this resonance confirmed its glueball
nature: the glueball descendant has the quark-antiquark component
$(q\bar q)_{glueball}=(u\bar u+d\bar d+\sqrt{\lambda}s\bar s)/
\sqrt{2+\lambda}$ \cite{alexei-glu},
which is defined by the probability to create
new $q\bar q$-pairs by the gluon field
$u\bar u:d\bar d:s\bar s=1:1:\lambda$,  the suppression parameter
for the production of strange quarks being in the interval
$\lambda\simeq 0.5-0.8$
\cite{lambda,lambda-hec}.
 In terms of mixing angle
for the $n\bar n$ and $s\bar s$ components, this means
that the glueball descendant must have
$\varphi_{glueball}\simeq 27^\circ -33^\circ $, and just these values
for $\varphi[f_0(1200-1600)]$ have been obtained in all variants of the
$K$-matrix fit \cite{K}. Such a value of mixing angle,
$\varphi[f_0(1200-1600)]=30^\circ\pm 3^\circ$, does not allow one to
determine the admixture of the  $q\bar q$
component in the broad state. This is
due to the fact that the $(q\bar q)_{glueball}$  and gluonium
components are coupled to the channels  $\pi\pi, K\bar K, \eta\eta$ and
$\eta\eta'$ in equal proportions.

Referring to the broad state $f_0(1200-1600)$  as a
resonance needs some comments. The observed spectra in the
channels $\pi\pi, K\bar K, \eta\eta,\eta\eta'$ demand
to introduce a broad bump,
and it occurred that this bump behaves in a universal
way, thus enabling us to describe it as a resonance state. Indeed, a
characteristic feature of the resonance is the
factorization  property: the resonance
amplitude may be represented in the form
$g_{in} (s-M^2)^{-1} g_{out}$, and universal constants
$g_{in}$ and  $g_{out}$ depend on the sort of initial and final states
only.  The description of a large number of reactions in \cite{K}
agrees well with the factorization property of the broad-state
amplitude. A
large width of the $f_0(1200-1600)$ does not allow us to
determine reliably its mass, but due to a strong production of the
$f_0(1200-1600)$ in a large number of reactions we can, with a
confidence, to find out the ratio of couplings to the channels
$\pi\pi, K\bar K, \eta\eta,\eta\eta'$, that is, to define its content
in terms of the $q\bar q$ and gluonium states.

The $K$-matrix \cite{K} analysis does
not point determinedly to the existence of the light
$\sigma$-meson which is actively discussed at present
(e.g. see \cite{van} and references therein),
 in particular in
connection with the recently reported signals in the $\pi\pi$ spectra
of  the decays
$D^+ \to \pi^+\pi^+\pi^-$ (pole in the amplitude at $M=(480\pm 40)\;
-i(160\pm 30)$ MeV \cite{D+}),
$J/\Psi\to\pi\pi\omega$ (pole at
 $M=(390\,^{+60}_{- 36})\;  -i(141\, ^{+38}_{- 25})$ MeV \cite{BES}),
$\tau\to\pi\pi\pi\nu$ (pole at
 $M\simeq 555-i270$ MeV \cite{tau}).
 Possible explanation of this
discrepancy may consist in a strong suppression of the light
$\sigma$-meson production in the
$p\bar p$ annihilation processes, like $p\bar p\to
\pi\pi\pi$, though, as one may think, there is no visible reason for
such a suppression. Recall that the statistics in the Crystal Barrel
reactions is by two orders of magnitude larger than in
\cite{D+,BES,tau}.  Alternative explanation can be associated with a
restricted applicability of the $K$-matrix approach at small invariant
energy squared $s$: the $K$-matrix amplitude does not describe properly
the left-hand singularities associated with meson exchanges in the
crossing channels. So one may suppose that the
$K$-matrix  analysis does not
reconstruct analytical amplitude correctly at ${\rm Re \,}s \la
4m^2_\pi$, i.e. at $({\rm Re \,}M)^2-({\rm Im \,}M)^2 \la 4m^2_\pi$,
$({\rm Re \,}M)^2-({\rm Im \,}M)^2 \la 4m^2_\pi$.
In numerous analyses, including those carried out in the dispersion
relation technique where left-hand cuts can be accounted for in one way
or another, the pole ascribed to the light $\sigma$-meson occurred just
at ${\rm Re \,}s \la 4m^2_\pi$, e.g. see
\cite{sigma_dis,sigma80,sigma,sigma_N/D}. In this connection let us
emphasize that the dispersion-relation $N/D$ analysis \cite{sigma_N/D},
where the low-energy $00^{++}$ amplitude was sewed with the $K$-matrix
amplitude \cite{YF}, leads to the pole at $M\simeq 430-i325$ MeV.

Therefore, the $K$-matrix amplitude analysis tells us that in
 the scalar--isoscalar sector there are two states,  the broad
resonance $f_0(1200-1600)$ and the light sigma-meson $f_0(300-500)$,
which are extra ones for the $q\bq$-systematics of mesons. In the final
Section 4, the status  of these possible exotic states is discussed as
well as two more candidates for exotics --- $\pi_2(1880)$ and the
$(J^{PC}= 1^{-+})$ state (recall that $q\bq$-system cannot have such
quantum numbers).

\section{Experimental data of the Crystal Barrel Collaboration and
systematics of mesons states}

In 1989-1997, the Crystal Barrel Collaboration studied the reactions of
the $p\bp$-annihilation at rest and in flight at LEAR (CERN), that
resulted in the accumulation of a huge statistics on multi-meson
states. However, initially  the analysis of experimental
data had not been carried out in due course.  The first attempt to
analyse pion spectra in the reaction $p\bp$ (at rest)
$\to\pi^0\pi^0\pi^0$ based on a simplified isobar model
led to a wrong identification of the peak near 1500 MeV; the peak had
been identified as the tensor meson AX(1515) \cite{AX}, and at the same
time no scalar states were seen in the $\pi\pi$ spectra over the range
1200--1700 MeV.

In reactions such as $p\bp$ (at rest) $\to\pi^0\pi^0\pi^0$ the
three-particle interaction characteristics reveal themselves in full
scale, that should be accounted for in the analysis of meson spectra.
The method based on the extraction of leading singularities (pole,
square root, logarithmic ones, and so on) in the production amplitude
of a few particles was developed in the papers
\cite{sing_thr,sing_tri,sing_ale}.   This method had been applied with
the purpose to re-analyse the Crystal Barrel data  for
the reaction $p\bp$ (at rest) $\to\pi^0\pi^0\pi^0$, and in the work
performed together with the Collaboration members, the resonances
 $f_0(1370)$ and $f_0(1500)$ had been discovered in
the region 1300--1500 MeV \cite{f1500,f1500cb,f1500PR}.
The data      on the reactions
 $p\bp\to\pi^0\pi^0\eta$ and $p\bp\to\pi^0\eta\eta$
involved into  combined analysis
enabled us to discover a group of tensor and scalar-isovector resonances
\cite{a0cb}. The masses and widths of these resonances obtained in the
latest analysis are shown in Table 1, and
 the analysed reactions are quoted in Table 2.

A combined fit to data collected in  various reactions and experiments
is a specific feature of the method used in the papers \cite{K,YF,YF99}.
The matter is that, because of  the presence of a considerable
reaction background, the resonance does not always reveal itself as a
peak in meson spectrum: due to interference effects the
resonance may appear as a dip in the spectrum, or it may be seen as a
"shoulder". Including into analysis the GAMS data  on the
two-meson spectra in the reactions
$\pi^-p\to\pi^0\pi^0n$, $\pi^-p\to\eta\eta n$,
$\pi^-p\to\eta\eta'n$ \cite{GAMS,GAMS-eta},   BNL data on $\pi^-p\to
K^+K^-n$ \cite{BNL}  and
CERN--M\"unich  data on $\pi^-p\to \pi^+\pi^-n$ \cite{C-M} allowed us not
only to determine reliably the ratios of different two-meson yields
(that is rather important for the resonance classification), but also
to conclude that the peak in the $\eta\eta$-spectrum in the reaction
$\pi^-p\to\eta\eta n$, which was previously claimed to be the resonance
$G(1590)$ \cite{GAMS-eta},  appeared in fact as a result of the
interference of the broad state $f_0(1200-1600)$ and $f_0(1500)$  resonance
\cite{APS}.

\begin{table}[h]
\caption{Resonances discovered in the analysis of Crystal Barrel data.
Masses and widths are presented in accordance with the latest combined
analysis [1,8] }

\begin{center}

\begin{tabular}{||l|l|l|l||} \hline \hline
&&&\\
Resonance  & $I^G J^{PC}$ & Mass, & Width, \\
 ~ & ~ & MeV & MeV  \\
\hline
$f_0(1300)$ (or $f_0(1370)$ [21]) & $0^+ 0^{++}$
& $1310\pm 20$        & $280\pm 30$ \\
$f_0(1500)$ & $0^+ 0^{++}$ & $1495\pm 6$         & $126\pm 5$   \\
$f_0(1200-1600)$ & $0^+ 0^{++}$ & $1400\pm200$ & $1200\pm400$ \\
$a_0(1450)$ & $1^- 0^{++}$ & $1520\pm 25$ & $240 \pm 20$ \\
$a_2(1660)$ & $1^- 2^{++}$ & $1670_{-20}^{+40}$ & $310 \pm 40$ \\
$f_2(1565)$ & $0^+ 2^{++}$ & $1580\pm 6$         & $160\pm 20$   \\
&&&\\
\hline\hline
\end{tabular}
\end{center}
\label{res_rest}
\end{table}

\begin{table}[h]
\caption{The list of reactions used in the combined analysis [1,8]}

\begin{center}
\begin{tabular}{||l|l|l|l||}
\hline \hline

Crystal Barrel & \multicolumn{2}{|c|}{Two-particle data}\\
\cline{2-3}
reactions& Reaction&  Collaboration \\
\hline
$\bar p p\to \pi^0\pi^0\pi^0$&$\pi^-p\to \pi^-\pi^+n$& CERN-
M\"unich\\
$\bar p p\to \eta\pi^0\pi^0$ &$\pi^-p\to \pi^0\pi^0n$& GAMS \\
$\bar p p\to \eta\eta\pi^0$  &$\pi^-p\to \eta\eta n$& GAMS \\
$\bar p p\to \pi^+\pi^-\pi^0$& $\pi^-p\to \eta\eta'n$& GAMS \\
$\bar p n\to \pi^+\pi^-\pi^-$ &$\pi^-p\to K\bar K   n$& BNL \\
$\bar p n\to \pi^-\pi^0\pi^0$ &$\pi^-p\to \pi^0\pi^0n$& E852  \\
$\bar p p\to \pi^0 K_SK_S$    & ~    & ~ \\
$\bar p p\to \pi^0 K^+K^-$    & ~    & ~ \\
$\bar p p\to \pi^+ K^-K_S$    & ~    & ~ \\
$\bar p n\to \pi^- K_SK_S$    & ~    & ~ \\
$\bar p n\to \pi^0 K^-K_S$    & ~    & ~ \\
&&\\
\hline\hline
\end{tabular}
\end{center}
\label{list_r}
\end{table}

\begin{table}[h]
\caption{Resonances discovered in the analysis of the
$p\bp$-annihilation reactions in flight [6]. One star means
that the resonance had been observed in one reaction only or it reveals
itself poorly. Two stars mean that the resonance was seen in two
reactions or in one where its contribution dominates. Three stars mark
well-established resonances by using several reactions.}

\begin{center}
{\footnotesize\begin{tabular}{||l|l|l|l|c||}
\hline
Resonance& $I^G J^{PC}$ & Mass, MeV & Width, MeV & Status of the state\\
\hline
$\pi$   & $1^- 0^{-+}$ & $2070\pm 35$        & $310\pm 80$ & * \\
$\pi$   & $1^- 0^{-+}$ & $2360\pm 30$        & $300\pm 80$ & * \\
$a_1$   & $1^- 1^{++}$ & $2270\pm 50$        & $300\pm 70$ & * \\
$\pi_2$ & $1^- 2^{-+}$ & $2005\pm 20$        & $210\pm 40$ & * \\
$\pi_2$ & $1^- 2^{-+}$ & $2245\pm 60$        & $320\pm 60$ & * \\
$a_2$   & $1^- 2^{++}$ & $1950\pm 40$        & $180\pm 40$ & ** \\
$a_2$   & $1^- 2^{++}$ & $2030\pm 20$        & $205\pm 30$ & *** \\
$a_2$   & $1^- 2^{++}$ & $2175^{+80}_{-30}$  & $310\pm 60$ & * \\
$a_2$   & $1^- 2^{++}$ & $2255\pm 20$        & $230\pm 15$ & *** \\
$a_3$   & $1^- 3^{++}$ & $2030\pm 20$        & $150\pm 20$ & ** \\
$a_3$   & $1^- 3^{++}$ & $2275\pm 40$        & $150\pm 20$ & *  \\
$a_4$   & $1^- 4^{++}$ & $2005\pm 30$        & $180\pm 30$ & *** \\
$a_4$   & $1^- 4^{++}$ & $2255\pm 40$        & $330\pm 70$ & **  \\
$\pi_4$ & $1^- 4^{-+}$ & $2250\pm 15$        & $215\pm 25$ & **  \\
\hline
$f_0$   & $0^+ 0^{++}$ & $2105\pm 15$        & $200\pm 25$ & ** \\
$f_0$   & $0^+ 0^{++}$ & $2320\pm 30$        & $175\pm 45$ & * \\
$f_2$   & $0^+ 2^{++}$ & $1920\pm 40$        & $260\pm 40$ & ** \\
$f_2$   & $0^+ 2^{++}$ & $2020\pm 30$        & $275\pm 35$ & *** \\
$f_2$   & $0^+ 2^{++}$ & $2240\pm 30$        & $245\pm 45$ & *** \\
$f_2$   & $0^+ 2^{++}$ & $2300\pm 35$        & $290\pm 50$ & ** \\
$f_4$   & $0^+ 4^{++}$ & $2020\pm 25$        & $170\pm 20$ & *** \\
$f_4$   & $0^+ 4^{++}$ & $2300\pm 25$        & $280\pm 50$ & ** \\
$\eta_2$  &$0^+ 2^{-+}$ & $2030\pm 40$        & $190\pm 40$ & ** \\
$\eta_2$  &$0^+ 2^{-+}$ & $2250\pm 40$        & $270\pm 40$ & *** \\
\hline
$\omega$  & $0^- 1^{--}$ & $2150\pm 20$        & $235\pm 30$ & ** \\
$\omega$  & $0^- 1^{--}$ & $2295\pm 50$        & $380\pm 60$ & * \\
$\omega_2$&$0^- 2^{--}$ & $1975\pm 20$        & $175\pm 25$ & * \\
$\omega_2$&$0^- 2^{--}$ & $2195\pm 30$        & $225\pm 40$ & * \\
$\omega_3$&$0^- 3^{--}$ & $1960\pm 30$      & $165\pm 30$ & ** \\
$h_1$     &$0^- 1^{+-}$ & $2000\pm 20$      & $205\pm 20$ & ** \\
$h_1$     &$0^- 1^{+-}$ & $2270\pm 15$      & $175\pm 30$ & ** \\

\hline
$\rho_1$  & $1^+ 1^{--}$ & $1980\pm 30$        & $165\pm 30$ & ** \\
$\rho_2$  & $1^+ 2^{--}$ & $1940\pm 40$        & $155\pm 40$ & * \\
$\rho_2$  & $1^+ 2^{--}$ & $2225\pm 35$        & $335\pm 75$ & * \\
$\rho_3$  & $1^+ 3^{--}$ & $1980\pm 15$        & $175\pm 20$ & ** \\
$\rho_3$  & $1^+ 3^{--}$ & $2260\pm 20$        & $200\pm 30$ & *  \\
$\rho_4$  & $1^+ 4^{--}$ & $2240\pm 25$        & $210\pm 40$ & ** \\
$b_1$     & $1^+ 1^{+-}$ & $1970\pm 40$        & $215\pm 60$ & ** \\
$b_1$     & $1^+ 1^{+-}$ & $2210\pm 50$        & $275\pm 45$ & *  \\
$b_3$     & $1^+ 3^{+-}$ & $2020\pm 15$        & $110\pm 20$ & ** \\
$b_3$     & $1^+ 3^{+-}$ & $2245\pm 50$        & $350\pm 80$ & *  \\
\hline\hline
\end{tabular}}
\end{center}
\label{res_flight}
\end{table}
\clearpage

The data by the Crystal Barrel Collaboration discussed above were
obtained  from the $p\bp$-annihilation at rest. In addition, the Crystal
Barrel Collaboration
has a huge statistics for the events of the $p\bp$-annihilation in
flight, the antiproton momentum covering the range 600--1900 MeV.
After the Crystal Barrel Collaboration
stopped its activity in 1999, these data were
provided to the PNPI group for further processing and  analysis of
spectra. In 1999--2001, together with the colleagues from
Queen Mary and Westfield College (London)
and Rutherford--Appleton Laboratory, the PNPI group
analysed  these data. More than thirty resonances have
been discovered in the mass range  1900--2400 MeV \cite{ral};  they are
shown in Table 3. The discovery of these resonances allowed us to
establish systematics of meson $q\bq$-states on the  $(n,M^2)$- and
$(J,M^2)$-planes \cite{pnpi,syst} ($n$ is radial quantum number
of the  $q\bq$-state
with mass $M$ and $J$ is its spin).

\subsection{Systematics of meson states}

In Figs. 1 and 2, the trajectories on the $(n,M^2)$- and
$(I,J^{PC})$-planes are shown for the states with negative and positive
charge parities:
\begin{eqnarray}
C\ =\ -: && b_1(11^{+-}),\ b_3(13^{+-}),\ h_1(01^{+-}),\
\rho(11^{--}),\ \rho_3(13^{--}), \nonumber \\
&& \omega/\phi(01^{--}),\quad \omega_3(03^{--})\ ; \\
C\ =\ +: && \pi(10^{-+}),\ \pi_2(12^{-+}),\ \pi_4(14^{-+}),\
\eta(00^{-+}),\ \eta_2(02^{-+}), \nonumber\\
&& a_0(10^{++}),\ a_1(11^{++}),\ a_2(12^{++}),\ a_3(13^{++}),\
a_4(14^{++}), \nonumber\\
&& f_0(00^{++}),\quad f_2(02^{++})\ .
\end{eqnarray}

In terms of $q\bq$-states, the mesons of the nonets $n^{2S+1}L_J$
fill in the following
$(n,M^2)$-trajectories for $M\la2400\,$MeV:
\begin{eqnarray}
&&^1S_0\ \to\ \pi(10^{-+}), \ \eta(00^{-+})\ ; \\
&& ^3S_1\ \to\ \rho(11^{--}),\ \omega(01^{--})/\phi(01^{--})\ ;
\nonumber\\
&&^1P_1\ \to\ b_1(11^{+-}),\ h_1(01^{+-})\ ; \nonumber\\
&& ^3P_J\ \to\ a_J(1J^{++}),\ f_J(0J^{++}),\ J=0,1,2\ ;\nonumber\\
&& ^1D_2\ \to\ \pi_2(12^{-+}),\ \eta_2(02^{-+})\ ;\nonumber\\
&& ^3D_J\ \to\ \rho_J(1J^{--}),\ \omega_J(0J^{--})/\phi_J(0J^{--}),\
J=1,2,3\ ; \nonumber\\
&& ^1F_3\ \to\ b_3(13^{+-}),\ h_3(03^{+-})\ ;\nonumber\\
&& ^3F_J\ \to\ a_J(1J^{++}),\ f_J(0J^{++}),\  J=2,3,4\ .\nonumber
\end{eqnarray}

Different orbital momenta can form the trajectories with the same $J$,
namely, $J=L\pm1$. Therefore, the number of such trajectories doubles;
these states are $(I,1^{--})$, $(I,2^{++})$, and so on. Isoscalar
states have two flavour components
$n\bar n=(u\bar u+d\bar d)/\sqrt2$ and $s\bar s$,
that also results in a doubling
of trajectories such as            $\eta(00^{-+})$, $f_0(00^{++})$,
and so on.

The trajectories with negative charge parity, $C=-$, can be  defined
practically unambigously (in Figs. 1 and 2,
black circles stand for the observed states
\cite{ral,K,PDG}, while open circles mark the states predicted by
trajectories).  The trajectories are linear
with a good accuracy:
\beq
M^2\ \simeq\ M^2_0+(n-1)\mu^2\ ,
\eeq
where $M_0$ is the mass of the basic meson with $n=1$, and the slope
parameter is
about $\mu^2\simeq1.3\,$GeV$^2$.  The trajectory slopes for
$b_1(11^{+-})$ and $b_3(13^{+-})$ are slightly lower: for them
$\mu^2\simeq1.2\,$GeV$^2$.

In the sector with $C=+$, the states $\pi_J$ belong definitely to
linear trajectories with $\mu^2\simeq1.2\,$GeV$^2$, with an exception
for
$\pi(140)$ that is not a surprise, for the pion is rather specific
state. The sector of the  $a_J$-states with
$J=0,1,2,3,4$ demonstrates a neat set of linear trajectories, with the
slopes $\mu^2\simeq1.15-1.20\,$GeV$^2$; the same slope is seen
for the $f_2$- and $f_4$-trajectories.

For the $f_0$-mesons the trajectory slope is
$\mu^2\simeq1.3\,$GeV$^2$. Let us emphasize that two states do not
lay on linear $q\bq$-trajectories, namely, the light $\sigma$-meson
$f_0(300-500)$ \cite{PDG} and the broad state
$f_0(1200-1600)$, this latter has been fixed by the  $K$-matrix analysis
\cite{K,YF,YF99}.

The picture of the  state location on
the $(n,M^2)$-plot is complemented by trajectories on the
$(J,M^2)$-plots: they are shown in Fig. 3.
To draw the daughter  $(J,M^2)$ trajectories, it was
important that leading meson trajectories $(\pi,\rho,,a_1, a_2$ and
$P')$ were well-known from the analysis of hadronic diffractive
scatterings at  $p_{lab}\sim5-50\,$ GeV/c.

Pion trajectories, leading and daughter ones, are linear with a good
accuracy, see Fig. 3. The other leading trajectories,
 $\rho,\eta,a_1,a_2,f_2$ or $P'$, can be also considered as linear,
with a good accuracy:
\beq
\alpha_X(M^2)\ \simeq\ \alpha_X(0)+\alpha'_X(0)M^2\ .
\eeq
The parameters of linear trajectories determined by masses of the
 $q\bq$-states are as follows:
\begin{eqnarray}
& \alpha_\pi(0)\simeq-0.015\ , & \alpha'_\pi(0)\simeq0.72 \mbox{
GeV}^{-2}; \nonumber\\
& \alpha_\rho(0)\simeq0.50\ , & \alpha'_\rho(0)\simeq0.83 \mbox{
 GeV}^{-2}; \nonumber\\
& \alpha_\eta(0)\simeq-0.24\ , & \alpha'_\eta(0)\simeq0.80 \mbox{
 GeV}^{-2}; \nonumber\\
& \alpha_{a_1}(0)\simeq -0.10\ , & \alpha'_{a_1}(0)\simeq0.72 \mbox{
 GeV}^{-2}; \nonumber\\
& \alpha_{a_2}(0)\simeq0.45\ , & \alpha'_{a_2}(0)\simeq0.91 \mbox{
 GeV}^{-2}; \nonumber\\
& \alpha_{P'}(0)\simeq0.71\ , & \alpha'_{P'}(0)\simeq0.83 \mbox{
 GeV}^{-2}.
\end{eqnarray}
The slopes of the $\alpha'_X(0)$ trajectories are approximately the
same. The inverse slope value, $1/\alpha'_X(0)\simeq1.25\pm0.15\,
$GeV$^2$,
is of the order of the slope  $\mu^2$ for the trajectories on the
$(n,M^2)$-plane:
\beq
\frac1{\alpha'_X(0)}\ \simeq\ \mu^2\ .
\eeq
The daughter trajectories for  $\pi$ (Fig. 3a), $a_1$ (Fig. 3b), $\rho$
(Fig.  3c), $a_2$ (Fig. 3d) and $\eta$ (Fig. 3e) are determined
unambigously.  On the $P'$-trajectories, leading and daughter ones
(Fig. 3f), there is no room for the states
$f_0(300-500)$ and $f_0(1200-1600)$:  this fact stresses once again that
these states are superfluous  for the
 $q\bq$-systematics, i.e. they are exotic states.

\subsection{Exotics --- the states which do not belong to the
quark-antiquark trajectories}

As is  seen, there are two states  in the scalar-isoscalar sector
which do not belong either to the $(n,M^2)$ or
$(J,M^2)$ trajectories; these are the broad state $f_0(1200-1600)$,
which was found in the  $K$-matrix analysis of the
$00^{++}$-wave \cite{K,YF,YF99}, and the light  $\sigma$-meson.

The problem of the light $\sigma$-meson is  discussed rather long ago
\cite{PDG,sigma_dis,sigma80,sigma,sigma_N/D},
yet, until now, there is no common opinion about the existence of
this state.

Recently there appeared indications that a signal from
$\sigma$-meson has been seen in the reactions $D^+\to\pi^+\pi^+\pi^-$
\cite{D+}, $J/\psi\to\pi\pi\omega$ \cite{BES}
and $\tau\to\pi\pi\pi\nu$ \cite{tau}. Nevertheless, one cannot state
that it was a reliable experimental identification of this state,
because in the three-particle spectra, in the region of small
$\pi\pi$-masses, an enhancement of the spectra can occur, due to both
reflected signals from other channels and the rescattering effects
associated with
the resonance production in other channels (triangle
singularity effect \cite{sing_tri,sing_ale}). To single out
these effects  rather large statistics is necessary, comparable with
that for the reactions investigated by Crystal Barrel; however, in the
works \cite{D+,BES,tau} such a level of statistics had not been
reached. Let us stress that the effects
of reflected signals and those of triangle singularities were seen,
 when the
Crysrtal Barrel reactions have been analysed, and for certain reactions
they occurred to be rather important. Indeed, a correct identification
of $f_0(1500)$ in the reactions
$p\bar p\to\pi^0\pi^0\pi^0$, $\pi^0\eta\eta$
\cite{f1500,f1500cb,f1500PR} became
plausible after accounting for the interference of the decay
$f_0(1500)\to\pi\pi,\eta\eta$  and reflected resonance signals
from other channels; the triangle-diagram effects were also studied by
analysing the reactions  $p\bar p\to\pi^0\pi^0\pi^0,\pi^0\eta\eta$, see
\cite{f1500,f1500PR}, but it occurred that they do not affect the
spectra directly, and they may be effectively described by
introducing complex-valued
parameters  for resonance  production. Still,
in certain cases the triangle diagrams cannot be taken into account in
such a simple way: another reaction investigated by Crystal Barrel,
that is,
$p\bar p\to\eta\pi^+\pi^-\pi^+\pi^-$ \cite{nana}, where singular terms
are located near the physical region, provided us with an example of
strongly affected spectra. One can expect noticeable corrections
coming from triangle diagramms also in the reactions
$D^+\to\pi^+\pi^+\pi^-$, $J/\psi\to\pi\pi\omega$ and
 $\tau\to\pi\pi\pi\nu$, for the effect of triangle diagrams is
enhanced just near the small  $\pi\pi$-masses, for more detail see
\cite{sing_tri,sing_ale}.

It is necessary to underline that in the reactions measured by Crystal
Barrel the bumps in the $\pi\pi$-spectra are observed, which may be
considered, owing to incorrect treatment, as an indication to the light
$\sigma$-meson. The reaction $d\bar p(\mbox{at
rest})\to\pi^0\pi^0\pi^-$ can serve us as an example (see \cite{K},
Fig.  5): in the $\pi^0\pi^0$-spectrum,  there is an obvious
enhancement over the phase space in the region
 $M_{\pi^0\pi^0}\sim400-500\,$MeV.  However, the Dalitz-plot analysis
proved that this enhancement is due to the reflected signal from
$\rho(1450)\to\pi^-\pi^0$. The existence of
a reflected signal in the $D^+\to\pi^+\pi^+\pi^-$
 is also possible \cite{sarantsev1}.

To summarize, for the experimental discovery of the light
$\sigma$-meson the data are needed, which
would exceed the present statistics
by one--two orders of value, i.e. it must be the statistics comparable
with that of the Crystal Barrel reactions.

In the paper  \cite{pi2}, the observation of the   resonance $\pi_2(1880)$
 was reported; this state does not belong to
 $(n,M^2)$- and $(J,M^2)$-trajectories.  Being an extra state for the
 $q\bar q$-systematics, the $\pi_2(1880)$ can be the hybrid: the
 $q\bar  qg$-system.

\section{Scalar-meson sector}

In this Section, the results of the $K$-matrix analyses
 performed in \cite{K,YF} for
the scalar sector are subsequently presented.
The sector of scalar mesons is of a particular interest; a
variety of opinions  exist concerning the properties of states
which belong to this sector,  e.g., see
\cite{ufn,ochs,montanet,shakin,klempt,lesniak,FEClose}.
The latest analysis \cite{K} is the most detailed investigation of the
$00^{++}$ wave, where the available data have been used in a full scale.

\subsection{The $K$-matrix analysis of the $(IJ^{PC}=00^{++})$-wave}

In a set of papers \cite{K,YF,YF99,km}, the
$K$-matrix analysis of the waves $IJ^{PC}=00^{++}$, $10^{++}$,
$02^{++}$, $12^{++}$ had been carried out  in the mass range
280--1900~MeV. For these states, the masses and widths of resonances
had been found.
  In the scalar--isoscalar sector the following states are
seen (see Fig. 4):
\begin{eqnarray}
00^{++}: \quad && f_0(980),\ f_0(1300),\ f_0(1500)\ , \nonumber\\
&& f_0 (1200-1600),\ f_0(1750)\ .
\label{1'}
\end{eqnarray}
The $f_0(980)$  is a well-known  resonance, its properties and its
nature are intensively discussed during several decades.
The $f_0(1300)$ resonance is denoted in the compilation \cite{PDG} as
$f_0(1370)$, however its mass following the most accurate
determination, as is stressed above, is near 1300 MeV --- so
the notation $f_0(1300)$ is used here. The $f_0(1500)$ resonance
had been discovered in \cite{f1500,f1500cb,f1500PR}, now
it is a well-established state.  A few years ago, there existed a
strong belief that in the region around 1700 MeV  a
comparatively narrow state, $f_J(1710)$,
was present, with $J=0$ or 2. The
$K$-matrix analysis \cite{K,YF,YF99,km} points to the existence of
$f_0(1750)$, with the width $\Gamma\sim 140-300$ MeV: the uncertainty
in the definition of the width of $f_0(1750)$ is due to a bad knowledge
of the $\pi\pi\pi\pi$ channel in this mass range and, correspondingly,
with two available solutions, with  $\Gamma\sim 140\,$MeV and
$\Gamma\sim 300\,$MeV.

The broad state $f_0(1200-1600)$,
with a half-width $500-900$ MeV, is definitely needed for the $K$-matrix
analysis. In the paper \cite{YF} this  state had been denoted as
$f_0(1530^{+90}_{-250})$: a large error in the definition of the mass
is due to the remoteness of the pole from the real axis (physical
region) as well as to the existence of several solutions given by the
$K$-matrix analysis.

A large number of reactions, which were succefully described with the
 $f_0(1200-1600)$, proved the valididty of a factorization
inherent in the resonance amplitude: near the pole the amplitude is
$g_{in}(s-M^2)^{-1}g_{out}$, where the universal coupling constants,
$g_{in}$ and $g_{out}$, depend on the type of the intial and final
states only. A strong production of the $f_0(1200-1600)$ in various
processes allows one to fix reliably these coupling constants.

In the scalar--isovector sector, the $K$-matrix analysis points to the
presence of two resonances:
\beq
10^{++}:\quad a_0(980),\ a_0(1520)\ ,
\label{2'}
\eeq
while in the tensor-meson sector the following states are seen:
\begin{eqnarray}
12^{++}\ : \quad && a_2(1320),\ a_2(1660)\ ; \nonumber\\
02^{++}\ :\quad  && f_2(1270),\ f_2(1525),\ f_2(1580)\ .
\label{3'}
\end{eqnarray}

\subsubsection{$K$-matrix amplitude}

In the $K$-matrix analysis, the fitting parameters are the $K$-matrix
elements which are represented as the sums of pole terms
$g^{(n)}_ag^{(n)}_b/(\mu^2_n-s)$ ($s\equiv M^2$ is the invariant
energy squared of  mesons,  $s\equiv M^2$)
 and a smooth $s$-dependent term
$f_{ab}(s)$. Namely,
\beq
K_{ab}\ =\ \sum \limits_n \frac{g^{(n)}_ag^{(n)}_b}{\mu^2_n-s}+f_{ab}(s)\ ,
\label{4'}
\eeq
where fitting parameters are $g^{(n)}_a$, $\mu_n$ and $f_{ab}$,
and the indices $a,b$ refer to the reaction channels:
$f_0\to \pi\pi, K\bar K, \eta\eta,\eta\eta'$ and $\pi\pi\pi\pi$.
 The
$K$-matrix poles are not the amplitude poles, i.e. they do not
correspond to real states. The amplitude in the $K$-matrix approach
is written as follows:
\beq
\wh A\ =\ \frac{\wh K}{1-i\hat\rho\wh K}\ ,
\label{6'}
\eeq
where $\hat\rho$ is the diagonal matrix of phase spaces for the
processes under
consideration,\\
$\hat\rho={\rm diag}\left(\rho_1(s),\rho_2(s),...\right)$.
Because of that, the amplitude poles correspond to the
zeros of the  determinant:
\beq
\det\left|1-i\hat\rho\wh K\right|\ =\ 0\ ,
\label{6''}
\eeq
while the $K$-matrix poles respond to the states with switched-off
decay channels. The states associated with the $K$-matrix poles
do not contain a cloud of real mesons which appear during resonance
decay: this circumastance allows one to call the $K$-matrix poles the
"bare states" \cite{YF,ufn,km,APS}.

\subsubsection{Partial amplitude for the $00^{++}$ wave:  unitarity,
analyticity and the problem of the left-hand cut}

The $K$-matrix representation of partial amplitude takes account of
the well-known fact that at low and moderate energies the inelastic
processes are dominantly two-particle ones.
Being applied to the $00^{++}$ wave
this means that, along with the elastic scattering
$\pi\pi\to\pi\pi$ (threshold at $\sqrt s=280\,$MeV), we have the
transitions $\pi\pi\to K\bar K$ (threshold at $\sqrt s=2m_K$),
$\pi\pi\to\eta\eta$ and $\pi\pi\to\eta\eta'$ (thresholds at $\sqrt
s=2m_\eta$ and $\sqrt s=m_\eta+m_{\eta'}$). Besides, in the
$00^{++}$-wave¥ at $\sqrt s\ga1300-1400\,$MeV, a
considerable four-pion production is observed, but this process can be
also treated, with a good accuracy, as a formation of  two
$\rho$'s or two effective $\sigma$-mesons:
$\pi\pi\to\rho\rho\to\pi\pi\pi\pi$ and
$\pi\pi\to\sigma\sigma\to\pi\pi\pi\pi$. The amplitude in the $K$-matrix
representation takes into account correctly both unitarity and
threshold singularities  for the two-meson processes. In this way,
the $K$-matrix representation of the amplitude serves us as a right
frame for the correct reconstruction of the amplitude above the $\pi\pi$
threshold (in our case, over the range 280--1900 MeV).

In the $K$-matrix amplitude, the threshold singularities are taken into
account by treating the phase spaces as analytical functions of
the total energy squared  $s$: above the threshold, the two-particle
phase space can be represented in a standard form
 $\rho_a(s)=\sqrt{1-(m_{1a}+m_{2a})^2/s}$, and below threshold it
should be expressed through analytical continuation:
$\rho_a(s)=i\sqrt{(m_{1a}+m_{2a})^2/s-1}$. The phase space of the
two-resonance state, of the type of $\rho\rho$  or $\sigma\sigma$, may
be expressed in the form which reproduces correctly the threshold
singularities, namely,  threshold singularity at
$\sqrt s=4m_\pi$ and the singularity in the lower complex half-plane
 $s$ (on the third sheet) at $\sqrt
s=2(m_\rho-i\Gamma_\rho/2)$, that is related to the production of
$\rho\rho$, see  \cite{K,YF,ufn} for more detail.

The singularities,  which are not  explicitly taken into account
within the $K$-matrix approach, are the so-called left-hand side
singularities of the partial amplitude.
These singularities are due to the exchange of particles in the crossing
channels (i.e. in the $t$- and $u$-channels), they determine the
interaction forces. In the $\pi\pi$-scattering amplitude, the nearest
left singularity is located at $s=0$; this singularity
is associated with the two-pion exchanges in the  $t$- and $u$-channels.
A large contribution is provided by the
$\rho$-meson exchange, that leads to the logarithmic singularity
of the partial amplitude at
$s=4m^2_\pi-m^2_\rho\simeq-0.5\,\rm GeV^2$. In this very region,
the two-pion exchange in the $00^{++}$ wave, that corresponds to
the effective $\sigma$-meson, contributes significantly too.
The contribution of tensor-meson exchanges
is important in the region  $s\simeq-1.5\,\rm GeV^2$.

The problem of a correct account for left-hand singularities becomes
even more complicated, for the contributions from various exchanges
cancel each other at small $s$ to a great extent
\cite{polyakov}, so the contribution from the
left-hand cut depends strongly on the details of the $t$- and
$u$-channel exchange mechanism, in particular on the structure of the
 $\rho,\sigma,f_2\to\pi\pi$  form factors.

Because of uncertainties in choosing the interaction forces,
it would be reasonable not to inflict
any hypothesis about the left-hand cut for the
$\pi\pi$-scattering amplitude,
but to allow a freedom for the fitting procedure. The
$K$-matrix technique, which was used in \cite{K,YF}, allows one to do
that: this technique may be easily generalized in a way that enables to
take account of the left-hand amplitude singularities.
This opportunity was used in \cite{K,YF}, according to what the data
require; this is an important item, let us discuss it in more detail.

By discussing the left-hand cut of the  $00^{++}$-amplitude, it is
sufficient to restrain our consideration by  one, the nearest,
$\pi\pi$   channel
only. In this case, partial amplitude
can be written as follows:
\beq A(s)\
=\ \frac{N(s)}{1-B(s)}\ ,
\eeq
where the pion loop diagram, which ensures the unitarity of
the partial amplitude, is
\beq B(s)\ =\
\int\limits^\infty_{4m^2_\pi} \frac{ds'}\pi\
\frac{N(s')\rho_{\pi\pi}(s')}{s'-s-i0}\ .
\label{312.2}
\eeq
At $s>4m^2_\pi$, the imaginary part of the loop diagram is
Im$\,B(s)=
\rho_{\pi\pi}(s)N(s)$ (half-residue in the pole $s'=s$), and
the real part is the principal value of the integral (\ref{312.2}). The real
part, Re$\,B(s)$, does not contain the
$\pi\pi$-threshold singularity; this singularity is defined by
the phase space
factor in the imaginary part,
$\rho_{\pi\pi}(s)$. So we may  re-write the
amplitude  $A(s)$ in the form
\beq A(s)\ =\
\frac{K(s)}{1-i\rho_{\pi\pi}(s)K(s)}\ ,
\eeq
where in the right-hand half-plane $s>0$ the $K$-matrix block
\beq
K(s)\ =\ \frac{N(s)}{1-\mbox{Re }B(s)}\ ,
\eeq
 may have the pole
singularities only,
and in the left-hand one,
$s\le0$, it contains a set of left-hand singularities.
By singling out the left-hand side singularities in the explicit form,
one can write the $K$-matrix block as follows:
\beq
K(s)\ =\ \sum_i\frac{g^2_i}{m^2_i-s}+f(s)\ ,
\eeq
where the pole positions,
$m^2_i$, are given by the equalities:
\beq
[1-\mbox{Re }B(s)]_{s=m^2_i}\ =\ 0\ ,
\eeq
and $f(s)$ contains all left-hand singularities:
\beq
f(s)\ =\ \frac{N(s)}{1-\mbox{Re }B(s)}-\sum_i
\frac{g^2_i}{m^2_i-s}\ =\int\limits^{-\infty}_{s_L} \frac{ds'}\pi\
\frac{\mbox{disc }f(s')}{s'-s}\ .
\label{left}
\eeq
The magnitude $s_L$ defines the location of the closest left-hand
singularity; for the considered $\pi\pi$-scattering amplitude,
$s_L=0$.

In the physical region, a suitable approximation for the amplitude
  is given by  substituting  the spectral integral
(\ref{left}) with the sum:
\beq
\int\limits^{-\infty}_{s_L}\frac{ds'}\pi\ \frac{\mbox{disc
}f(s')}{s'-s}\ \longrightarrow\ \sum_n \frac{f_n}{s+s_n}\ ,
\eeq
where $s_n>-s_L$. In the multichannel analyses
\cite{K,YF}, the approximation had been used, when the left-hand cut in
the $K$-matrix terms $K_{ab}$ was fitted to the one-pole term as
follows:
\beq
f_{ab}(s)\ \to\ \frac{f_{ab}}{s+s_0}\ .
\eeq
The
parameter $s_0$ in different variants of the fit \cite{K} appeared
in the interval $0.5\la s_0\la1.5\,\rm GeV^2$, that points to a large
contribution of the $t$- and $u$-channel $f_2$, $\rho$ and $\sigma$
exchanges.  The use of the two-pole approximation for the multichannel
 $K$-matrix deteriorates the convergence of the fitting
procedure, for the available experimental data are insufficient to
fix unambigously a large number of the left-hand-cut parameters.

More fruitful for the reconstruction of analytical structure of
the amplitude near $s\sim 0$ is
the fitting to the  $00^{++}$ wave in the region
$s<1000$ MeV, when one may restrict oneself by the one-channel
approximation. In Section 3.8, we present more detailed narration
of the study of  analytical structure of the
amplitude at the region $s\sim 0-4m_\pi^2$,  with the restoration of
the left-hand singularities obeying equations (20) and (21); this topic
is related to the light $\sigma$-meson problem.

\subsubsection{Three-meson production in the reactions of the
$p\bar p$ and  $n\bar p$ annihilation}

The $K$-matrix representation of the amplitude may be applied to
the description of the production of resonances in the three or more
particle production. The uniformity of the amplitude representation is
important in the combined analysis of the two-particle, like
 $\pi\pi\to \pi\pi,K\bar K $, and multiparticle, like
$p\bar p \to \pi\pi\pi,\pi\eta\eta,\pi K\bar K$,
reactions. The $K$-matrix approach to multiparticle processes is based
on the fact that the denominator of the $K$-matrix two-particle
amplitude (\ref{6'}), $[1-\hat \rho \hat K]^{-1}$, describes pair
interactions of mesons in the final state as well.

Let us clarify this statement using as an example the amplitude of the
$p\bar p$ annihilation from the $^1S_0$ level:
$p\bar p(^1S_0)\to three\, mesons$. Let the produced mesons be labelled
by the indices 1,2,3; then the production amplitude for the resonance
with the spin $J=0$ in the channel (1,2) is as follows:
\beq
A_3(s_{12})= K^{(prompt)}_3(s_{12})[1-i\rho_{12}
\widehat K_{12}(s_{12})]^{-1},
\label{subsub1}
\eeq
where the matrix factor $[1-i\hat \rho_{12}\widehat
K_{12}(s_{12})]^{-1}$ depends on the invariant energy squared of
the  mesons 1 and 2 and it coincides with matrix factor of the
two-particle amplitude (\ref{6'}). The factor
$\widehat K^{(prompt)}_3(s_{12})$ stands for the prompt production
of particles 1,2 and resonances in this channel:
\beq
\left (K^{(prompt)}_3(s_{12}) \right )_{ab}  =\ \sum \limits_n
\frac{\Lambda^{(n)}_a g^{(n)}_b}{\mu^2_n-s_{12}}+
\varphi_{ab}(s_{12})\ ,
\label{subsub2}
\eeq
where $\Lambda^{(n)}_a$ and $\varphi_{ab}$ are the parameters of the
prompt-production amplitude, and $g_b^{(n)}$ and $\mu_n$  are defined
by the two-meson scattering amplitude, see (\ref{6'}).

It is also appropriate to mention that the description of the
two-particle interactions by using the $K$-matrix factor
$(1-\hat \rho \hat K)^{-1}$ is nothing else than a generalization
of the well-known Watson--Migdal formula used for the two-nucleon
interactions at low energies in nuclear reactions with multiple
production of nucleons \cite{W-M}.

The whole amplitude for the production of the $(J=0)$-resonances is
defined by the sum of  contributions from  all channels:
\beq
A_3(s_{12})+A_2(s_{13})+A_1(s_{23})
\label{subsub3}
\eeq
To account for resonances with the
nonzero spin $J$ one needs to substitute in
(\ref{subsub1}):
\beq
A_3(s_{12}) \to \sum\limits_J A_3^{(J)}(s_{12})
X_{\mu_1\mu_2...\mu_J}^{(J)}(k_{12}^\perp)
X_{\mu_1\mu_2...\mu_J}^{(J)}(k_{3}^\perp),
\label{subsub4}
\eeq
where the $K$-matrix amplitude $A^{(J)}_3(s_{12})$ is
determined by the expression similar to  (\ref{subsub1}), and
$X^{(J)}_{\mu_1 \mu_2...\mu_J}$ stands for meson states
with the angular momentum $J$.
The angular-momentum operators depend on the perpendicular components
of meson relative momenta: $k^\perp_{12}$ and $k^\perp_3$. Here
$k^\perp_{12}$ is the component of relative momentum of particles
1 and  2, $k_{12}=(k_1-k_2)/2$, which is orthogonal to the total
momentum of particles 1 and 2, $p_{12} =(k_1+k_2)$, namely,
$(k^\perp_{12}p_{12})=0$; likewise, $(k^\perp_{3}p)=0$, where
$p=k_1+k_2+k_3$.

For the lowest angular momenta, the operators
$X^{(J)}_{\mu_1 \mu_2...\mu_J}$
can be easily written; for example, for $J=1,2$ we have, up to the
normalization factor, $X^{(1)}_{\mu }(k^\perp)\sim k^\perp_\mu$ and
$X^{(2)}_{\mu_1 \mu_2}(k^\perp)\sim (k^\perp_{\mu_1}k^\perp_{\mu_2}
-\frac13 k^{\perp 2} g^\perp_{\mu_1\mu_2})$, where
$g^\perp_{\mu_1\mu_2}$ is
the metric tensor in the space orthogonal to
the total momentum. The construction of operators for arbitrary
 $J$ may be found in \cite{operator}.

The covariant operators $X^{(J)}_{\mu_1 \mu_2...\mu_J}$
are determined in the four-momentum space; this is a  relativistic
generalization of the 3-dimensional angular-momentum
operators of  Zemach  \cite{Zemach}.
The use of the 4-dimentional operators is
suitable for the analysis of multiparticle final states, for in this
case one does not need  numerous  Lorentz boosts, which are necessary
in the Zemach technique or by using spherical functions.

The amplitude expansion with respect to the states with different
angular momenta by using relativistic covariant operators has been
carried out for all the reactions
 $p\bar p\to three\, mesons$, in the analysis of  which the PNPI group
took part \cite{f1500,f1500cb,f1500PR,a0cb,ral,K}. Such an expansion,
apart from being simple and compact, as was mentioned above, has an
advantage of taking correct account of kinematical factors, which are
significant for the reconstruction of the correct behaviour of the
threshold singularities of multiparticle amplitude. In the other
orbital-moment decomposition techniques this is a subject of a special
care; in particular, in \cite{chung} a special method was suggested to
take account of kinematical factors  by using the expansion with
respect to spherical functions.

In the paper \cite{operator}, the operators are also constructed  for
spins and total angular momenta in case of fermion and photon systems,
which are used for the analysis of the reactions
$p\bar p \to mesons$ and $\gamma\gamma\to mesons$.

The formula (\ref{subsub1}) serves us for singling out the pole
singularities of the amplitude $p\bar p(^1S_0)\to three\,
mesons$, which are the leading ones.  The next-to-leading logarithmic
singularities are related to the rescattering of mesons produced by the
decaying resonances  (that is, triangle-diagram singularities
\cite{sing_tri}).  The analysis performed in
\cite{f1500,f1500PR} showed that in the reactions $p\bar p$(at\,
rest)$\to \pi^0\pi^0\pi^0$, $\pi^0\pi^0\eta$, $\pi^0\eta\eta$, to
determine  parameters of a  resonance produced in the two-meson
channels one may not take into consideration explicitly the triangle
diagram singularities --- it is important to account only for the
complex-valuedness
 of the prompt production  amplitude, that is due to final-state
interactions (that is, the complexity of parameters
 $\Lambda^{(n)}_a $ and $\varphi_{ab}$ in
(\ref{subsub2})).  Note that it is not a universal rule for the meson
production processes in the $p\bar p$ annihilation --- for
example, in the reaction
$p\bar p \to \eta\pi^+\pi^-\pi^+\pi^-$ \cite{nana},
the explicit form of the triangle singularity is important.

A complete account for unitarity and analyticity in the three-meson
production amplitude is related to the
consideration of a full set of meson interactions in the final
state. In the reactions $p\bar p$(at\, rest)$ \to \pi\pi\pi$,
$\pi\pi\eta$, $\pi\eta\eta$, $\pi K\bar K$,  the connected system of
the dispersion-relation $N/D$ equations, with all pair-meson equations
accounted for, has been written in  \cite{alex-3}. The three-meson
 production amplitudes being related to each other by the $N/D$
equations leave less freedom for the fitting than formula
(\ref{subsub1}) and, in principle, they provide more information about
meson--meson amplitudes.  However, the fitting  on the basis of the
$N/D$ equations is much more complicated procedure than the  $K$-matrix
analysis.

\subsubsection{Peripheral two-meson production in meson--nucleon
collisions at high energies}

The two-meson production reactions $\pi p\to\pi\pi n$, $K\bar Kn$,
$\eta\eta n$, $\eta\eta'n$ at high energies and small momentum
transfers to the nucleon, $t$, provide us with a direct information
about the amplitudes $\pi\pi\to\pi\pi$, $K\bar K$, $\eta\eta$,
$\eta\eta'$, for at  $|t|<0.2\,(\rm GeV/c)^2$
the reggeized $\pi$ exchange dominates the production reactions.
At larger $|t|$, the change of the regime occurs:
at $|t|\ga0.2\,(\rm GeV/c)^2$ a significant contribution of
other reggeons become plausible ($a_1$-exchange, daughter
$\pi$- and $a_1$-exchanges).  Despite a vague knowledge of the exchange
structure, the study of the two-meson production processes at
$|t|\sim0.5-1.5\,(\rm GeV/c)^2$ looks rather attractive, for at
such momentum transfers the broad resonance vanishes, so the production
of the $f_0(980)$ and $f_0(1300)$ appears practically without
background, that is important for finding parameters of these
 resonances.

The amplitude of the peripheral production of two
$S$-wave mesons reads:
\beq
(\bar\psi_N\hat0_R\psi_N)R(s_{\pi N},t)\widehat K_{\pi R(t)}
\left[1-\hat\rho\widehat K\right]^{-1}\ ,
\label{periph}
\eeq
where the factor
 $(\bar\psi_N\hat0_R\psi_N)$ stands for the reggeon--nucleon
vertex, and $\hat0_R$ is the spin operator; $R(s_{\pi
N},t)$ is the reggeon propagator depending on the total energy squared
of colliding particles, $s_{\pi N}$, and momentum transfer squared $t$,
and the factor $\widehat K_{\pi R(t)}[1-i\hat\rho\widehat K]^{-1}$ is
related to the block of the two-meson production --- one may find the
detailed description of the  amplitude (\ref{periph}) in
\cite{K,bnl-gams}.

The factor
$\widehat K_{\pi R(t)}[1-i\hat\rho\widehat K]^{-1}$ describes
the transitions $\pi R(t)\to\pi\pi$, $K\bar K$, $\eta\eta$, $\eta\eta'$:
the block $\widehat K_{\pi R(t)}$
is associated with the prompt meson production §®­®¢, and
$[1-i\hat\rho\widehat K]^{-1}$ is a standard factor for meson rescattrings,
see  (\ref{6'}). The prompt-production block is parametrized
as follows:
\beq
\left(\widehat K_{\pi R(t)}\right)_{\pi R,b}\ =\ \sum_n
\frac{G^{(n)}_{\pi R}(t)g^{(n)}_b}{\mu^2_n-s}+f_{\pi R,b}(t,s)\ ,
\eeq
where $G^{(n)}_{\pi R}(t)$ is the $f_0$ production vertex,
and $f_{\pi R,b}$  stands for the background production of mesons,
while the parameters $g^{(n)}_b$ and  $\mu_n$ are the same as in the
transition amplitude $\pi\pi\to\pi\pi$, $K\bar K$, $\eta\eta$,
$\eta\eta'$, see  (\ref{6'}).

At the early stage of the analysis of the $S$-wave two-pion production,
 $\pi^-p\to(\pi\pi)_S\,n$, the mechanism of the reggeized-$\pi$ exchange
was suggested both at small and moderately large $|t|$
\cite{PL95}. The change of the regime at  $|t|\sim0.2-0.4\,(\rm
GeV/c)^2$ has been described by including the
effective $\pi {\bf \rm P}$-exchange ($\bf\rm P$ is the pomeron):
the amplitude of the $\pi {\bf\rm P}$-exchange has another sign than the
$\pi$-exchange, thus leading to the amplitude zero and,
correspondingly, to a zero
in the  $t$-distribution at $|t|\sim0.3\,(\rm GeV/c)^2$,
where the regime changes. The filling-in of the dips  in the
$t$-distributions at  $|t|\sim0.3\,(\rm GeV/c)^2$  can appear due to
exchanges of other reggeons, namely, reggeized $a_1$-exchange and
contributions from the daughter
trajectories, $\pi_{(daughter)}$ and $a_{1(daughter)}$. Practically,
all these terms do not   interfere in the reaction
$\pi^-p\to(\pi\pi)_S\,n$: the $\pi$- and $a_1$-exchanges do not
interfere due to different spin structures in
reggeon--nucleon vertices,
($\hat0_\pi\sim\mbox{\boldmath$\sigma$}_\perp$ and $\hat0_{a_1}
\sim\mbox{\boldmath$\sigma$}_{\|}$, where {\boldmath$\sigma$}$_\perp$
and {\boldmath$\sigma$}$_{\|}$ are the transverse and longitudinal
components of the Pauli matrix operating in the spin space of the
nucleon). At the same time
 the contributions of leading and daughter trajectories
do not interfere practically
due to a phase shift in reggeon propagators
(for example, the propagator of
the leading  $\pi$-trajectory may be considered, with
a good accuracy, as real magnitude, while the propagator of the
$\pi_{(daughter)}$-trajectory is nearly imaginary), see
\cite{bnl-gams} for more detail.

 The $a_1$ exchange for the reaction $\pi p\to(\pi\pi)_S\,n$ was
considered in \cite{YF99,achasov}. In \cite{YF99}, the calculations of
the  $\pi\pi$ spectra were performed both with the
$a_1$ exchange and within an effective $\pi$ exchange. As a result,
it occurred that
the $a_1$ exchange affects weakly the parameters of the
 $f_0(980)$ and $f_0(1300)$: the matter is that the $\pi\pi$ spectra
measured by
GAMS group \cite{GAMS} were averaged over large $t$-intervals,
thence the details of the  $t$-distributions are not
significant for fixing  resonance parameters.

New data on the production of the $\pi\pi$ system at
$|t|\le1.5\,\rm(GeV/c)^2$ \cite{BNL-new} gave rise to the discussion
about the role of the $t$ exchange in the definition of
parameters of the $f_0(980)$ and
 $f_0(1300)$.  The data of the E852 Collaboration obtained at lower
energies $(p_{lab}=18\,\rm GeV/c)$ as compared to the GAMS energy
 $(p_{lab}=38\,\rm GeV/c)$ point definitely to the fact that
the description of the peripheral pion production
$\pi^-p\to(\pi\pi)_Sn$ in terms of the leading $t$ exchanges, $\pi$ and
 $a_1$, is not complete:  the alteration of spectra with energy in the
region $|t|\sim0.3-0.4\,\rm(GeV/c)^2$ and
$M_{\pi\pi}\sim1300\,$MeV proved a considerable weight of daughter
trajectories: $\pi_{(daughter)}$ and/or $a_{1(daughter)}$.

Combined analysis of the $(\pi\pi)_S$ spectra by  GAMS
\cite{GAMS} and E852 \cite{BNL-new} at $|t|\le1.5\,\rm(GeV/c)^2$ has
been performed in \cite{bnl-gams}. The analysis \cite{bnl-gams}
showed that, though the data do not allow us to find out the
$t$ exchange  mechanism unambigously,  this circumstance affects
weakly the definition of parameters of the produced
$f_0(980)$ and $f_0(1300)$ resonances:  in all variants of the fit
(including
 various combinations of the  $\pi_{(leading)}$,
$a_{1(leading)}$, $\pi_{(daughter)}$, $a_{1(daughter)}$
exchanges, the consideration of the effective $\pi P$ and $a_1P$
exchanges at  $|t|\ga0.2\,\rm(GeV/c)^2$, with
$\bf {\bf\rm P}$ being the pomeron, or the Orear mechanism \cite{orear})
the parameters of $f_0(980)$ and $f_0(1300)$ are almost the same.
This circumstance allows us to restrain
ourselves in the $K$-matris fit \cite{K}
by  the two variants of the $t$-channel mechanism, namely,
$\pi_{(leading)}$, $a_{1(leading)}$, $\pi_{(daughter)}$ or
$\pi_{(leading)}$, $a_{1(leading)}$, $a_{1(daughter)}$.

An opposite statement, according to which the choice of the
$t$-exchange mechanism influences signficantly the definition of the
parameters of $f_0(980)$ and $f_0(1300)$, was claimed in
 \cite{achasov1} (though without performing the fitting to resonance
parameters). However, the analysis \cite{bnl-gams} does not confirm
the statements of  \cite{achasov1}.

\subsection{Classification of scalar bare states}

The systematics of scalar bare states has been carried out in
\cite{nonet}, where bare $K$-mesons were found which are
needed to fix two quark-antiquark nonets,
 $1^3P_0q\bq$ and
$2^3P_0q\bq$.
The $q\bq$ nonet contains two scalar--isoscalar states,
$f^{bare}_0(1)$ and $f^{bare}_0(2)$, scalar--isovector meson
$a^{bare}_0$ and scalar kaon $K^{bare}_0$.  The decay couplings
to pseudoscalar mesons for  these four states,
\begin{eqnarray}
&& f^{bare}_0(1),\ f^{bare}_0(2)\ \to\ \pi\pi,K\bar K, \eta\eta,
\eta\eta'\ , \nonumber\\ && a^{bare}_0\ \to\ \pi\eta,\ K\bar K\ ,
\nonumber\\
&& K^{bare}_0\ \to\ \pi K,\ \eta K \ ,
\label{7'}
\end{eqnarray}
are determined, in the leading-order terms of the $1/N$-expansion
\cite{t'Hooft}, by three parameters only. They are the common decay
coupling $g$, parameter $\lambda$ for the  probability
to produce  strange quarks in the
decay process (in the limit of a precise $SU(3)_{flavour}$ symmetry
we have $\lambda=1$) and mixing angle for
$s\bar s$ and $n\bar n=(u\bar u+d\bar d)/\sqrt2$ components in the
$f_0$-mesons,
$\psi_{flavour}(f_0) =n\bar n \cos\varphi+s\bar s\sin\varphi_i$.
For the nonet partners,
$\varphi[f_0^{bare}(1)]-\varphi[f_0^{bare}(2)]=90^\circ$.

The rigid constraints on the decay couplings of bare states (\ref{7'})
are imposed by the quark-combinatorics relations. The rules of quark
combinatorics were first suggested for the high-energy hadron
production \cite{as-bf}
and then extended for hadronic $J/\Psi$ decays
\cite{wol}.  The quark-combinatorics relations were used for the decay
couplings of the scalar-isoscalar states in the analysis of the
quark-gluonium content of resonances in \cite{glp} and later on in
the $K$-matrix analyses \cite{K,YF,YF99,ufn,km,nonet}.

These
constraints being imposed on
the decay couplings (\ref{7'}) provide us with an opportunity
to fix unambigously the states belonging to the basic nonet
\cite{nonet}, also see \cite{K,YF}:
\begin{eqnarray} 1^3P_0q\bq\ : \quad &&
 f^{bare}_0(700\pm100),\ f^{bare}_0(1220\pm40)\ , \nonumber \label{8'}
  \\ \quad && a^{bare}_0(960\pm30),\
K^{bare}_0 \left(1220^{+50}_{-150}\right)\ ,
\end{eqnarray}
as well as the mixing angle for $f^{bare}_0(700)$  and
$f_0^{bare}(1220)$,
\begin{eqnarray}
\varphi[f_0^{bare}(700)]\ &=&\ -70^\circ \pm 10^\circ \ ,
\\ \nonumber
\varphi[f_0^{bare}(1220)]\ &=&\ 20^\circ \pm 10^\circ \ ,
\label{9'}
\end{eqnarray}
To establish the nonet of the first radial
excitation, $2^3P_0q\bq$, is more complicated task.
The $K$-matrix analysis gives us two scalar-isoscalar
bare states at 1200--1650 MeV,
$f^{bare}_0(1230\pm40)$ and $f_0^{bare}(1580\pm40)$, whose decay
couplings (\ref{7'}) obey the relations appropriate to the
glueball. The matter is that the relations between couplings for
the glueball decay, $glueball \to \pi\pi,
K\bar K, \eta\eta, \eta\eta'$, and the decay couplings of the
quark-antiquark flavour singlet, $(q\bar q)_{singlet} \to \pi\pi, K\bar
K, \eta\eta, \eta\eta'$, are almost the same (they are exactly the same
in the limit $\lambda=1$). Because of that, it is impossible, by using
hadronic decay couplings only, to distinguish between the glueball and
 flavour singlet.

Systematics of bare  $q\bq$-states on the $(n,M^2)$-plane helps us
to resolve the dilemma which one of these states is the glueball.
This systematics (which is discussed in more detail in Section
3.8) definitely tells us that the state $f_0^{bare}(1580\pm40)$
not being on the $q\bq$-trajectory is an extra one, so
furthermore we accept
that this state is the glueball. Indeed,
\beq
0^{++}\ {\rm glueball}: \quad f^{bare}_0(1580\pm40)\ .
\label{10'}
\eeq
Lattice calculations \cite{lattice} agree with this statement:
gluodynamical glueball should be in the mass range 1550--1750 MeV.

Having  accepted the $f_0^{bare}(1580\pm40)$ to be non-$q\bq$-state,
we construct the nonet $2^3P_0q\bq$ in a unique way:
\begin{eqnarray}
 2^3P_0q\bq\ : \quad &&  f^{bare}_0 (1230\pm40),\
f^{bare}_0(1800\pm30)\ , \nonumber
 \label{11'} \\
 \quad && a^{bare}_0(1650\pm50),\ K^{bare}_0\left(1885^{+50}_{-100}
\right)\ .
\end{eqnarray}

The kaonic bare states were defined in the $K$-matrix analysis of the
$\pi K$-spectrum performed in \cite{nonet} on the basis of data
\cite{Kpi}.  This analysis provided us with a few solutions. In
(\ref{8'}) and (\ref{11'}) the values of
$K^{bare}_0(1220^{+50}_{-150})$ and $K^{bare}_0(1885^{+50}_{-100})$
were  used,
which were the average ones for these solutions: a large error in the
determination of masses of bare states resulted from the variaty of
mass values coming from different solutions.

After switching on the decay channels, the bare states  (\ref{8'}),
(\ref{10'}), (\ref{11'}) turn into real resonances. For
scalar--isoscalar states we have,
after  the decay onset,  the transformation as follows:
\begin{eqnarray}
&&
f^{bare}_0(700\pm100)\ \longrightarrow\ f_0(980)\ , \nonumber\\
&& f^{bare}_0(1220\pm40)\ \longrightarrow\ f_0(1300)\ , \nonumber\\
&& f^{bare}_0(1230\pm40)\ \longrightarrow\ f_0(1500)\ ,
\nonumber\\
&& f^{bare}_0(1580\pm40)\ \longrightarrow\ f_0(1200-1600)\ ,
\nonumber\\
&& f^{bare}_0(1800\pm40)\ \longrightarrow\ f_0(1750)\ .
\label{12'}
\end{eqnarray}
This evolution of states is illustrated by Fig. 5, where
the shift of amplitude poles
into comlex plane is
shown depending on gradual onset of the decay channels.
Technically, it is not difficult to switch on/off the decay channels
for the  $K$-matrix amplitude:
one should substitute in
the $K$-matrix elements  (\ref{4'}):
\beq
g^{(n)}_a\ \to\ \xi_n(x)g^{(n)}_a\ , \qquad f_{ab}\ \to\
\xi_f(x)f_{ab}\ ,
\label{13'}
\eeq
where the parameter-functions for  switching on/off the decay
channels, $\xi_n(x)$ and $\xi_f(x)$,  satisfy the
following  constraints:
$\xi_n(0)=\xi_f(0)=0$ and $\xi_n(1)=\xi_f(1)=1$, and $x$ varies
in the interval $0\le x\le1$. Then, at $x=0$, the amplitude $\wh A$
turns into the $K$-matrix,
 $\wh A(x\to0)\to\wh K$, and the amplitude poles occur on the
real axis, that corresponds to the stable $f^{bare}_0$-states. At $x=1$,
we deal with real resonance; varying $x$ from $x=0$ to $x=1$ we
observe the shift of poles into the complex  $M$-plane.

The $x$-dependence of $\xi_n(x)$ and $\xi_f(x)$
is governed by
the dynamics of the state mixing, and the $K$-matrix solution does not
clear up this dynamics. In \cite{aas-z}, the mixing has been modelled
within the two-component approach for the decay processes, when the
resonance widths are due to the transitions $f_0\to q\bar q$ and
$f_0\to gg$.  This model proved that just the glueball accumulated the
widths of neighbouring $q\bar q$ states. The dominant accumulation of
widths by the glueball occurs by virtue of two reasons. First, mutual
transitions {\it $q\bar q$-mesons} $\leftrightarrow glueball$ are not
suppressed within the $1/N$ expansion rules, e.g. see \cite{ufn}.
Second, the orthogonality of the $q\bar q$-states suppresses direct
mixing of the $q\bar q$-mesons.

\subsection{Ovelapping $f_0$-resonances in the region \newline
            1200--1700~MeV:  the accumulation of widths of
            quark-antiquark states by the glueball}

The formation of the broad state is not an accidental phenomenon. The
broad state appeared as a result of  mixture of  resonances, due to the
transitions
$f_0(1) \to real\; mesons \to f_0(2)$. Such transitions in case of
overlapping resonances, result in a specific phenomenon:
when several resonances with common decay channels overlap,
one of them accumulates the widths of neighbouring  resonances. So a
broad resonance appears together with several narrow ones.

Initially, the effect of accumulation of widths had been discussed in
nuclear physics \cite{Shapiro,Okun,Stodolsky}.  Concerning the scalar
$00^{++}$-mesons, the accumulation of widths by one of overlapping
resonances was observed in
\cite{km,APS,glueball}. In \cite{glueball,glueball_model},
the following scheme has been suggested: the broad state
$f_0(1200-1600)$ is the glueball descendant; this state was formed
because of
the glueball mixing with neighbouring  $q\bq$-states, that was
accompanied by the accumulation of widths of neighbouring states by the
glueball descendant.  As a result, comparatively narrow states
$f_0(1300)$, $f_0(1500)$, $f_0(1750)$ have had considerable admixtures
of the glueball component, while the broad state got a large
$q\bq$-component.
The quark-antiquark component in the glueball should be close
to the flavour singlet, namely \cite{alexei-glu}:
\beq
(q\bar q)_{glueball}=(u\bar u+d\bar
d+\sqrt{\lambda}s\bar s)/ \sqrt{2+\lambda},
\eeq
where the parameter $\lambda$ is nearly the same as that in the decay
processes, $\lambda\simeq 0.5-0.8$.

In meson physics, the accumulation of widths can play a decisive role
for the destiny of exotic states which are beyond the
$q\bq$ systematics. Indeed, the exotic state, after
appearing in a set of $q\bq$-states, creates a group of
overlapping resonances. If the transition of the exotic state
into $q\bq$-state is not suppressed (as in case of the glueball,
where, according to the
$1/N$-expansion rules \cite{t'Hooft}, the transition $glueball\to q\bar
q$-{\it meson} is allowed in the leading-order terms \cite{ufn,ufn95}),
then just the exotic meson has an advantage to accumulate the
widths:  the wave functions of neighbouring $q\bq$-states are
orthogonal to each other but not to the exotic state.
 Therefore, the
existence of the broad state together with comparatively narrow ones
must serve as a signature  of  exotics in this mass region
\cite{PR-exot}.

The broad state may play rather constructive role in the formation of
the confinement barrier. The broad state, after accumulating the
widths
of resonances--neighbours
in the mass scale, plays the role of a
locking state. The evaluation of radii of the broad state
$f_0(1200-1600)$ and two narrow neighbouring ones,
$f_0(980)$ and $f_0(1300)$, which was performed in
\cite{YF99,rad_pl}, tells us that the radius of the broad state is
considerably larger than the radii of $f_0(980)$ and $f_0(1300)$: this
fact agrees well with the assumption that $f_0(1200-1600)$ plays the
role of a locking state for its resonance-neighbours. Recent
measurement of the $t$-distributions in the reaction
$\pi^-n\to(\pi\pi)_S\,n$ \cite{BNL-new} confirmed  relatively large
magnitude of the broad state radius: with the increase of $|t|$
  (momentum square transferred to the resonance),  the broad state
vanishes much faster than $f_0(980)$ and $f_0(1300)$.

\subsection{Evolution of couplings of the $00^{++}$-states to
channels $\pi\pi$, $\pi\pi\pi\pi$, $K\bar K$, $\eta\eta$ $\eta\eta'$
with the onset of decay processes}

The $K$-matrix analysis does not allow one to determine
partial widths of the $f_0$-resonances
 directly. To find out partial widths for
the decays $f_0\to\pi\pi$, $\pi\pi\pi\pi$, $K\bar K$, $\eta\eta$,
$\eta\eta'$ it is necessary to calculate the residues of the amplitude
poles corresponding to resonances. Near the resonance, the
transition amplitude  $a\to b$ (indices $a$ and $b$ stand for the
resonance channels $\pi\pi$, $K\bar K$, $\eta\eta$, $\eta\eta'$,
$\pi\pi\pi\pi$)
takes the form:
\beq \label{1''} A_{ab}\ \simeq\
\frac{g^{(n)}_ag^{(n)}_b}{\mu^2_n-s}\
e^{i(\theta^{(n)}_a+\theta^{(n)}_b)} + B_{ab}\ .
\label{coupling}
\eeq
The   pole position  defines  the mass and width of the resonance
$\mu_n=M_n-i(\Gamma_n/2)$, and the real-valued coupling constants to
channels,
 $g^{(n)}_a$ and $g^{(n)}_b$, allow us to find partial widths of
resonances. Factorized residues are complex-valued magnitudes,
their complexity is determined in (\ref{coupling}) by the phases
$\theta^{(n)}_a$ and $\theta^{(n)}_b$; in (\ref{coupling}),
a smooth, non-pole, term $B_{ab}$ is also written.

On the basis of the results of the $K$-matrix fit \cite{K},
the decay coupling constants $g^{(n)}_a$ were calculated for
$f_0(980)$, $f_0(1300)$, $f_0(1500)$, $f_0(1750)$ and the broad state
$f_0(1200-1600)$, and the comparison was done for the obtained values
with corresponding couplings of bare states, which are predecessors of
the resonances under discussion.

\begin{table}
\caption{Coupling constants squared (in GeV$^2$) of
scalar--isoscalar resonances
decaying to the hadronic channels $\pi\pi$, $K\bar
K$, $\eta\eta$, $\eta\eta'$ and $\pi\pi\pi\pi$ for different $K$-matrix
solutions.}
\label{table5}
\begin{tabular}{l|ccccc|c}
\hline
Pole position&$\pi\pi$&$K\bar K$&$\eta\eta$&$\eta\eta'$&$\pi\pi\pi\pi$
&Solution\\
\hline
$f_0(980)$ &      &       &        &     &           &    \\
$ 1031-i32 $  &0.056 & 0.130 &  0.067 &  -- &   0.004   &  I \\
$ 1020-i35 $  &0.054 & 0.117 &  0.139 &  -- &   0.004   &II\\
\hline
$f_0(1300)$&      &        &        &       &       &    \\
$ 1306-i147$   &0.036 &  0.009 &  0.006 & 0.004 & 0.093 &I   \\
$ 1325-i170$   &0.053 &  0.003 &  0.007 & 0.013 & 0.226 &II\\
\hline
$f_0(1500)$&      &        &        &          &        &\\
$ 1489-i51 $   & 0.014 &  0.006 &   0.003 &  0.001 &  0.038 &I\\
$ 1490-i60 $   & 0.018 &  0.007 &   0.003 &  0.003 &  0.076 &II\\
\hline
$f_0(1750)$&      &        &        &          &        &\\
$ 1732-i72 $   & 0.013 &  0.062 &   0.002 &  0.032 & 0.002 &I\\
$ 1740-i160$   & 0.089 &  0.002 &   0.009 &  0.035 & 0.168 &II\\
\hline
$f_0(1200-1600)$&      &        &        &          &        &\\
$ \;1480-i1000$   & 0.364 &  0.265 &   0.150  &0.052&  0.524 &I\\
$ 1450-i800 $  & 0.179 &  0.204 &   0.046  &  0.005 &  0.686 &II\\
\hline
\end{tabular}
\end{table}

The values of couplings calculated in \cite{K} are shown in Table 4.
The comparison reveals a significant difference between
the decay couplings for bare states and their descendant--resonances.
This undoubtedly proves a strong effect of the mixing
of $q\bar q$-states with the glueball: the real resonance is a mixture
of these states.

Figure 6 demonstrates
the evolution of coupling constants at the onset of the decay channels:
following \cite{content}, relative changes of coupling constants
are shown for
 $f_0(980)$, $f_0(1300)$, $f_0(1500)$ and $f_0(1750)$
after switching on/off the decay channels (recall that the value
$x=0$ corresponds to the amplitude poles on the real axis and the value
$x=1$ stands for the resonance observed experimentally).

Let us bring our attention to a rapid increase of the coupling
constant $f_0\to K\bar K$ at the trajectory $f_0^{bare}(700)-f_0(980)$
in the region $x\sim 0.8-1.0$, where
$\gamma^2(x=1.0)-\gamma^2(x=0.8)\simeq 0.2$, see Fig. 6a.
Actually this increase is the upper limit of the
possible admixure of the long-range $K\bar K$
component in the $f_0(980)$: it cannot be greater than $20\%$.

\subsection{Evaluation of the glueball component in the
resonances $f_0(980)$,
$f_0(1300)$, $f_0(1500)$, $f_0(1750)$ and  broad state
$f_0(1200-1600)$ based on the analysis of hadronic-decay channels}

The evolution of couplings observed in the transition from bare states
to real resonances is due to the mixture of the  $q\bq$-states with the
glueball that is a consequence of the transitions $f_0(1)\to\,real\
 mesons\,\to f_0(2)$. One can evaluate the quark-antiquark and glueball
components in $f_0(980)$, $f_0(1300)$, $f_0(1500)$,
$f_0(1750)$ and  $f_0(1200-1600)$ using the  rules of quark
combinatorics for the decay couplings.

In the leading-order terms of the  $1/N$-expansion the vertices of
hadronic decays of the resonances are defined by planar diagrams,
the examples of the planar diagram for the decay of
$q\bq$-state and the glueball into two mesons are presented in Fig.
7a,b, respectively. In the course of the $q\bq$-state decay, the gluons
produce a new $q\bq$-pair; in the glueball decay, a subsequent
production of  two pairs occurs.

For the decay
couplings squared for $f_0\to\pi\pi,K\bar K,\eta\eta,\eta\eta'$,
the quark-combinatorics rules,
in case  when the $f_0$ state is  the mixture of the quarkonium and
gluonium components,  give us \cite{K,content}:
\begin{eqnarray}
&&
g^2_{\pi\pi}\ =\ \frac32\left(\frac g{\sqrt2}\cos\varphi+\frac
G{\sqrt{2+\lambda}}\right)^2, \nonumber\\
&& g^2_{K\bar K}=\ 2\left(\frac g2(\sin\varphi
+\sqrt{\frac\lambda2}\cos\varphi)+G\sqrt{\frac\lambda{2+\lambda}}
\right)^2 , \nonumber\\
&& \hspace*{-0.5cm} g^2_{\eta\eta}=\frac12\left(g(
\frac{\cos^2\Theta}{\sqrt2}
\cos\varphi+\sqrt\lambda\sin\varphi\sin^2\Theta)+\frac
G{\sqrt{2+\lambda}}(\cos^2\Theta+\lambda\sin^2\Theta)\right)^2,
\nonumber\\
&& g^2_{\eta\eta'}=\
\sin^2\Theta\cos^2\Theta\left(g(\frac1{\sqrt2}\cos\varphi
-\sqrt\lambda\sin\varphi)+G\frac{1-\lambda}{\sqrt{2+\lambda}}
\right)^2.
\label{3''}
\end{eqnarray}
The terms proportional to $g$ stand for the transitions
$q\bq\to\,two\ mesons$, while those with $G$ respond to transitions
$glueball\to two\ mesons$. Accordingly, $g^2$ and
$G^2$ are proportional to the probability to find in the
considered $f_0$-meson
the quark-antiquark and glueball componets. Recall that the angle
$\varphi$ stands for the content of the $q\bq$-component in the
decaying state, $q\bq=\cos\varphi\,n\bar n+\sin\varphi\,s\bar
s$, and the angle $\Theta$ for the contents of $\eta$ and $\eta'$
mesons: $\eta=\cos\Theta\,n\bar n-\sin\Theta\,s\bar s$ and
$\eta'=\sin\Theta\, n\bar n+\cos\Theta\,s\bar s$; we use
$\Theta=38^\circ$ \cite{ABMN}.

One may believe that the decay of the glueball is going in two steps:
initially, one  $q\bar q$ pair is produced, then with the production of
the next $q\bar q$ pair a fusion of quarks into mesons occurs.
Therefore, at the intermediate stage of the $f_0$ decay, we deal with a
mean quantity of the quark-antiquark component,
 $\langle q\bq\rangle$, which later on turns into hadrons. The equation
(\ref{3''}), under the condition  $G=0$, defines the content of this
intermediate state
$\langle q\bq\rangle=n\bar n
\cos\langle\varphi\rangle+s\bar s\sin\langle\varphi\rangle$.

As was said above, the $K$-matrix analysis  \cite{K} gave us two
Solutions, {\bf I} and {\bf II}, which differ mainly by the
parameters of the resonance $f_0(1750)$.
Fitting to the decay couplings squared for these Solutions leads to
the values of $\langle\varphi\rangle$ as follows:

Solution {\bf I}:

\bea
f_0(980):&\quad \langle\varphi \rangle\simeq -68^\circ\ ,
&\lambda\simeq 0.5-1.0\ ,\\
\nonumber
f_0(1300):&\quad \langle\varphi \rangle\simeq (-3^\circ)-4^\circ\ ,
&\lambda\simeq 0.5-0.9\ ,\\
\nonumber
{\rm Broad\;\;  state\;\;}
f_0(1200-1600):&\quad \langle\varphi \rangle\simeq 27^\circ\ ,
&\lambda\simeq 0.54\ , \\
\nonumber
f_0(1500):&\quad \langle\varphi \rangle\simeq 12^\circ-19^\circ\ ,
&\lambda\simeq 0.5-1.0\ ,\\
\nonumber
f_0(1750):&\quad \langle\varphi \rangle\simeq -72^\circ\ ,
&\lambda\simeq 0.5-0.7\ ,\\
\nonumber
\label{sol1}
\eea

Solution {\bf II}:

\bea
f_0(980):&\quad \langle\varphi \rangle\simeq -67^\circ\ ,
&\lambda\simeq 0.6-1.0\ ,\\
\nonumber
f_0(1300):&\quad \langle\varphi \rangle\simeq (-16^\circ)-
(-13^\circ)\ ,
&\lambda\simeq 0.5-0.6\ ,\\
\nonumber
{\rm Broad\;\;  state\;\;}
f_0(1200-1600):&\quad \langle\varphi \rangle\simeq 33^\circ\ ,
&\lambda\simeq 0.85\ ,\\
\nonumber
f_0(1500):&\quad \langle\varphi \rangle\simeq 2^\circ-11^\circ\ ,
&\lambda\simeq 0.6-1.0\ ,\\
\nonumber
f_0(1750):&\quad \langle\varphi \rangle\simeq -18^\circ\ ,
&\lambda\simeq 0.5\ .\\
\nonumber
\label{sol2}
\eea

In both Solutions, the average values of mixing angle for $f_0(980)$
coincide with one another,
with a good accuracy. Still, one should inderline that
equations (\ref{3''}) allow us to get one more magnitude for the mixing
angle of the $f_0(980)$, namely, $\langle \varphi
[f_0(980)]\rangle\simeq 40^\circ$.
The fact that the decay constants for
$f_0(980)\to\pi\pi$ and $f_0(980)\to K\bar K$ accept the solution with
 $\langle\varphi[f_0(980)]\rangle\simeq 40^\circ$ was underlined in
\cite{Anis_Mont}.
 However, this value of mixing angle does not suit
the classification of bare states provided by the $K$-matrix solutions.
Indeed, the $f_0(980)$ is the descendant of the bare state
$f_0^{bare}(700\pm100)$ which is close to the flavour octet.  The
evolution of coupling constants (see Fig.  6) tells us that $f_0(980)$
by its content remains  close  to its predecessor,
$f_0^{bare}(700\pm100)$.
Because of that, in Eqs.
(39) and (40) only
solutions with $\langle \varphi [f_0(980)]\rangle\simeq - 67.5^\circ$
are kept.

The values of average mixing angles for $f_0(1300)$ are stable
negative for both Solutions {\bf I} and {\bf II}, they differ slightly,
so we may accept
$\langle \varphi [f_0(1300)]\rangle= - 10^\circ\pm 6^\circ$.

Also the mean mixing angle for the $f_0(1500)$ does not differ
noticeably for Solutions {\bf I} and {\bf II}, so we may adopt
$\langle \varphi [f_0(1500)]\rangle=  11^\circ\pm 8^\circ$.

For the $f_0(1750)$, Solutions {\bf I} and {\bf II} provide different
mean values of  mixing angle. In Solution {\bf I},    the resonance
$f_0(1750)$ is dominantly $s\bar s $ system; correspondingly,
$\langle \varphi [f_0(1750)]\rangle= -72^\circ\pm 5^\circ$.
In Solution {\bf II}, the absolute value of mixing angle is much less,
$\langle \varphi [f_0(1750)]\rangle= -18^\circ\pm 5^\circ$.

For the broad state,  both Solutions give proximate values of mixing
angle, namely, $\langle \varphi [f_0(1200-1600)]\rangle= 30^\circ\pm
3^\circ$. This magnitude favours the opinion that the broad state can
be treated as the glueball descendant, because such a value of the mean
mixing angle corresponds to $\varphi_{glueball}=\sin^{-1}
\sqrt{\lambda/(2+\lambda)}$ at $\lambda\sim 0.50- 0.85$.

Let us emphasize that the coupling  magnitudes for the
$f_0$-resonances found in \cite{K} do not provide us any alternative
variants for the glueball descendant. Indeed, the value which
is the closest to the $\varphi_{singlet}$ is the limit value of the mean
angle for $f_0(1500)$ in Solution {\bf I}:  $\langle \varphi
[f_0(1500)]\rangle= 19^\circ$. Such a magnitude
being used for the definition of $\varphi_{glueball}$ corresponds to
$\lambda=0.24$, but this suppression parameter is much lower than those
observed in other processes: for  the decaying processes we have
$\lambda=0.6\pm0.2$ \cite{ufn,lambda}, while
for  the high-energy multiparticle production it is
 $\lambda\simeq0.5$ \cite{lambda-hec}.
In this way, the quark combinatorics points to the one candidate only,
that is, the broad state $f_0(1200-1600)$; we come back to this important
statement later on.

Generally, the formulae (\ref{3''}) allow us to find $\varphi$ as
a function of the coupling constant ratio  $G/g$ for the decays
$glueball\to mesons$ and $q\bar q$-$state \to mesons$. The
results of the fit for
 $f_0(980)$, $f_0(1300)$, $f_0(1500)$, $f_0(1750)$ and the broad state
$f_0(1200-1600)$ are shown in Fig. 8.

First, consider the results for
 $f_0(980)$, $f_0(1300)$, $f_0(1500)$, $f_0(1750)$ shown in Fig. 8a for
Solution {\bf I} and in Fig. 8c for Solution {\bf II}. The bunches of
curves in the $(\varphi, G/g)$-plane demonstrate  correlations
between mixing angle values and the $G/g$ ratios, for which the
description of couplings given in Table 4 is satisfactory. A vague
dissipation of curves, in particular noticeable for $f_0(1300)$ and
$f_0(1500)$, is due to the uncertainty of $\lambda$ in Eqs.
(39)  and (40).

The correlation curves in Fig. 8a,c allow one to see, on a qualitative
level, to what extent the admixture of the gluonium component in
 $f_0(980)$, $f_0(1300)$, $f_0(1500)$, $f_0(1750)$ affects the
quark-gluonium content, $q\bar q=n\bar n\cos\varphi+s\bar
s\sin\varphi$ determined from hadronic decays.
The magnitudes $g^2$ and $G^2$ are proportional to the probability to
find out, respectively, the quarkonium and gluonium components,
$W_{q\bar q}$ and  $W_{gluonium}$
 in a considered resonance:
\beq
g^2=g^2_{q\bar q} W_{q\bar q}\; , \qquad
G^2=G^2_{gluonium}W_{gluonium}\ .
\eeq

The results of the $K$-matrix
fit obtained in \cite{K} tell us that  the coupling
constants $g^2_{q\bar q}$  and    $G^2_{gluonium}$  are of the same
order of magnitude (also  see the discussion in
\cite{ufn,aas-z,ufn95}),
 therefore  we accept as a
qualitative estimate:
\beq G^2/g^2\simeq W_{gluonium}/W_{q\bar q}\ .
\label{G/g}
\eeq

The figures 8a,c show the following permissible scale of values
$\varphi$ for the resonances
$f_0(980)$, $f_0(1300)$, $f_0(1500)$, $f_0(1750)$,
after mixing with the  gluonium component.

Solution {\bf I}:
\bea
&W_{gluonium}[f_0(980)]\la 15\%\; :& \quad
-93^\circ \la \varphi [f_0(980)] \la -42^\circ , \\
\nonumber
&W_{gluonium}[f_0(1300)]\la 30\% \; :& \quad
-25^\circ \la \varphi [f_0(1300)] \la 25^\circ\ , \\
\nonumber
&W_{gluonium}[f_0(1500)]\la 30\% \; :& \quad
-2^\circ \la \varphi [f_0(1500)] \la 25^\circ\ , \\
\nonumber
&W_{gluonium}[f_0(1750)]\la 30\% \; :& \quad
-112^\circ \la \varphi [f_0(1750)] \la -32^\circ\ . \\
\nonumber
\label{SolutionI}
\eea

Solution {\bf II}:
\bea
&W_{gluonium}[f_0(980)]\la 15\%\; :& \quad
-90^\circ \la \varphi [f_0(980)] \la -43^\circ , \\
\nonumber
&W_{gluonium}[f_0(1300)]\la 30\% \; :& \quad
-42^\circ \la \varphi [f_0(1300)] \la 10^\circ\ , \\
\nonumber
&W_{gluonium}[f_0(1500)]\la 30\% \; :& \quad
-18^\circ \la \varphi [f_0(1500)] \la 23^\circ\ , \\
\nonumber
&W_{gluonium}[f_0(1750)]\la 30\% \; :& \quad
-46^\circ \la \varphi [f_0(1750)] \la 7^\circ\ . \\
\nonumber
\label{SolutionII}
\eea

 The $\varphi$-dependence  of $G/g$ is linear
for $f_0(980)$, $f_0(1300)$, $f_0(1500)$, $f_0(1750)$.
 Another
type of the correlation takes place for the state which is the
glueball descendant: the correlations curves for this case form in the
$(\varphi,G/g)$-plane a typical cross. Just this cross appeared for the
broad state $f_0(1200-1600)$ for both
Solutions {\bf I} and {\bf II}, see Fig. 8b,d.

The appearance of  glueball cross by correlation curves
in the $(\varphi,G/g)$-plane is due to the formation mechanism of
the quark-antiquark component in the gluonium state: in the transition
$gg\to (q\bar q)_{glueball}$ the state $(q\bar q)_{glueball}$ is fixed
by the value of $\lambda$. So the glueball descendant is the
quarkonium-gluonium composition as follows:
$$gg\cos\gamma+(q\bar q)_{glueball}\sin\gamma\ , $$
where
$$
(q\bar q)_{glueball}\ =\ n\bar n\cos\varphi_{glueball}+s\bar s
\sin\varphi_{glueball}\ ,
$$
and   $\varphi_{glueball}=\tan^{-1}
\sqrt{\lambda/2}\simeq26^\circ-33^\circ$ for $\lambda\simeq0.50-0.85$.
The ratios of
couplings for the transitions
$gg\to\pi\pi,K\bar K,\eta\eta,\eta\eta'$ are the same as for the
quarkonium $(q\bar q)_{glueball}\to\pi\pi,K\bar K,\eta\eta,\eta\eta'$,
so the study of
  hadronic decays only do not permit to fix the mixing
angle $\gamma$. This property -- similarity of hadronic decays for the
states $gg$ and $(q\bar q)_{glueball}$ -- implies a specific form of
the correlation curve in the $(\varphi, g/G)$-plane: the gluonium
cross.  Vertical component of the gluonium cross means that the
glueball descendant has a considerable
admixture  of the quark-antiquark component
$(q\bar q)_{glueball}$. Horizontal line of the cross corresponds to
the dominant $gg$ component. The value of $\lambda$ which affects the
cross-like correlation on the $(\varphi, g/G)$-plane is denoted from
now on as $\lambda_{glueball}$. For Solution {\bf I}, we have
$\lambda_{glueball}=0.55$, while for  Solution {\bf II}
$\lambda_{glueball}=0.85$.

At not a large shift of $\lambda$ from its mean value
$\lambda_{glueball}$, the coupling constants $f_0(1200-1600)\to
\pi\pi,K\bar K, \eta\eta,\eta\eta'$ can be also described, with a
reasonable accuracy, by Eq. (\ref{3''}); in this case correlation
curves on the $(\varphi, g/G)$-plane take the form of hyperbola.
Shifting the value of $\lambda$ in $|\lambda-\lambda_{glueball}|\sim
0.2$ breaks the description of couplings of the broad state by formulae
(\ref{3''}).

The cross-type correlation on the $(\varphi, g/G)$-plane in the
description of coupling constants
$f_0\to \pi\pi,K\bar K, \eta\eta,\eta\eta'$  by formula (38) is a
characteristic signature of the glueball or glueball descendant. And
{\it vice versa}: the absence of the cross-correlation should point to
the quark-antiquark nature of resonance. Therefore, the $K$-matrix
analysis proves definitely that $f_0(1200-1600)$ is
 the gluonium  descendant, while
 $f_0(980)$, $f_0(1300)$, $f_0(1500)$, $f_0(1750)$ cannot pretend to be
the glueballs.

The analysis proves that $f_0(1300)$, $f_0(1500)$ are dominantly the
$n\bar n$-systems. Still, in Solution {\bf II} the $q\bar q$ component
of the resonance $f_0(1300)$ may contain rather large $s\bar s$
component in the presence of the 30\% gluonium admixture in this
resonance. As to the $f_0(1500)$, the mixing angle
$\langle \varphi [f_0(1500)]\rangle$ in the
$q\bar q$ component may reach $25^\circ$ at $G/g\simeq -0.6$ (Solution
{\bf I}) that is rather close to $\varphi_{glueball}$. However, in this
case the description of coupling constants $g^2_a$ (Table 4) is
attained as an effect of the strong destructive interference of the
amplitudes $(q\bar q)\to two\,pseudoscalars$ and $gg\to
two\,pseudoscalars$. This fact tells us that one cannot be tempted to
interprete $f_0(1500)$ as the glueball descendant.

\subsection{The light $\sigma$-meson: Is there a pole of the
$00^{++}$-wave amplitude?}

Effective $\sigma$-meson is needed in nuclear physics as well as in
effective theories of the low-energy strong interactions --- and such
an object exists in a sense that there exists
rather strong interaction,
which is realized by the scattering
phase passing through
the value $\delta^0_0=90^\circ$ at
$M_{\pi\pi}\simeq 600-1000$ MeV. In the naive Breit--Wigner-resonance
interpretation, this would correspond to an amplitude pole; but the
low-energy $\pi\pi$ amplitude is a result of the interplay of singular
contributions of different kinds (left-hand cuts as well as poles
located highly, $f_0(1400-1600)$ included) , so a straightforward
interpretation of the $\sigma$-meson as a pole may fail.

The question is whether the
$\sigma$-meson exists as a pole of the  $00^{++}$-wave amplitude.
(See also \cite{penn}, where this problem was particularly underlined.)
However,
until now there is no definite answer to this question, though
 this point is crucial for meson systematics.

The consideration of the
partial $S$-wave $\pi\pi$ amplitude, by accounting for left
singularities associated with  the $t$- and $u$-channel interactions,
favours the idea of the pole at ${\rm Re}\, s \sim 4m_\pi^2$.
The arguments are based on the analytical continuation of the
$K$-matrix solution to the region $s\sim0-4\mu^2_\pi$ \cite{sigma_N/D}.

In  \cite{sigma_N/D}, the $\pi\pi$-amplitude of the
$00^{++}$ partial
wave was considered in the region $\sqrt s<950\,$MeV. The
account for the left-hand singularities had been done within the
method described in Section 3.10.2, namely, the $K$-matrix
amplitude was fitted in the form
\beq
K(s)\ =\ \sum^6_{n=1}
\frac{f_n}{s+s_n}+f+\frac{g^2}{M^2-s}\ ,
\label{37.1}
\eeq
where $f_n,f,s_n,g^2,M^2$ are parameters. The left-hand cut was fitted
to six pole terms;  the pole at $s=M^2$ ($M\sim 900$ MeV)
corresponds to $\delta^0_0=90^\circ$.
The fitting was performed to the low-energy scattering
phases, $\delta^0_0$, at $\sqrt{s}<450$ MeV,
and the scattering length, $a^0_0$. In addition
at $450\le\sqrt s\le950\,$MeV the value $\delta^0_0$ was sewn with
those found in the  $K$-matrix analysis \cite{YF}: from this point of
view the solution found in \cite{sigma_N/D} may be treated as
analytical continuation of the $K$-matrix amplitude to the region
$s\sim 0-4m^2_\pi$. The analytical continuation of the $K$-matrix
amplitude of such a type accompanied by simultaneous reconstruction of
the left-hand cut contribution provided us with the characteristics of
the amplitude as follows. The amplitude has a pole at
\beq
\sqrt s\
\simeq\ 430-i325\ \mbox{ MeV },
\label{37.2}
\eeq
the scattering length,
\beq
a^0_0\ \simeq\ 0.22\, m^{-1}_\pi\ ,
\label{37.3}
\eeq
and the Adler zero at
\beq
\sqrt s\ \simeq\ 50\mbox{ MeV }.
\label{37.4}
\eeq
The errors in
the definition of the pole in solution (\ref{37.2}) are large,
and unfortunately they are poorly controlled, for they are governed,
in the main, by uncertainties when left-hand singularities
are restored. As
to experimental data, the position of pole is rather sensitive to the
scattering length value, which in the fit \cite{sigma_N/D} was taken in
accordance to the paper \cite{Ke4-1}:
$a^0_0=0.26\pm0.06\,m^{-1}_\pi$. As one can see, the solution
\cite{sigma_N/D} requires a small scattering length value:
$a^0_0\simeq0.22\,m^{-1}_\pi$. New and much more precise measurements
of the $K_{e4}$-decay \cite{Ke4-2} provided
$a^0_0=(0.228\pm0.015)m^{-1}_\pi$, that agrees completely with the
value (\ref{37.3}) obtained in \cite{sigma_N/D}. Such a coincidence
favours undoubtedly  the pole position (\ref{37.2}).

So, the $N/D$-analysis of the low-energy
$\pi\pi$-amplitude sewn with the $K$-matrix one  \cite{YF},
provides us with the arguments for
 the existence of the light $\sigma$-meson. In a set of papers,
by modelling the left-hand cut of the  $\pi\pi$-amplitude (namely,
by using interaction forces or the $t$- and $u$-channel
exchanges), the light  $\sigma$-meson had been also obtained
\cite{sigma_dis,sigma80,sigma}, but the mass values are widely
scattered, e.g. in \cite{sigma_h} the pole has been obtained
at essentially larger masses, $\sqrt s\sim600-900\,$MeV.

The observation of the light $\sigma$-meson is aggravated by the
existence of the Adler zero in the $\pi\pi$-amplitude near the
$\pi\pi$-threshold. Therefore, as was emphasized above,
for the reliable
determination of the $\sigma$-meson it is necessary to study
the $\pi\pi$-production in the annihilation and decay reactions (of
the type of $p\bar p\to\pi\pi\pi$ and $D\to\pi\pi\pi$), where the Adler
zero is absent. However, one should
emphasize again, in the $p\bar p$-annihilation, where the statistics is
rather high, the $\sigma$-meson is not seen, while in the reaction
$D^+\to\pi^+\pi^-\pi^+$  \cite{D+} the
statistics is not sufficient for the reliable analysis of the
low-energy $\pi\pi$-spectrum.

By discussing the search for the light $\sigma$-meson in the
three-particle reactions, it is necessary to accentuate a requirement,
had it been fulfilled, one can confidently confirm the discovery of
this resonance. I mean the rescattering of pions in the final state.
The matter is that the fitting to the near-threshold state, that is
actually the $\sigma$-meson, should be carried out under the correct
account for the unitarity at small masses of the $\pi\pi$-system, while
the contribution of crossing channels (for example, resonance
production) leads to the violation of unitarity. The account for
rescatterings in the $\pi\pi$-channel reconstructs the two-particle
unitarity. Besides, the rescatterings restore logarithmic singularities
of the three-particle amplitude, that may affect significantly the
region of small $\pi\pi$ masses, in particular, in the presence of
heavy resonances.  The effects of pion rescatterings were studied in
\cite{f1500PR}, where the combined analysis of the Dalitz-plot has been
done for the reactions $p\bar p$(at rest)$\to \pi^0 \pi^0 \pi^0,
\pi^0 \eta\eta$
 --- it appeared that they
act rather strongly upon the region of small $\pi\pi$-masses, though
they are not important for the analysis of resonances at 1300--1500
MeV, that was the main goal of the investigation performed in
\cite{f1500PR}. An opposite point of view, namely, the statement about
the possibility to single out the signal from $\#$-meson without taking
account of rescatterings, was claimed in \cite{Ishida}.

Concerning the problem of search for the light $\sigma$-meson in the
three-particle reactions, an important point should be
emphasized, which being fulfilled would allow us to speak with
confidence about reliable determination of this state. This is
rescattering of pions in the final state. The matter is that fitting to
a near-threshold state, such as $\sigma$-meson, should be carried
out with correct account for unitarity at small
$\pi\pi$-masses; at the same time the contribution of crossing channels
(for example, the production of resonances)  violates the unitarity.
The rescatterings in the $\pi\pi$-channel
being  accounted for restore the
two-particle unitarity near the $\pi\pi$-threshold.  Besides,
rescatterings lead to logarithmic singularities of the
three-particle amplitude, which in the presence of heavy resonances
affect significantly the low-mass region.  The effect of pion
rescatterings were considered in \cite{f1500PR} at simultaneous
analysis of Dalitz-plots in the reactions $p\bar p(at\,rest)\to
\pi^0\pi^0\pi^0,\pi^0\eta\eta$:  it occurred that they affect strongly
the low-mass region though being not important for the analysis of
resonances at 1300--1500 MeV --- the main scope of investigation in
\cite{f1500PR}. An opposite viewpoint, namely, one can reliably detect
the $\sigma$-meson signal without accounting for
$\pi\pi$-rescatterings, is presented in \cite{Ishida}.

\subsection{Systematics of scalar states on the $(n,M^2)$- and
$(J,M^2)$-planes and the problem of basic multiplet $1^3P_0q\bar q$}

The figures 1,2 and 3 demonstrate  linear behaviour of the $q\bar
q$-meson trajectories in the $(n,M^2)$- and $(J,M^2)$-planes for a
variaty of states. In this way, it would be instructive
to juxtapose the nonet classification given by the  $K$-matrix
analysis \cite{K,YF,nonet}  with the systematics of the $q\bar
q$-states in the $(n,M^2)$- and $(J,M^2)$-planes.

\subsubsection{The $K$-matrix classification of scalars and
$q\bar q$-trajectories in the  $(n,M^2)$-planes}

Consider the variant which follows directly from the
$K$-matrix calssification of bare states:  let us accept that
the light $\sigma$-meson does not reveal itself as the
amplitude pole or, if it exists, it is an extra state for the  $q\bar
q$-systematics. For this case, the location of scalar states
$(f_0,a_0,K_0)$ on the $(n,M^2)$ trajectories is shown in Fig. 9.

Figure 9a demonstrates the trajectories for the resonance states
 $00^{++}$, $9^{++}$ and $\frac120^+$, while
in Fig. 9b the trajectories for corresponding bare states
are shown (recall that the doubling of the
$f_0$-trajectories occurs due to the existence of two components,
 $s\bar s$ and $n\bar n$). It is seen that the
trajectory slopes for bare and real states are nearly the same (in
Fig. 9 the trajectory slope is $\mu^2=1.3\,$GeV$^2$).

The state $f_0^{bare}(1580\pm50)$ certainly does not belong  to any of
$f^{bare}_0$-trajectories shown in Fig. 9b, the trajectories of the
real $f_0$-states does not need the state $f_0(1200-1600)$ too. This is
natural, provided the $f_0^{bare}(1580\pm50)$ is the glueball
and $f_0(1200-1600)$ the glueball descendant.

Gluodynamical lattice calculations tell us that the lightest scalar
glueball is located in the mass region 1550--1750 Β' \cite{lattice}.
There are also other arguments in  favour of
  just in this mass region one may
encounter the gluonium: the  estimate of the mass of soft (or
effective) gluon points to the gluon mass
$m_{gluon}\sim700-1000\,$MeV.  Experimental estimations
of the effective-gluon mass are based on
the study of hadronic spectra in radiative decays
 $J/\psi\to\gamma+hadrons$ and $\Upsilon\to\gamma
+hadrons$ \cite{parisi,field}; they give for the gluon mass the value
800--1000~MeV.  The close values were obtained within model calculations
of the quark-gluon interactions in the soft
region giving the value  700--800~MeV   for the effective-gluon
 mass \cite{cornwell,gerasyuta}. The lattice calculations of
the effective
gluon agree reasonably with these estimates: accordingly,
$m_{gluon}\simeq700\,$MeV \cite{gluon_lattice}.  It is natural to
believe that the mass of the lightest scalar glueball is of the order
of the double effective gluon mass, $M_{glueball}\sim2m_{gluon}.$

One should pay attention to the fact that
the resonances $f_0(980)$, $f_0(1500)$ and
$f_0(1300)$, $f_0(1750)$ lay neatly on linear trajectories,
 with a slope which nearly coincides with slopes of
isovector trajectories: $\rho_J$, $a_J$ and $\pi_J$. This
circumstance  argues in favour of the
 admixture of the glueball components
 not leading to strong violation of the linear behaviour of
trajectories, at least for the  $f_0$-meson sector.

It would be pertinent to recall that in the $K$-matrix analyses
\cite{K,YF} three Solutions have been found denoted as {\bf I},
{\bf II-1} and {\bf II-2}. Solutions {\bf II-1} and {\bf II-2} occurred
to be close to each other in resonance characteristics; yet, in
Solution {\bf II-1} the  $f_0^{bare}(1580)$ is the
 $q\bar q$-state, and the glueball mass is around 1250 MeV.
We reject this Solution, for it contradicts both the linearity of
$f_0^{bare}$-trajectories and gluodynamical calculation results.

Therefore, from the point of view of the  classification of states on
the $(n,M^2)$-trajectories, the scheme where the
$f_0^{bare}(700\pm100)$ and its descendant, $f_0(980)$, are the
lightest states in the $1^3P_0q\bar q$ nonet looks  self-consistent.
This scheme agrees with the coupling constants for the transitions
$f_0^{bare}$, $a_0^{bare}$, $K^{bare}_0\to
\pi\pi,K\bar K,\eta\eta,\eta\eta'$ \cite{K,ufn,nonet}.
Moreover, model calculations of masses of the $q\bar q$-mesons also
point to a possible location of the  $1^3P_0q\bar q$-nonet in
 the range 900--1300 MeV \cite{petry}. Still,
by considering the nonet classification, the question arises about
large  $s\bar s$-component in the
lightest $q\bar q$ state: 67\% in
$f^{bare}_0(700\pm100)$ and $(45-75)\%$ in $f_0(980)$.
But it would be appropriate to recall that we have already faced
similar situation in the pseudoscalar sector:
 the $\eta$-meson
also contain a large $s\bar s$-component, about 40\%.
Both states, $\eta$-meson and $f^{bare}_0(700\pm100)$, are close to
the flavour octet, and strong interaction in the flavour octet states
may cause the large $s\bar s$ component in light mesons. In other
words, the value of the $s\bar s$ component in lightest mesons depends
on the structure of the short-range forces, and to determine the
structure of  forces is the goal of the analysis and
 the motivation to perform the systematics of meson states.

So the root of the matter is whether the $\sigma$-meson, if it
exists, is the  standard $q\bar q$-state, a partner to $f_0(980)$. Then
the nonet of scalar mesons looks as follows:  $f_0(300-500)$,
 $f_0(980)$, $a_0(980)$ and the low-lying $\kappa$ meson,
$\kappa(800-1000)$. The argument against this scheme is the
systematics of the $K$-mesons in the $(J,M^2)$ plane, that is
discussed below.

\subsubsection{Systematics of kaons in the $(J,M^2)$ plane}

The existence of the
scalar $\kappa$-meson, with the mass
800-1000~MeV, is under discussion  as long as the discussion of
$\sigma$-meson, e.g. see \cite{van} and references therein.  The
low-energy data on the $\pi K$-scattering are poor, and this is the
main reason of uncertainty in the $\kappa$-meson
problem. The $K$-matrix
analysis of the  $\pi K$-amplitude \cite{nonet} does not supply us with
 unambigous answer, for there exist solutions without  amplitude
pole at 800-1000~MeV but there are also solutions with the
pole at $\sim1000\,$MeV \cite{nonet}.  The $K$-matrix analysis of the
high-statistics Crystal Barrel data on the reactions $p\bar p\to K\bar
K\pi$ and $n\bar p\to K\bar K\pi$ does not require the $\kappa$-meson 
\cite{K}.

Let us turn to the systematics of kaon states on the
$(J,M^2)$ plane: such systematics of kaons brings additional
argumets to the discussion of whether $\sigma$ and $\kappa$ mesons are
the standard $q\bar q$ states, members of the $1^3P_0q\bar q$
multiplets, or not.

In the kaonic sector, the experimental data are poor, and a complete
picture of the kaon disposition on the trajectories with
$J^P=0^-,2^-,4^-,...$ and $J^P=1^+,3^+,5^+,...$ cannot be unambigously
reconstructed: in Fig. 10a,d a conventional picture is presented for
the location of these trajectories. More definite is the situation in
the sectors with $J^P=0^+,2^+,4^+,...$ and $J^P=1^-,3^-,5^-...$, see
Fig.  10b,c:  here
we know the states  laying on the leading and
first daughter trajectories --- and just on these trajectories the
$\kappa$ meson should  lay, provided it is the standard $q\bar
q$-state.

In the sectors with $J^P=0^-,2^-,4^-,\ldots$ and
$J^P=1^+,3^+,5^+,\ldots$ (Fig. 10a,d) the doubling of
trajectories takes place because of the presence of two spin
states: for example, two states with $J^P=1^+$ are defined by the
quantum numbers $(L=1,S=1)$ and $(L=1,S=0)$; likewise, two states
$J^P=2^-$ are formed by two sets of quantum numbers,
$(L=2,S=1)$ and  $(L=2,S=0)$. Figures 10a,d
represent a conjectural  disposition
of trajectories and demonstrate how
many unknown states (open circles) are present in the groups
$J^P=0^-,2^-,4^-$ and $J^P=1^+,3^+,5^+$.

More reliably determined are the trajectories for the groups with
$J^P=0^+,2^+,4^+$ (Fig. 10b) and $J^P=1^-,3^-,5^-$ (Fig.
10c), where, as was said above, the leading and first daughter
trajectories are fixed firmly. The leading and first daughter
trajectories in these groups are degenerate: the states
$K_{2^+}(1430)$, $K_{4^+}(2045)$ and
$K^*(890)$, $K_{3^-}(1780)$, $K_{5^-}(2380)$ lay on the common leading
trajectory $\alpha_{leading}(J)\simeq0.25+0.90\,M^2$ (the slope is in
GeV units), and
the states  $K_{0^+}(1430)$, $K_{2^+}(1980)$ and
$K^*(1680)$ lay on the common trajectory
$\alpha_{daughter}(J)\simeq-2.0+0.90\,M^2$, where
$J^P=0^+,1^-,2^+,3^-$.

Figure 10b demonstrates also the location of the $\kappa$ meson: one
can see that this state does not lay either on  leading or
daughter trajectories.  In order to put the $\kappa$ meson onto the
$q\bar q$ trajectory, the existence of the $2^+$ kaon with the mass
$\sim 1600$ MeV is needed, but
  the  analysis \cite{Kpi} does not point to the presence of
tensor resonance in this mass region.
 The fact that there is no room
for the $\kappa$ meson on the  $(J,M^2)$  $q\bar q$-trajectories
 is the argument against the construction of the lowest scalar
nonet by using the states $f_0(300-500)$, $f_0(980)$,
$a_0(980)$ and $\kappa(800-1000)$, that is actually an indication that
the $\kappa$ meson does not exist.

\subsection{Exotic scalar states,
$f_0(1200-1600)$ and $f_0(300-500)$}

Two scalar states, the broad resonance  $f_0(1200-1600)$
and the light $\sigma$-meson $f_0(300-500)$, remain beyond the
considered here $q\bar q$ classification that is based on the
$K$-matrix analysis \cite{K,YF}; they should be treated as exotic
states.

The $K$-matrix analysis gives us an unambigous interpretation of the
$f_0(1200-1600)$. This state is the descendant of a pure glueball
which, by accumulating of widths of neighbouring
scalar--isoscalar states, turned into the broad resonance.
All comparatively narrow resonances from the region
$\sim1500\,$MeV (they are $f_0(1300)$, $f_0(1500)$ and $f_0(1750)$)
fit well the $q\bar q$-trajectories in the $(n,M^2)$ plane,
so just the broad resonance  $f_0(1200-1600)$ is an extra one from
the point of view of the $q\bar q$-systematics.

The fact that the glueball turned into the broad state allows us to say
that the glueball is "melted". The idea that the bound state of gluons
can disappear due to strong interaction in the soft region was brought
out rather long ago \cite{QCD-20}.  However, the "melting"
observed in the $K$-matrix analysis is a specific one: the
transformation of the glueball into the broad state occurs as a result
 of the decay processes (transitions of resonances into real mesons)
that takes place at large distances, of the order or greater than
$R_{confinement}$. Besides, the amplitude pole associated with
$f_0(1200-1600)$ did not go far away from the physical region:
the resonance half-width, that defines the imaginary
value of mass related to
the pole, is of the order of 500--1000~MeV, so the pole occurs inside
the hemicircle, where the analysis of experimental data allowed us to
reconstruct the analytical amplitude (supposing
 the threshold singularities be correctly taken into account).

The transition of the lightest scalar glueball into the broad resonance
$f_0(1200-1600)$ gives rise to a number of questions. The
glueball tranformation into the broad resonance, is it a unique event
(in a sense,  an occasional event) or is this a common
phenomenon for exotic states? The resonance
$f_0(1200-1600)$, after having accumulated the widths of
neighbouring $q\bar q$-states plays, with respect to them, the
role of a locking state, does this lead to the increase of the proper
size of $f_0(1200-1600)$?  These questions were put in
\cite{YF99,PR-exot}; still, to have a veracious response
 one needs much more accurate data. Qualitative evaluation
of the radius of $f_0(1200-1600)$ had been carried out in
\cite{YF99,PR-exot,rad_pl}, on the basis of the GAMS data for the
$M_{\pi\pi}$-distributions at different momenta transferred to the
nucleon in the reaction $\pi^-p\to n\pi^0\pi^0$ at $p_{lab}=38\,$GeV/c
\cite{GAMS}.  According to this estimate, the broad state is much more
loosely-bound system than its neighbours-resonances. Recent
measurements performed by the E852 Collaboration \cite{BNL-new} support
the fact that comparatively narrow resonances $f_0(980)$ and $f_01300)$
are more compact than the broad state $f_0(1200-1600)$; the
discussion of these data can be found in \cite{K,bnl-gams}.

The situation with the light $\sigma$-meson, $f_0(300-500)$, is less
definite as concern its experimental status
and the understanding of its nature.
The nature of the $\sigma$-meson, if it exists as an amplitude pole, is
rather enigmatic.  The light $\sigma$-meson is hardly the glueball-like
formation.  Also it is difficult to imagine the light $\sigma$-meson to
be the standard $q\bar q$-system: the arguments against such an
interpretation were formulated in Section 3.8.

It was suggested in \cite{eye} that the existence of the light
$\sigma$-meson may be due to a singular behaviour of the
forces between quark and antiquark
at large distances
(in quark models they are conventionally
called "confinement forces"). The scalar confinement
potential, which defines
the spectrum of the $q\bar q$-states in the region
1000--2000 MeV, behaves at large hadronic distances as
$V^{(c)}_{confinement}(r)\sim\alpha r$, where
$\alpha\simeq0.19\,$GeV$^2$. In the momentum representation,
such a growth
of the potential is associated with the singular
behaviour at small  $q$:
\beq \label{A}
V^{(c)}_{confinement}(q)\ \sim\ \frac1{q^4}\ .
\eeq
In the colour space, the main contribution comes from the
component $c=8$, i.e. the confinement forces should
 be the octet ones.   The question that is  crucial
for the structure of $\sigma$-meson is as follows:
is there a component with the colour singlet,
$V^{(1)}_{confinement}(q)$, in the singular
 potential (\ref{A})?
If the singular component with $c=1$ exists, then it must reveal itself
in hadronic channels as well, that is, in the $\pi\pi$-channel. In
hadronic channels, this singularity should not be exactly the same as in
the colour octet ones, because
the hadronic unitarization of the amplitude
(which is absent in the channel with  $c=8$) should modify somehow
the low-energy amplitude. One may believe that, as a result of the
unitarization in the channel  $c=1$, i.e. due to the account for
hadronic rescatterings, the singularity of $V^{(1)}_{confinement}(q)$
may appear in the $\pi\pi$-amplitude
 on the second sheet and one may believe that this singularity
is  what we call "the light $\sigma$-meson".

Therefore, the main question consists in the following:
the $V^{(1)}_{confinement}(q^2)$, does it have the same singular
behaviour as $V^{(8)}_{confinement}(q^2)$?
The observed linearity of the $(n,M^2)$-trajectories, up to the
large-mass region, $M\sim2000-2500$ MeV, see Figs. 1 and 2, favours
the idea of the universality in the behaviour of potentials
$V^{(1)}_{confinement}$ and $V^{(8)}_{confinement}$
at large $r$, or small $q$. To see
that (for example, in the process
$\gamma^*\to q\bar q$, Fig. 11a)
 let us
discuss the colour neutralization mechanism of outgoing quarks as
a break of the gluonic string by newly born $q\bar q$-pairs. At
 large distances, (2.0--2.5~fm), that corresponds to
the formation of states with masses 2000--2500 MeV, two--three new
$q\bar q$-pairs should be formed.
 This follows from the value
$\langle n_{ch}(W)\rangle$ in the  $e^+e^-$-annihilation: at
$W\simeq2$ GeV we have $\langle n_{ch}\rangle\simeq3$.
It is natural to suggest that a convolution of the quark--gluon combs
governs the interaction forces of quarks at large distances,  see Fig.
11bc.
This means that
the potential $V^{(c)}_{confinement}$  working at such large distances
 contains two or three
$t$-channel $q\bar q$-pairs. The mechanism of
the formation of new $q\bar q$-pairs to neutralize
colour charges does not have a
selected  colour component, see e.g.
\cite{book}. In this case all colour  components
$3\otimes\bar3=1+8$ behave similarly, that is, at small $q^2$
the singlet and octet components of the potential
 are uniformly singular, $V^{(1)}_{confinement}(q^2)\sim
V^{(8)}_{confinement}(q^2) \sim1/q^4$.
This is seen in Fig. 11a: the quark--gluon ladder ensures the
$t$-channel flow of colour charge $C=3$, so quark--antiquark
interaction block being the convolution of ladder diagrams
$3\otimes \bar 3=1+8$ contains
two equivalent, singlet and octet,  components.
This points to a
similarity of $V^{(1)}_{confinement}$ and
$V^{(8)}_{confinement}$.

So, from the point of view of  mechanisms of colour
neutralization
related to the ideas of the $t$-channel production
of new quark--antiquark pairs,
the existence of the light $\sigma$-meson is well-grounded. Therefore,
it is of utmost importance to discover whether there exist
singularities at low energies in the $00^{++}$-amplitude, which we call
the light $\sigma$-meson, and what is the type of these singularities,
if any.

\section{Conclusion}

Recent analysis of meson spectra in the  $p\bar p$-annihilation in
flight \cite{ral} resulted in the opportunity to
systematize $q\bar q$-mesons, practically leaving too small a room for
speculations about the existence of exotic states. First of all, this is
related to the $00^{++}$ sector, where all comparatively narrow
resonances $f_0(980)$, $f_0(1300)$, $f_0(1500)$, $f_0(1750)$
lay on  linear  $q\bar q$-trajectories on the $(n,M^2)$- and
$(J,M^2)$ planes, that does not provide us with a ground for hypotheses
on their exotic origin.

In quark models of 60--70's the idea of dominant
${\bf LS}$-splitting was pushed forward \cite{Zeld,Glashow} --- within
such a structure of forces the $P$-wave $q\bar q$-multiplet with
$J=0$ is lighter than multiplets with $J=1,2$. Recently the models
became popular, where $f_0(980)$ and $a_0(980)$  are considered as
exotic states (see  \cite{J,I,Cheng} and references therein): in this
case the $1^3P_0q\bar q$-nonet should be placed in the region
1300--1700 MeV. It is seen that the $K$-matrix analysis [8,11,12,13]
renders us back to a former picture:  the $1^3P_0q\bar q$ multiplet is
the lightest one among the $P$-wave $q\bar q$-states.

Beyond the $00^{++}$ sector,
the candidate for the exotics is the $\pi_2(1880)$  \cite{pi2}
--- for sure
this state does not belong definitely to the $q\bar q$-trajectory,
and it could be a hybrid state. Of course, the existence of this
resonance should be confirmed by its observation in  other
reactions.
Besides, if $\pi_2(1880)$ was the $q\bar q g$-state, there should
exist the other nonet members near 1900 MeV in addition to those that
lay on standard $(n,M^2)$ trajectories. However, no
$2^{-+}$ mesons have been observed yet, which could be extra ones with
respect to the $q\bar q$ classification.

In the literature, there are indications to a plausible existence of the
non-quark-antiquark mesons with $J^{PC}=1^{-+}$ (see \cite{PDG} and
references therein), though experimental evidence for these states is
rather poor.

In the $00^{++}$ meson sector, two states are beyond the $q\bar q$
systematics, namely, the light $\sigma$-meson and the broad state
$f_0(1200-1600)$.

Concerning the problem of the light $\sigma$-meson, though important
and having a long discussion history, there is no direct
 evidence of its existence --- on the level of fixing a pole in the
$00^{++}$ amplitude on the basis of the experimental data with
large statistics.  Such a situation existed a few years ago, see the
discussion in \cite{penn}, yet there is no noticeable
experimental progress till
now.

Nevertheless there are reasons which make us to concern
about the problem of the light $\sigma$-meson: this is singular
behaviour of the quark--antiquark interaction block at small momentum
transfers,  $1/q^4$,  in the coordinate space
this corresponds to the linear growth of the potential,
$\sim \alpha r$, at large $r$ (in quark models such a potential is
conventionally called "confinement potential"). Linear behaviour of
meson trajectories in the $(n,M^2)$-plane at large masses (Figs. 1,2)
tells us that such a behaviour occurs actually at rather large
 $r$, up to $r\sim 2.0-2.5$ fm. Assuming that both
colour components of the "confinement potential", octet and singlet
ones, behave at large $r$ similarly,
$V^{(1)}_{octet}(r)\sim V^{(8)}_{singlet}(r)$,  we gain singular
behaviour at small masses in the white (hadronic) channel.
Hadron unitarization of the amplitude (the account for the
$\pi\pi$-rescatterings), which is necessary in the white channel but is
absent in colour one, is capable to "hide" the white-channel singularity
under the $\pi\pi$ branching cut
thus re-creating the picture
of the light $\sigma$-meson. The mechanism of colour neutralization,
where the  $t$-channel formation of new
$q\bar q$ pairs is of principal importance, favours the hypothesis about
similar growth of potentials
$V^{(1)}_{confinement}(r)\sim  V^{(8)}_{confinement}(r)$.

As concern the broad state $f_0(1200-1600)$, we know neither its mass
nor the width reliably. Still, we do know that this broad state is
strongly produced  in a number of reactions. Also we know that, in
terms of the $q\bar q$  and gluonium components, this state is for sure
the mixture of
the gluonium with the quark-antiquark state which is close, by its
content, to the flavour singlet
$(q\bar q)_{glueball}=(n\bar n \sqrt{2}+s\bar s\sqrt{\lambda})/
\sqrt{2+\lambda}$ with $\lambda\simeq 0.5-0.8$:
 this fact is proved rather confidently by the
relations between the couplings
$f_0(1200-1600)\to\pi\pi,K\bar K, \eta\eta,\eta\eta'$, which are
reliably found from the  data.

Another reliably determined feature of the broad state $f_0(1200-1600)$
is that it is a more loosely-bound system than the surrounding
$f_0$-resonances; this fact points to a  possibility to create it
as a result of the accumulation of widths of neighbouring
resonances. If so, and the $K$-matrix analyses of the $00^{++}$ waves
\cite{K,YF} performed on an extended data basis favours this scenario,
then this circumstance explains us where the lightest scalar glueball
vanishes, though it promised to be around $\sim1500\,$MeV but it is
definitely missing among the observed narrow $f_0$-mesons. In this way,
let us emphasize that the possibility for a new comparatively narrow
$00^{++}$ resonance near 1000-1800 MeV is strictly excluded by
the available experimental data.

A scenario for $f_0(1200-1600)$ being the glueball descendant, which
lost a part of its gluonium component due to the mixing with
neighbouring resonances, was suggested in
\cite{glueball,glueball_model}, on the basis of data which were much
poorer than the present ones. The nowaday data which are richer not
only in statistics  but also in a number of investigated reactions, fit
to this picture rather well, without noticeable contradictions to it.
One more check of the picture with $f_0(1200-1600)$ as a glueball
descendant is expected in the nearest future for the high-statistics
radiative decays $J/\Psi\to\gamma+hadrons$:
in  hadronic spectra, the production of comparatively narrow
resonances $f_0(1300)$ and $f_0(1500)$ in the $00^{++}$ wave should be
accompanied by a strong "background" due to the production of the broad
state $f_0(1200-1600)$.

At the time being the mixing of $f_0$-mesons at 1300--1700 MeV is
intensively discussed, see recent papers \cite{CL,Weingarten} as an
example. However, it should be specially stressed that the mixing
considered in the $K$-matrix technique is in principle different from
mixing discussed in
\cite{CL,Weingarten}. In the $K$-matrix approach,
the mixing occurs due to transitions of bare states to real mesons ---
in other words, due to imaginary parts of hadron loop diagrams.
As a result of such a mixing the amplitude poles "move" in the complex
mass plane, a loss or accumulation of widths take place, that is,
mixed states "push" each other in the complex plane by sinking deeply
into complex plane or springing out  to
real axis; as a result, real mesons-resonances are created.
In this way the shift of real part of resonance mass may be
considerable --- of the order of resonance width.
The mixing discussed in the papers
 \cite{CL,Weingarten} happens without decay processes, that is,
without imaginary parts of hadronic loop diagrams, so corresponding
amplitude poles are on real axis: by mixing the levels repulse each
 other, with the increase or decrease of the mass but not width; after
such a mixing the bare states should be created but not real resonances,
as was suggested in papers
\cite{CL,Weingarten}. Once again: in the $K$-matrix technique the
formation of the observed resonances takes place by the onset of
the decay processes only.

In accordance with gluodynamical lattice calculations, the second scalar
glueball should exist at 2100--2200 MeV \cite{MP}. In
\cite{BPZ}, it was suggested that $f_0(2105)$ is either this second
glueball or strongly mixed with it. However there is not enough  data
to judge about the glueball being in this mass region: the discovered
resonances lay on the $q\bar q$ trajectories quite comfortably. It
is rather probable that the second scalar glueball turned into the
broad state too by mixing with the neighbouring $q\bar q$ mesons --- in
this case to fix it experimentally  the measurement
of a large variety of spectra is needed  as well as
the evaluation of  ratios for the $\pi\pi$, $K\bar K$,
$\eta\eta$, $\eta\eta'$, $\eta'\eta'$ yields in the region 2100--2200
MeV.

\section*{Acknowledgement} I am indebted to A.V. Anisovich,
Ya.I. Azimov, D.V. Bugg, L.G. Dakhno,
Yu. S. Kalashnikova, D.I. Melikhov, V.A. Nikonov and
A.V. Sarantsev for helpful discussions of problems involved.
This work is supported by the RFBR Grant N 0102-17861.

\begin{figure}[p]
\centerline{\epsfig{file=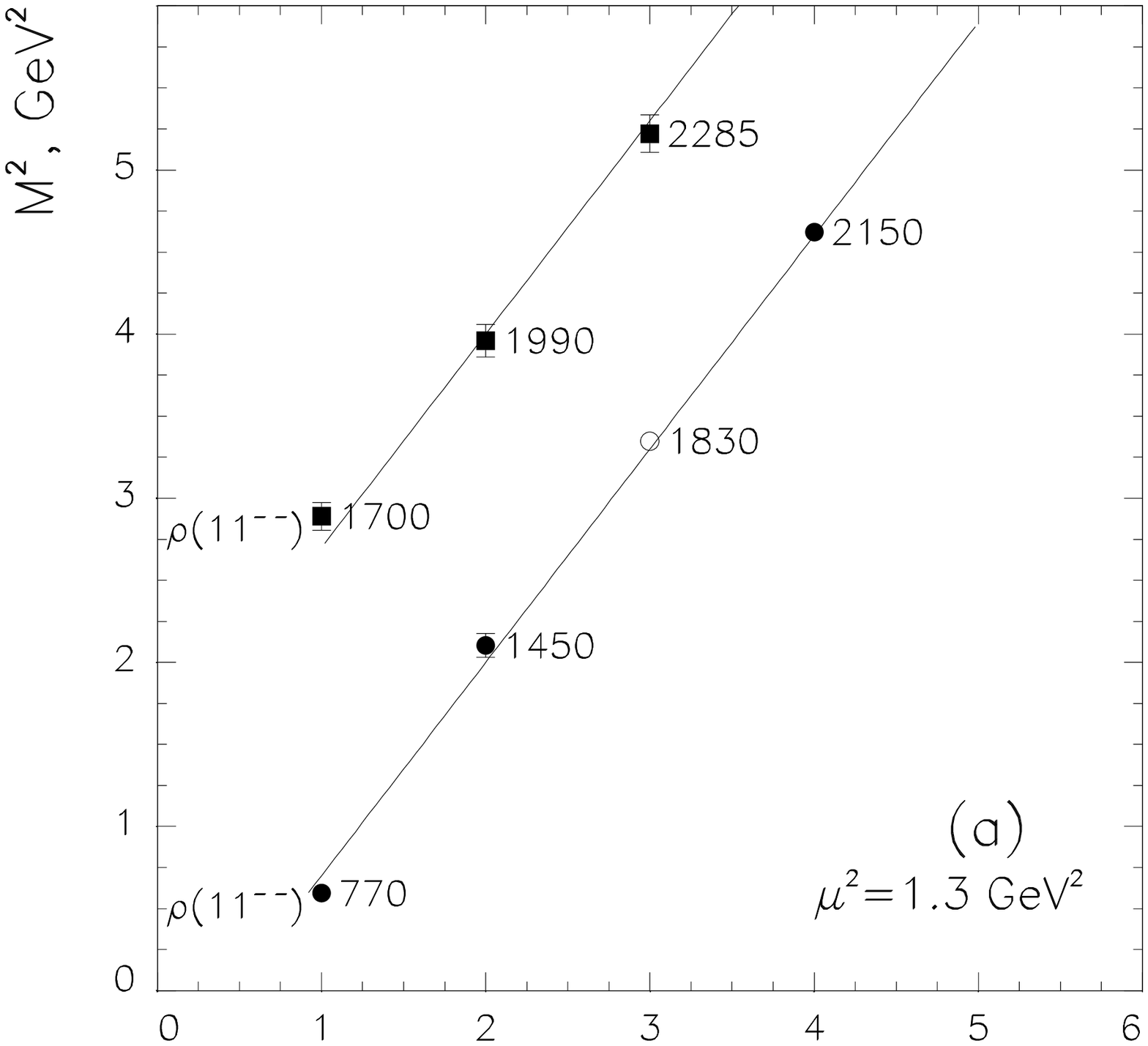,width=6cm}\hspace{-1.5cm}
            \epsfig{file=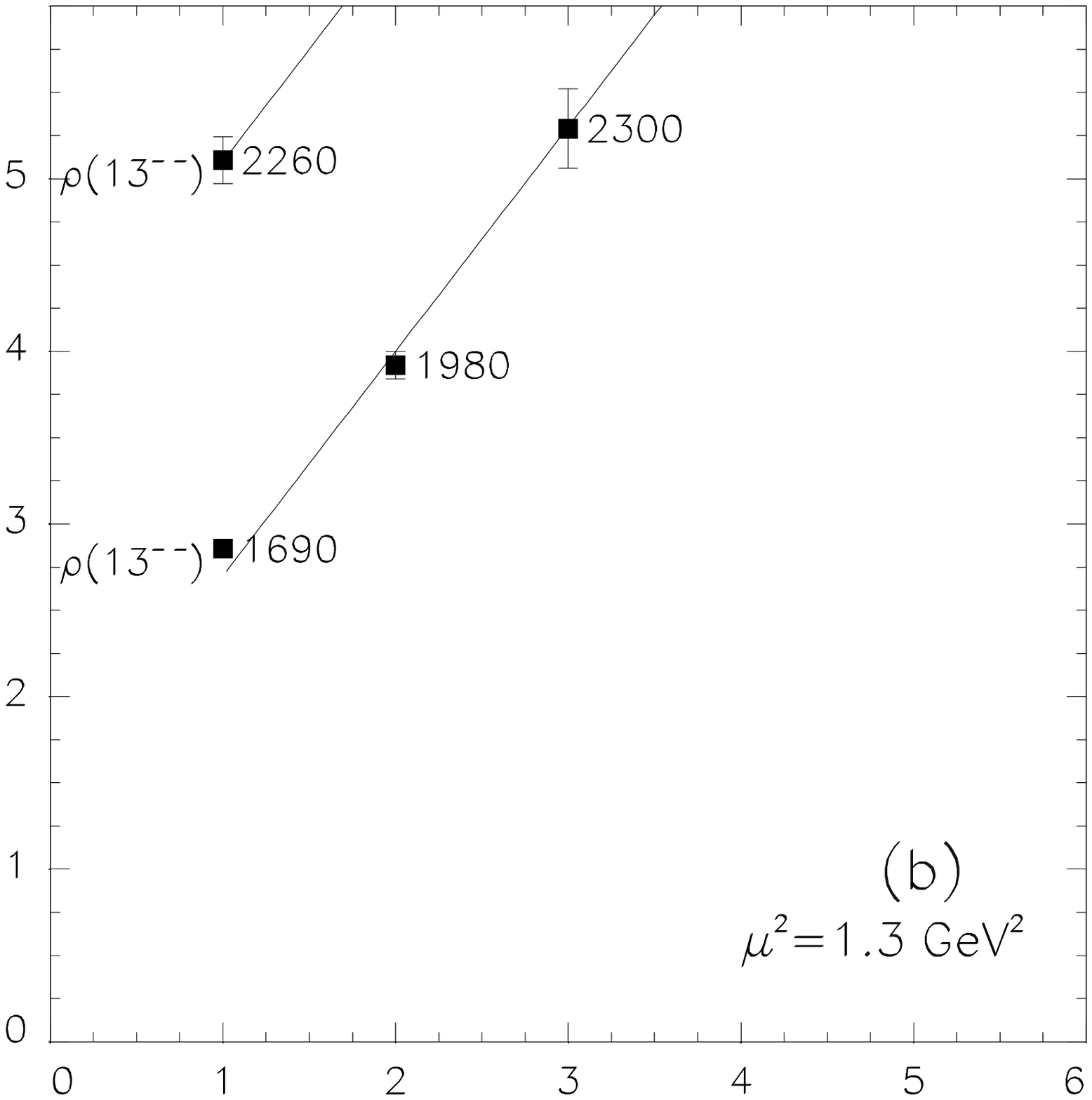,width=6cm}}
\vspace{-1.5cm}
\centerline{\epsfig{file=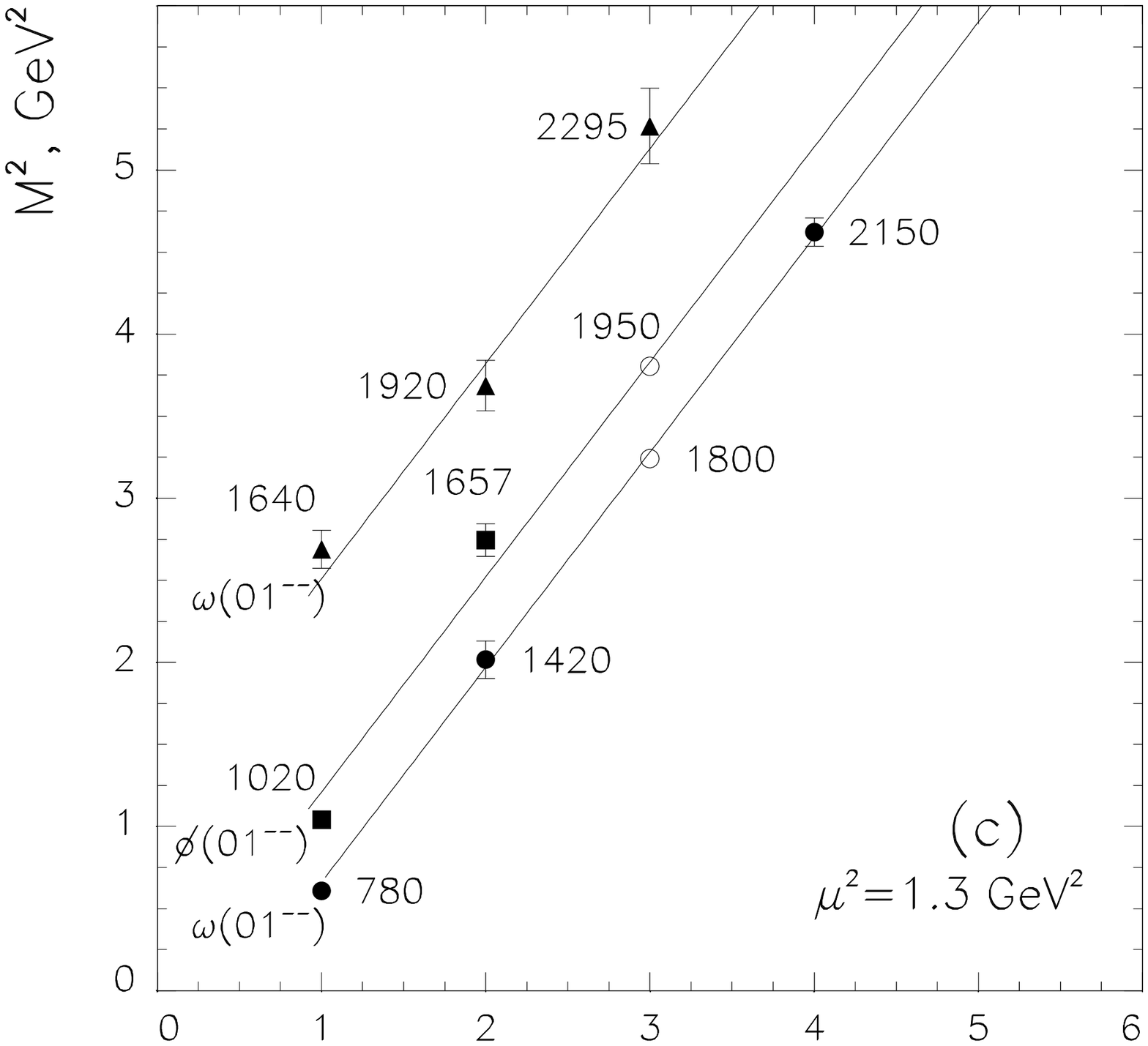,width=6cm}\hspace{-1.5cm}
            \epsfig{file=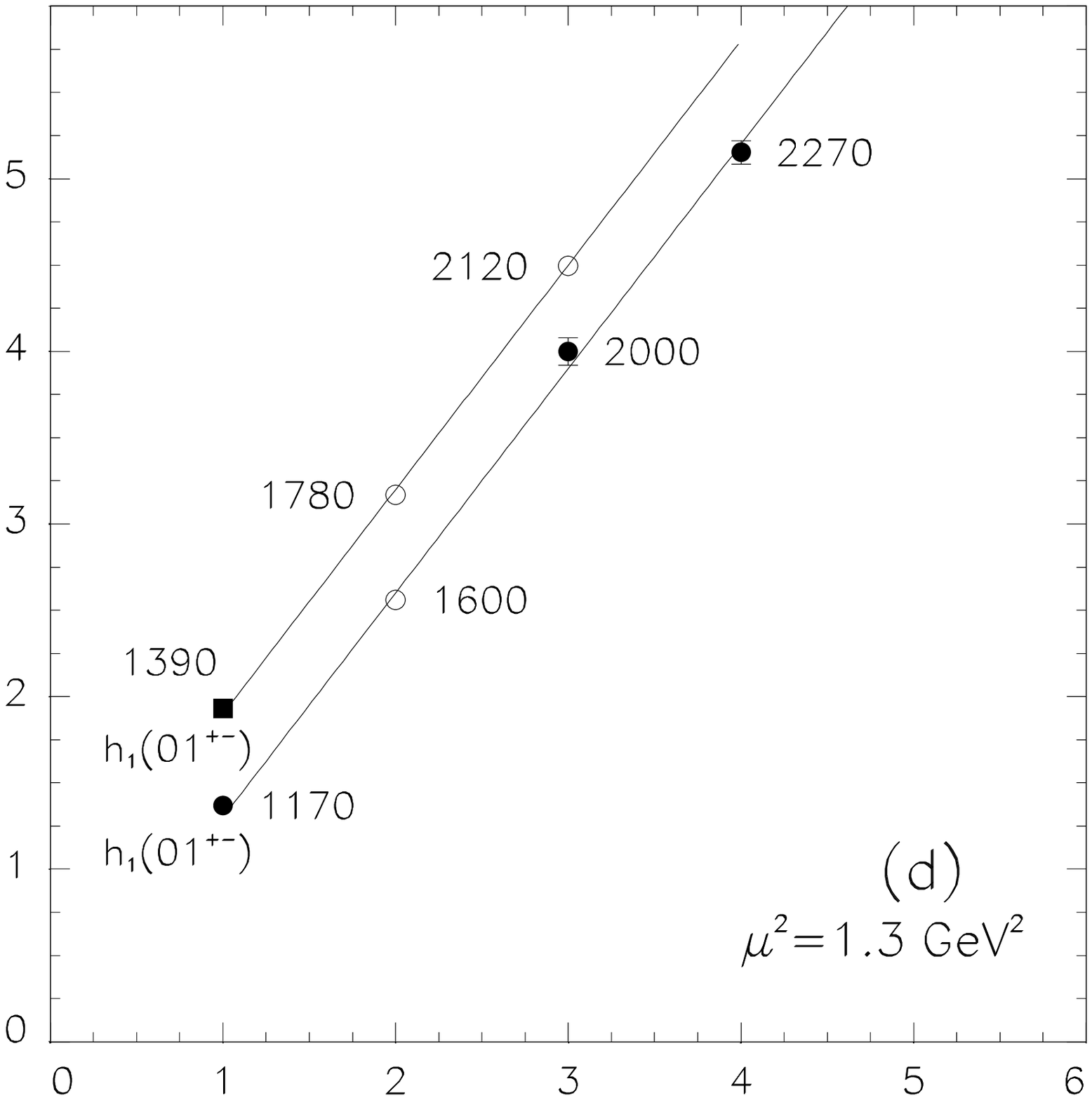,width=6cm}}
\vspace{-1.5cm}
\centerline{\epsfig{file=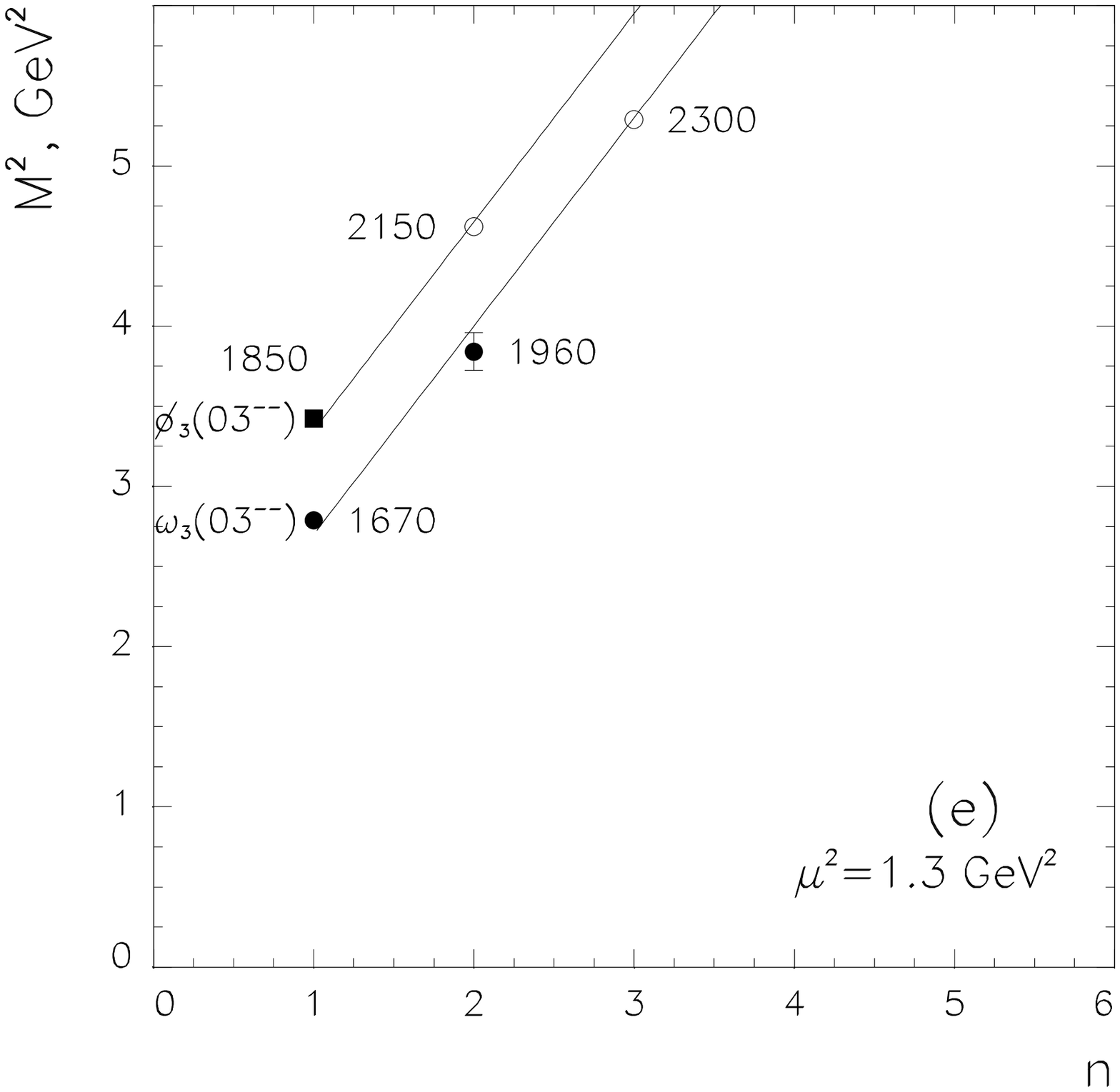,width=6cm}\hspace{-1.5cm}
            \epsfig{file=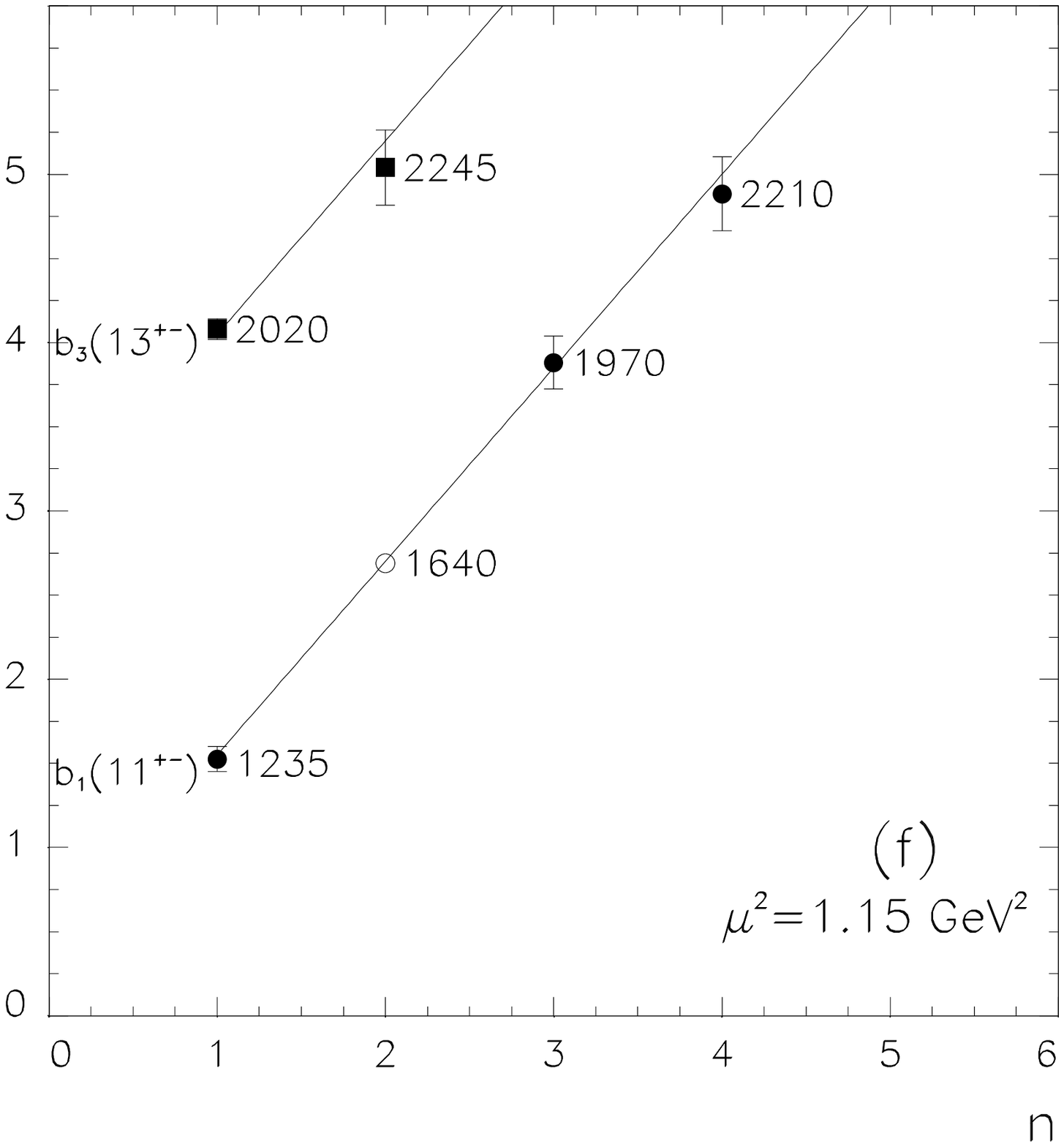,width=6cm}}
\caption{\footnotesize
Trajectories on the $(n,M^2)$ planes for the states with $(C=-)$.
Open circles stand for the predicted states.}
\end{figure}
\clearpage


\begin{figure}[p]
\centerline{\epsfig{file=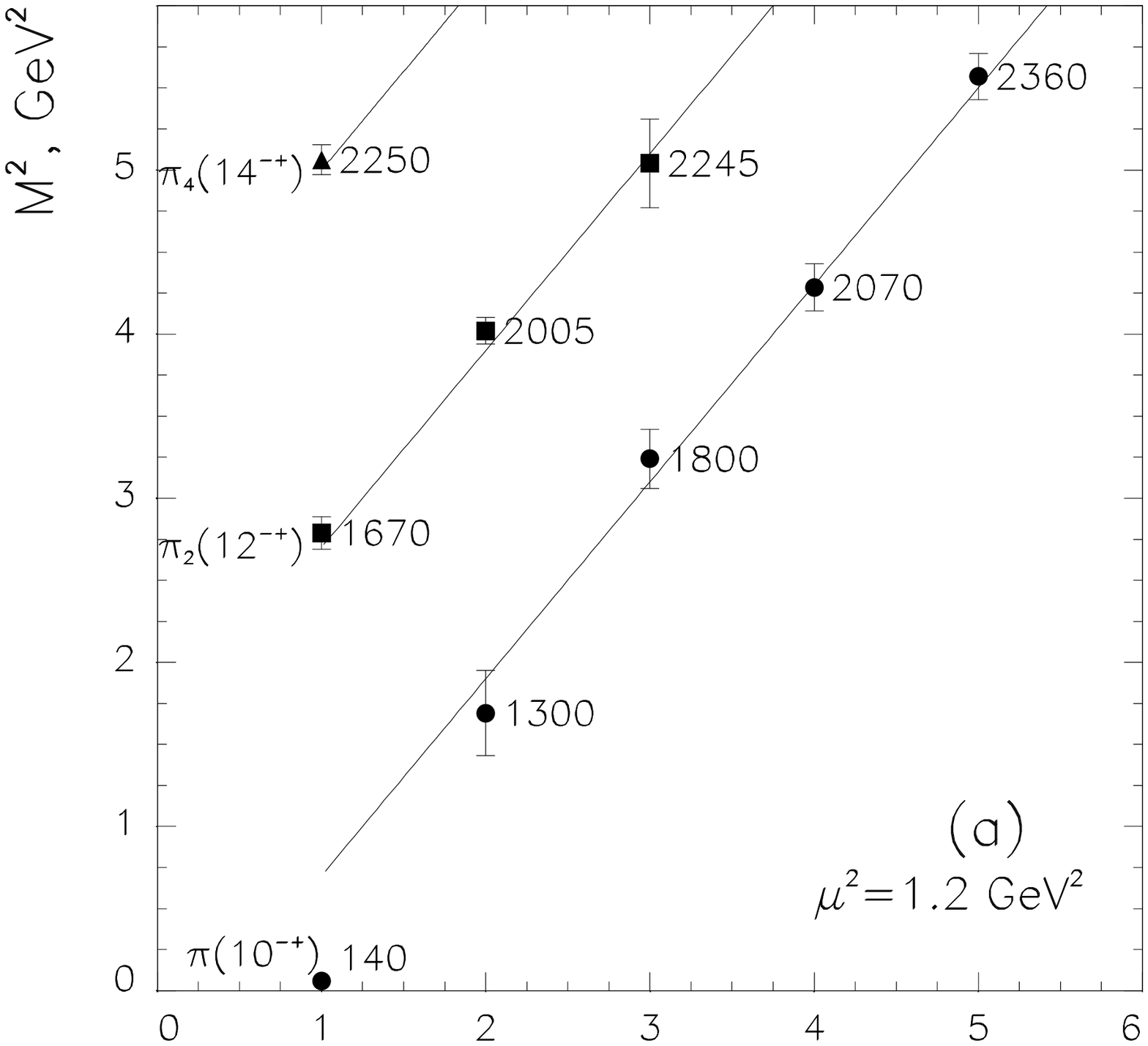,width=6cm}\hspace{-1.5cm}
            \epsfig{file=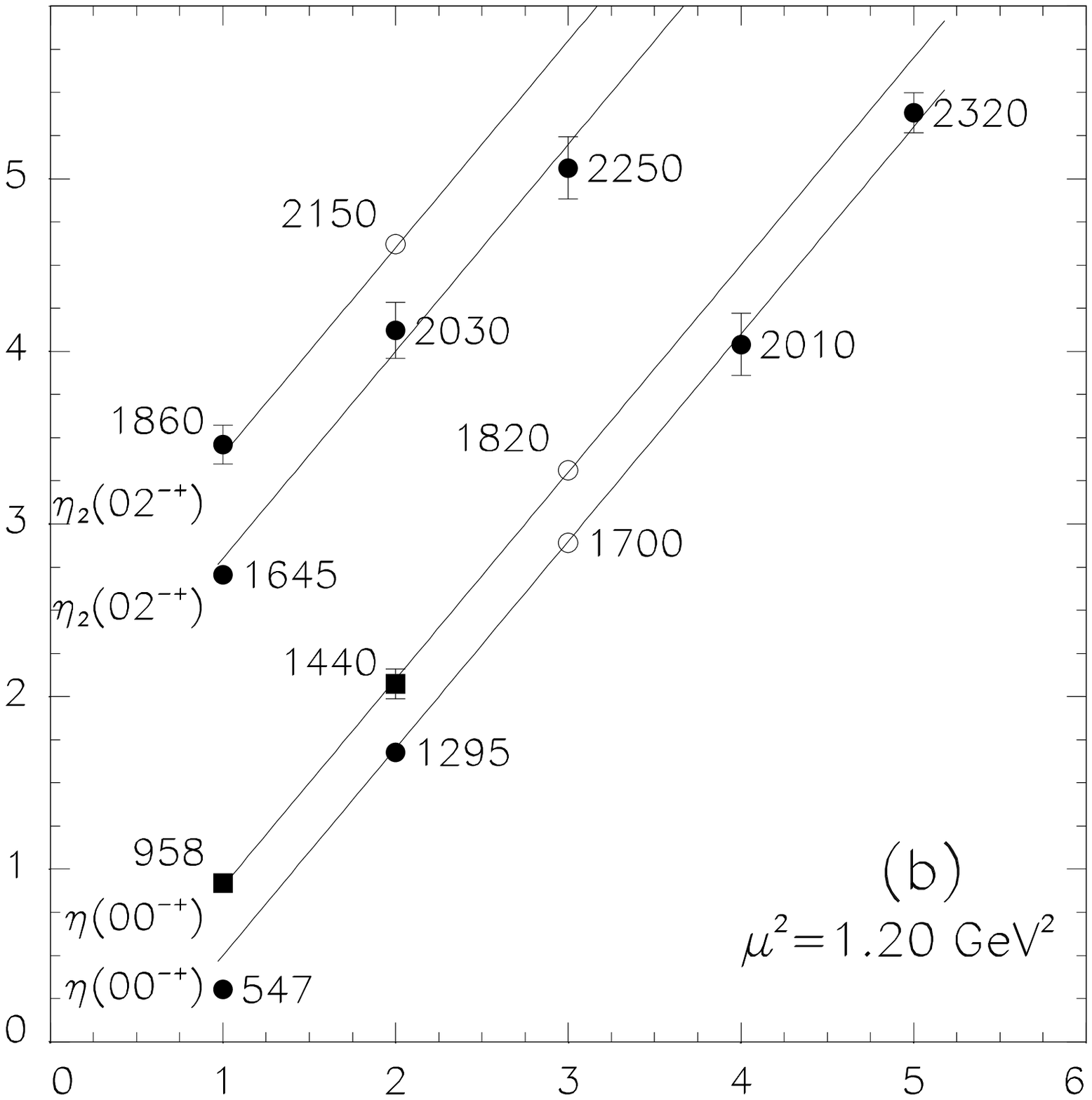,width=6cm}}
\vspace{-1.5cm}
\centerline{\epsfig{file=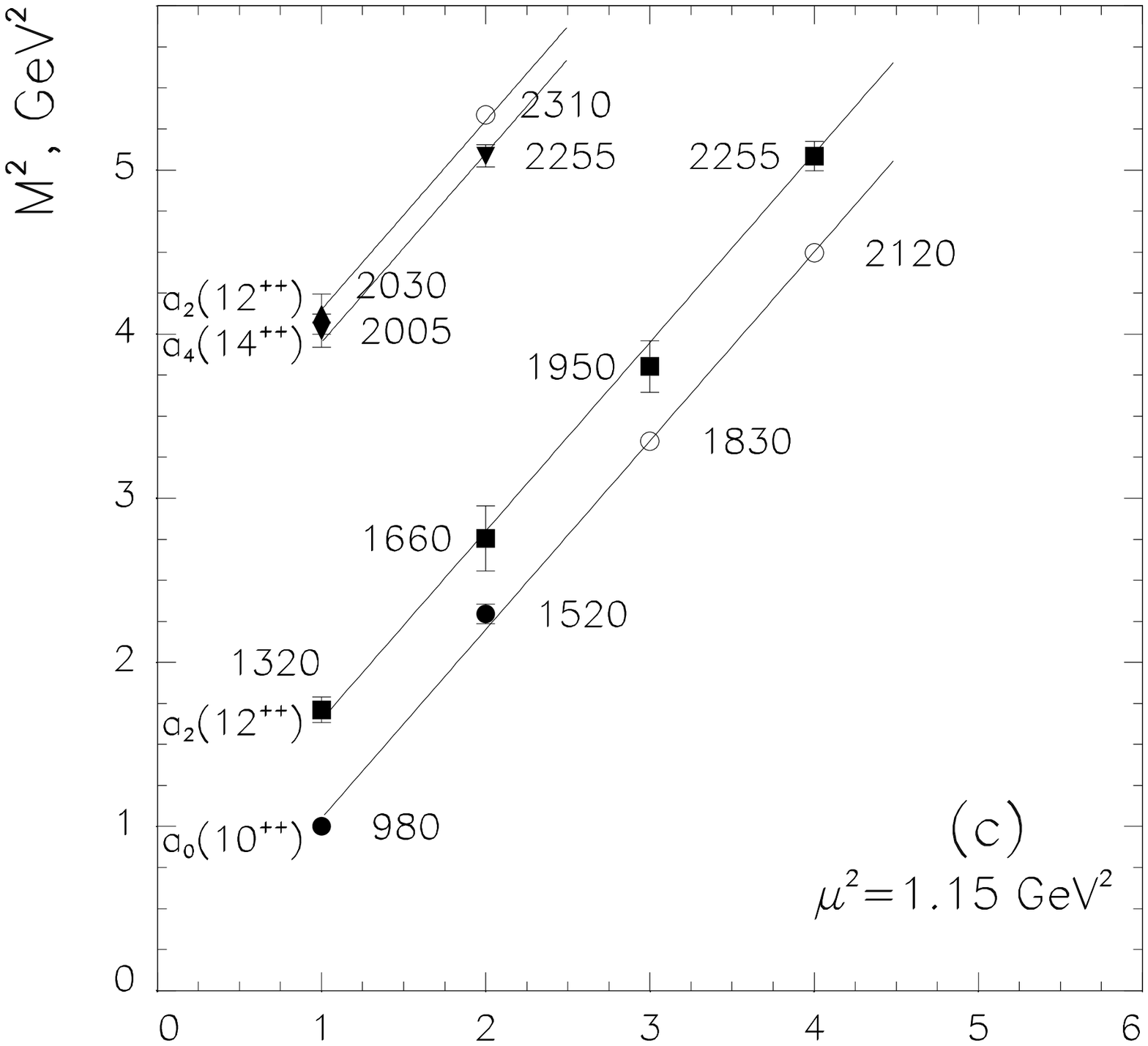,width=6cm}\hspace{-1.5cm}
            \epsfig{file=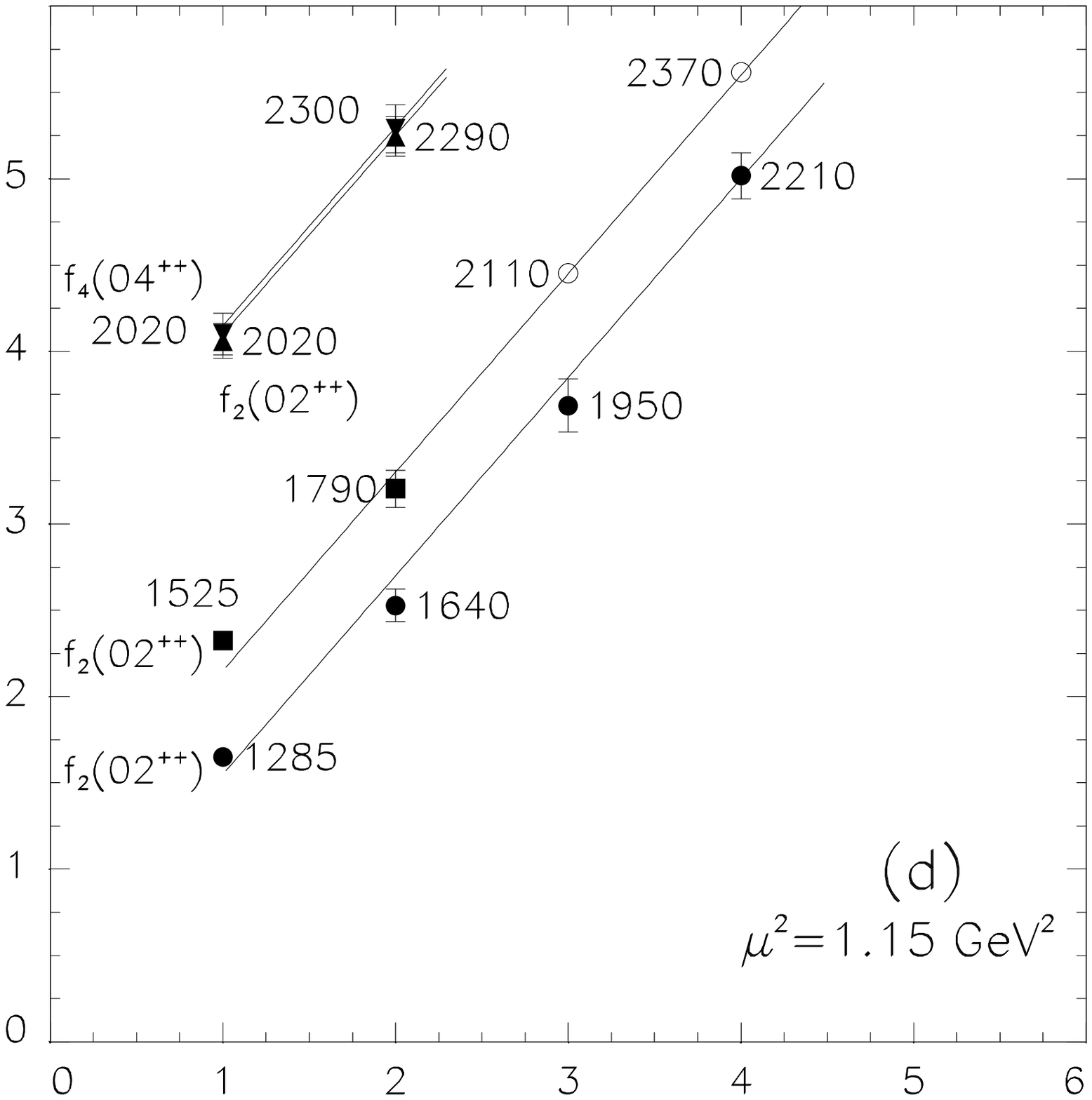,width=6cm}}
\vspace{-1.5cm}
\centerline{\epsfig{file=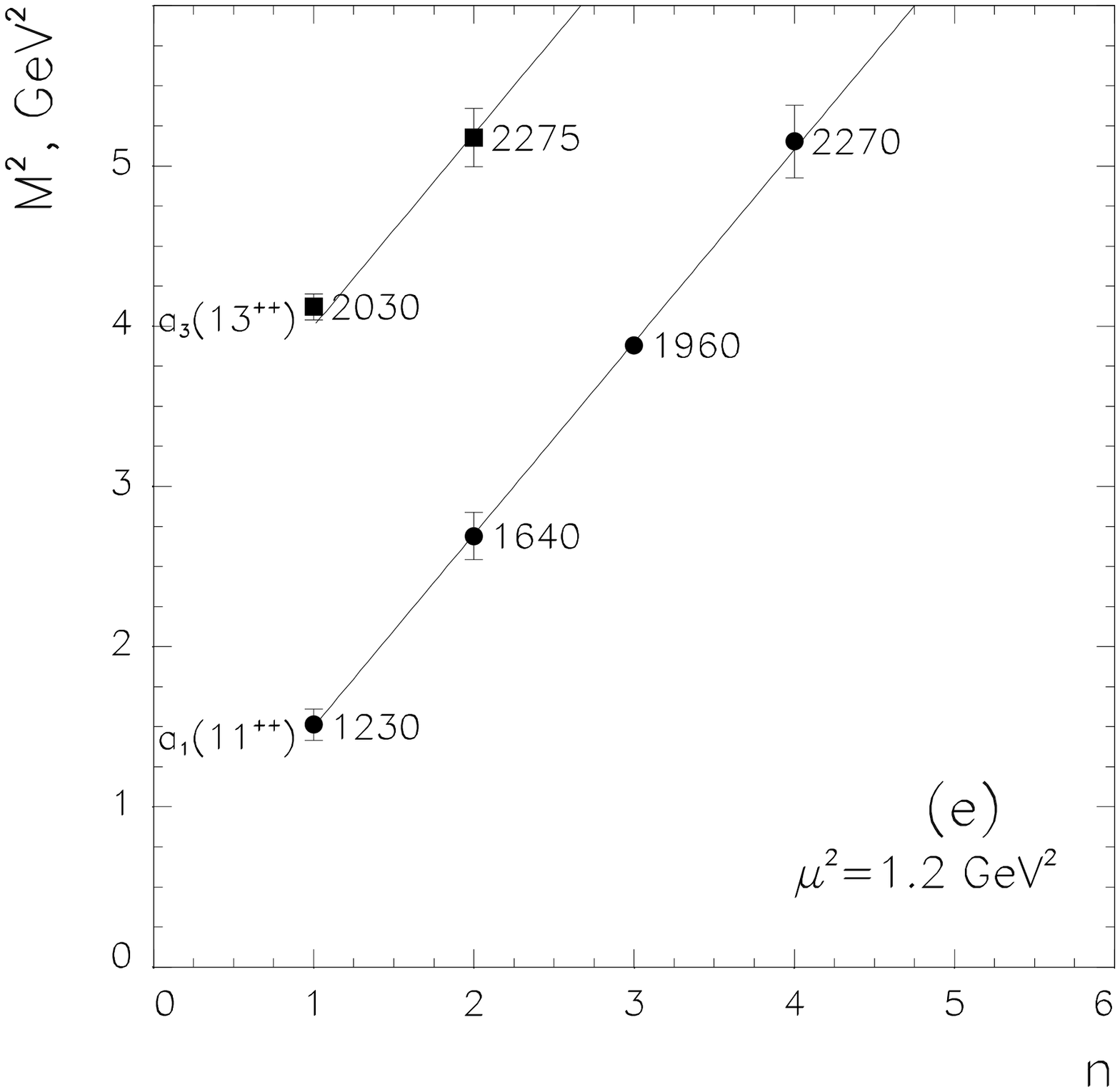,width=6cm}\hspace{-1.5cm}
            \epsfig{file=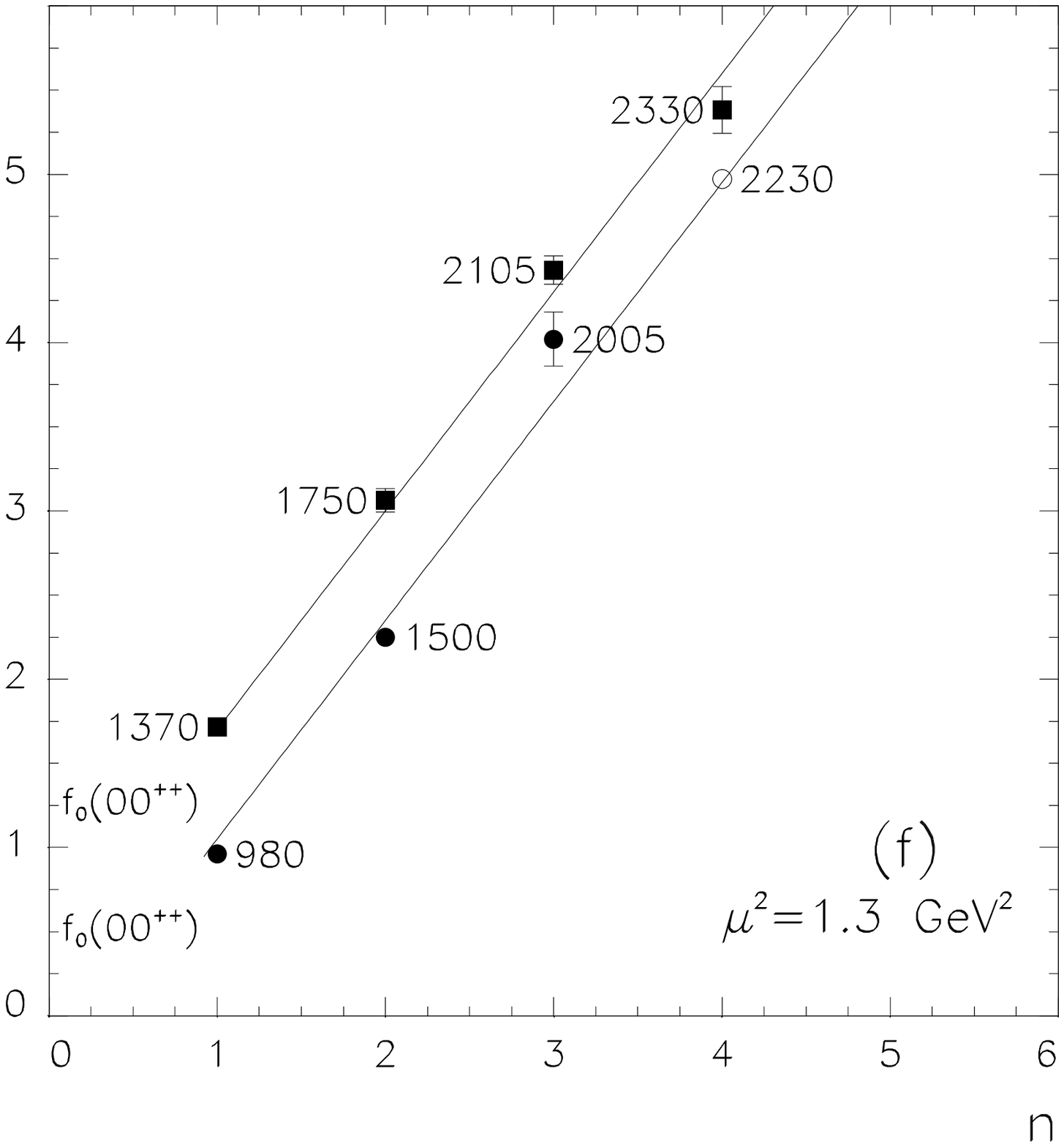,width=6cm}}
\caption{\footnotesize Trajectories on the $(n,M^2)$ planes for
the states with  $(C=+)$.}
\end{figure}
\clearpage

\begin{figure}[p]
\centerline{\epsfig{file=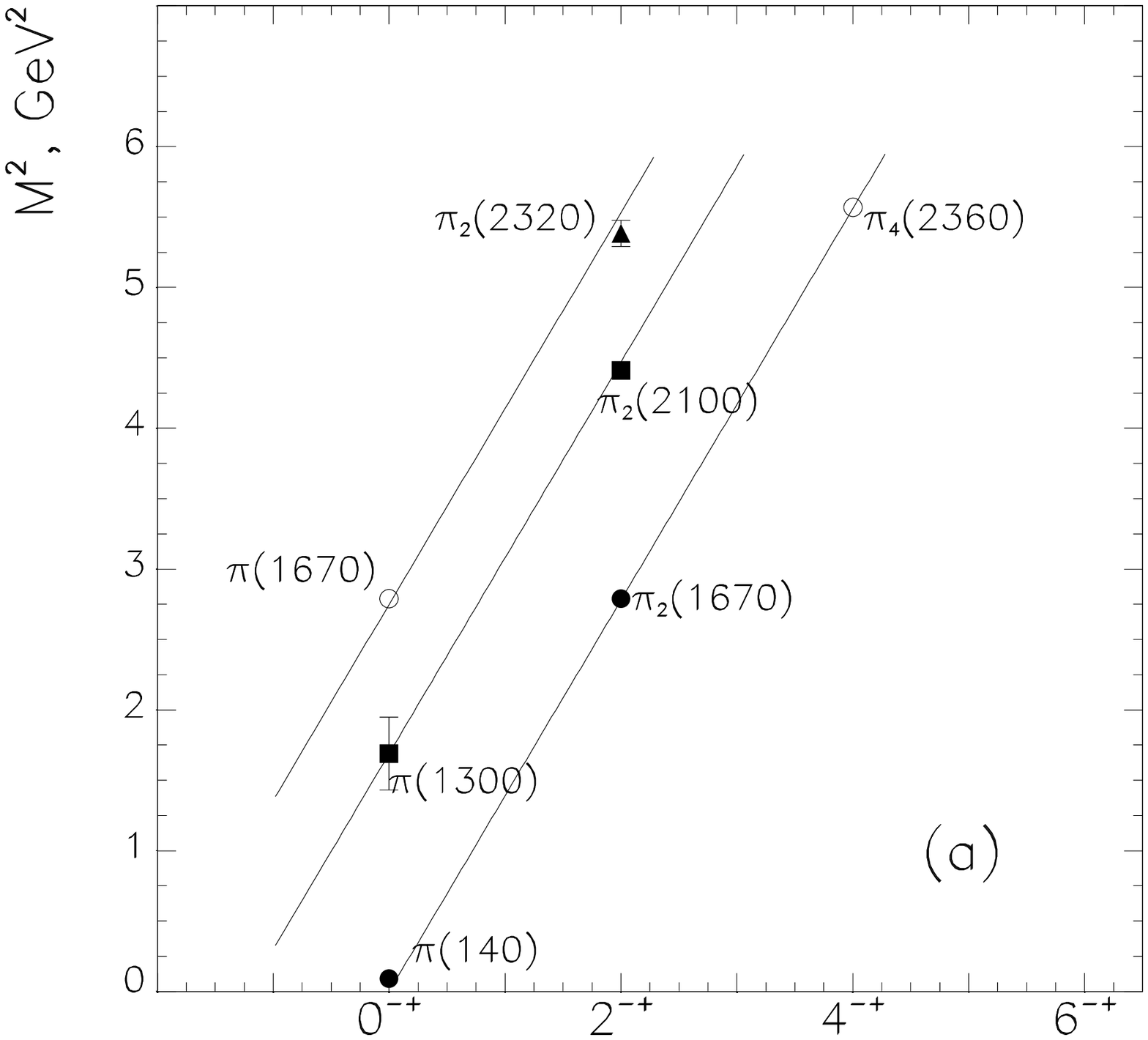,width=6cm}\hspace{-1.5cm}
            \epsfig{file=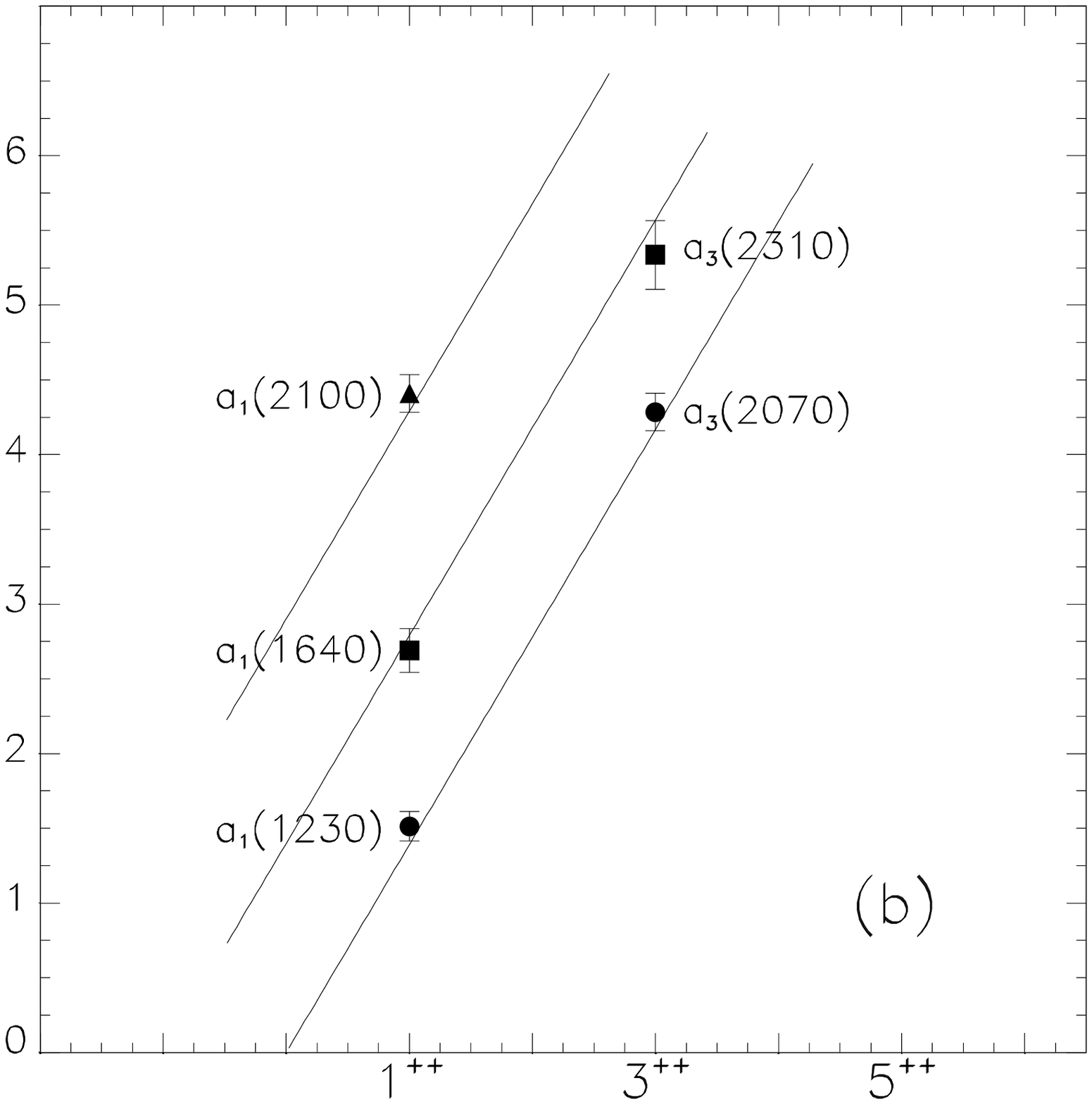,width=6cm}}
\vspace{-1.5cm}
\centerline{\epsfig{file=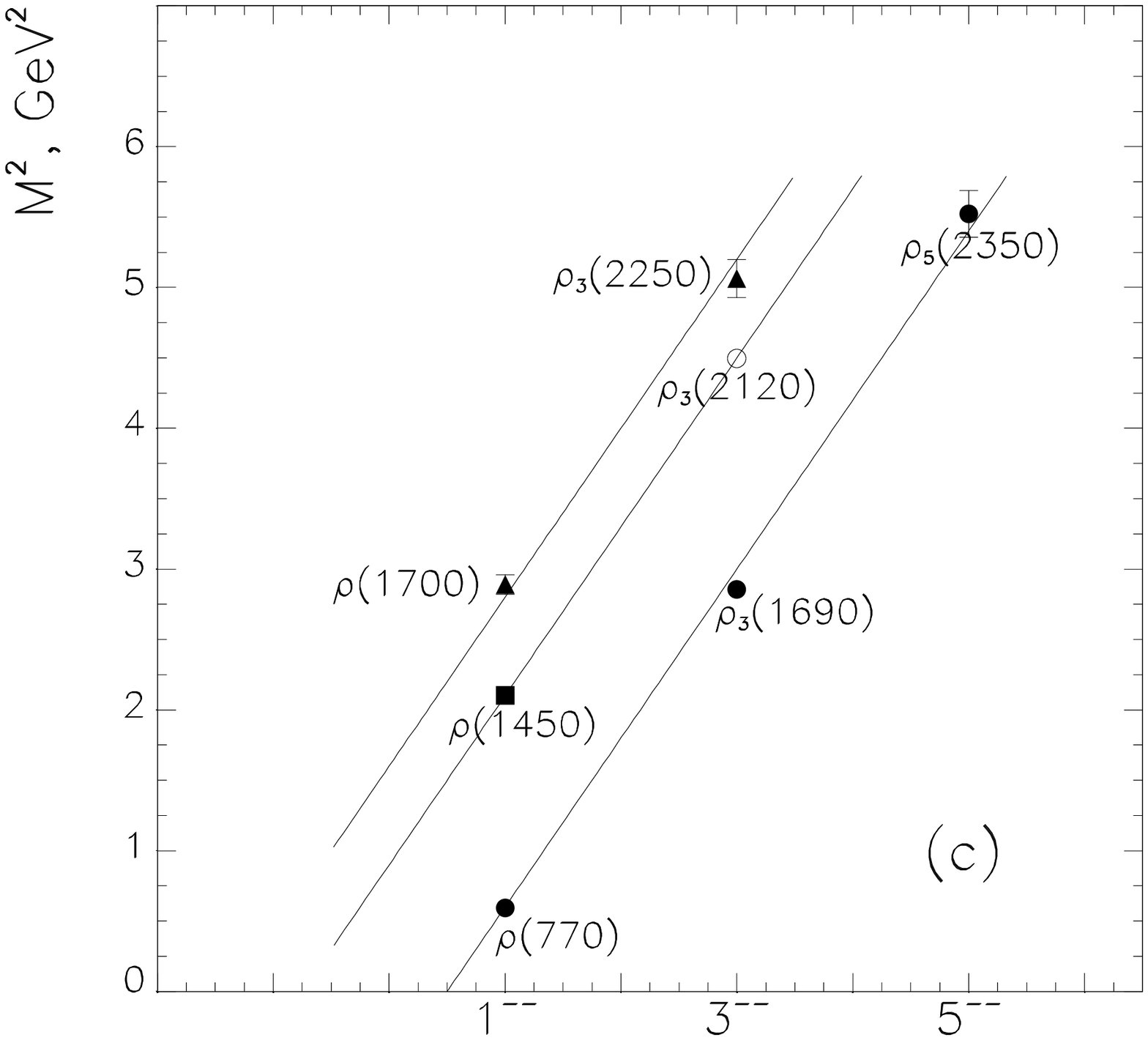,width=6cm}\hspace{-1.5cm}
            \epsfig{file=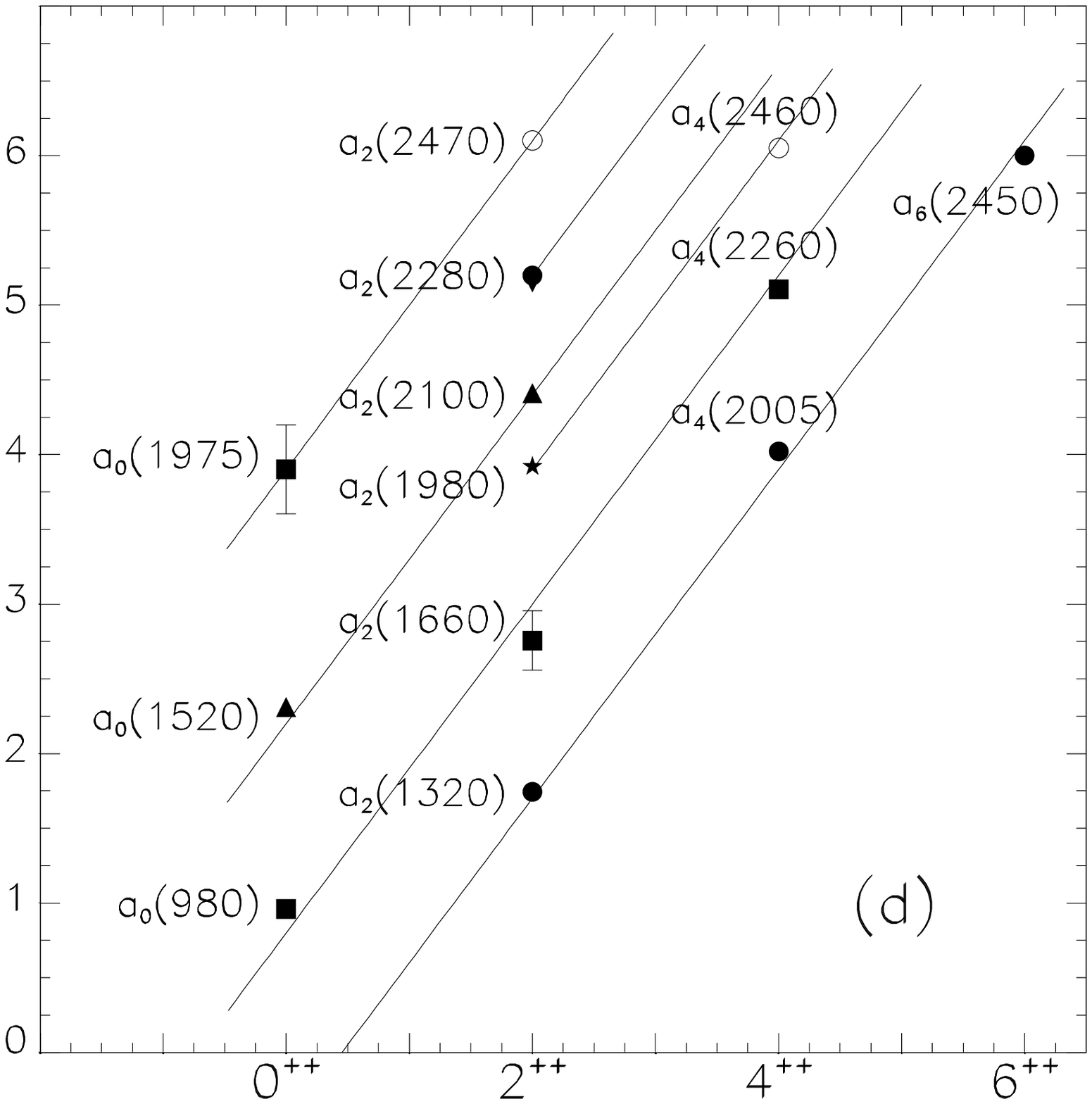,width=6cm}}
\vspace{-1.5cm}
\centerline{\epsfig{file=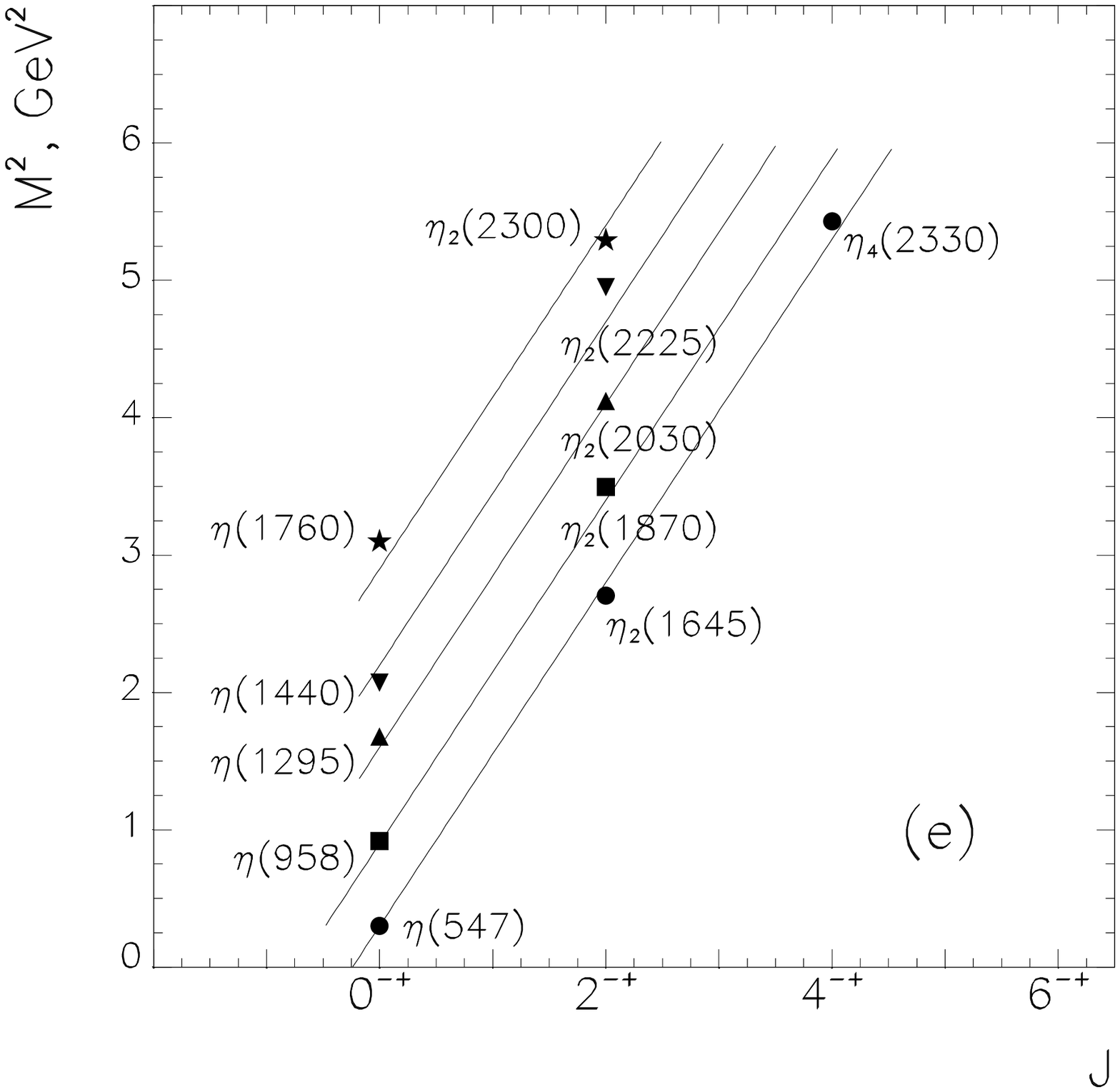,width=6cm}\hspace{-1.5cm}
            \epsfig{file=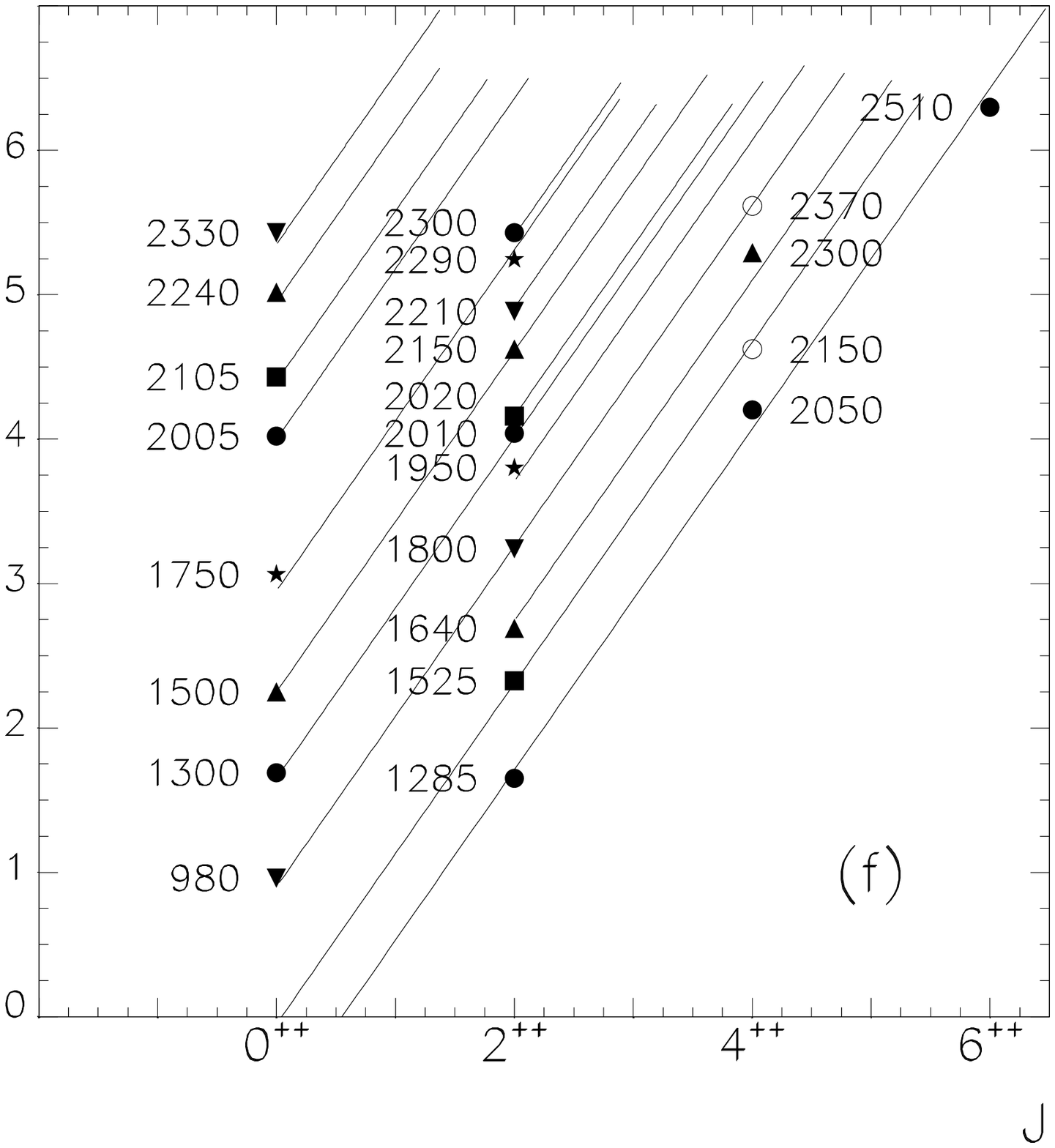,width=6cm}}
\caption{\footnotesize
 $(J, M^2)$ planes for leading and daughter trajectories:
a)~$\pi$-trajectories, b)~$a_1$-trajectories,
c)~$\rho$-trajectories, d)~$a_2$-trajectories,
e)~$\eta$-trajectories, f)~$P'$-trajectories.}
\end{figure}
\clearpage

\newpage
\begin{figure}
\centerline{\epsfig{file=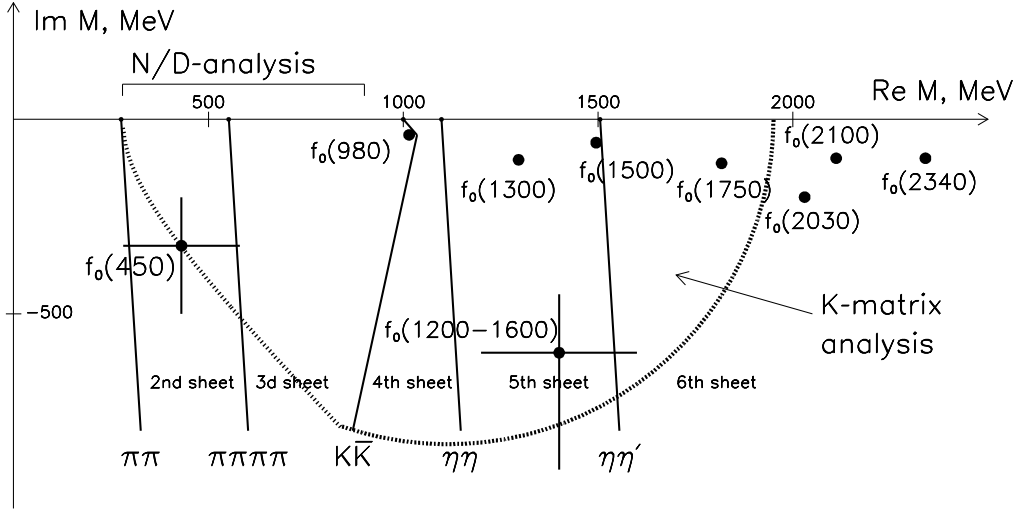,width=14cm}}
\caption{ Complex $M$ plane in the $(IJ^{PC}=00^{++})$ sector. Dashed
line encircle the part of the plane where the $K$-matrix analysis [8]
reconstructs the analytic $K$-matrix amplitude: in this area
the poles corresponding to resonances
$f_0(980)$, $f_0(1300)$, $f_0(1500)$,
$f_0(1750)$ and the broad state $f_0(1200-1600)$ are
located. On the border of this area
 the light $\sigma$-meson denoted as
$f_0(450)$ is shown (the position of pole corresponds to that found
 in the $N/D$ method [31]). Beyond the $K$-matrix analysis area, there
 are resonances
$f_0(2030),f_0(2100),f_0(2340)$ [6]. }
\end{figure}

\begin{figure}[h]
\centerline{\epsfig{file=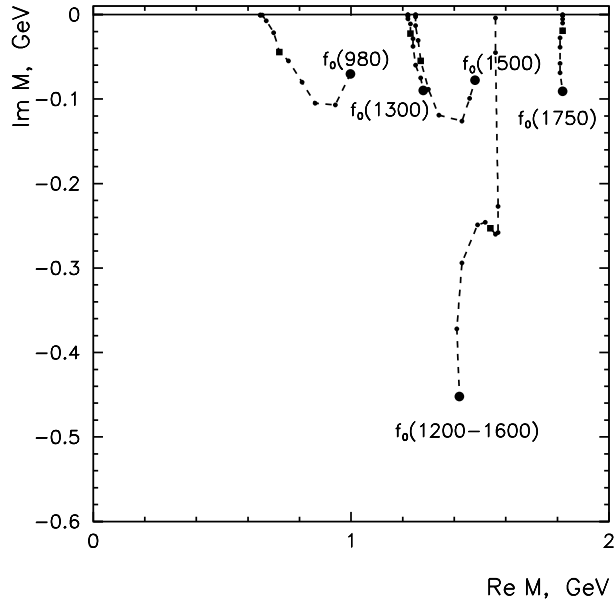,width=9cm}}
\caption{\footnotesize Complex $M$ plane: trajectories of poles
corresponding to the states $f_0(980)$, $f_0(1300)$, $f_0(1500)$,
$f_0(1750)$, $f_0(1200-1600)$ within a uniform onset of the decay
channels.}
\end{figure}

\newpage
\begin{figure}
\centerline{\epsfig{file=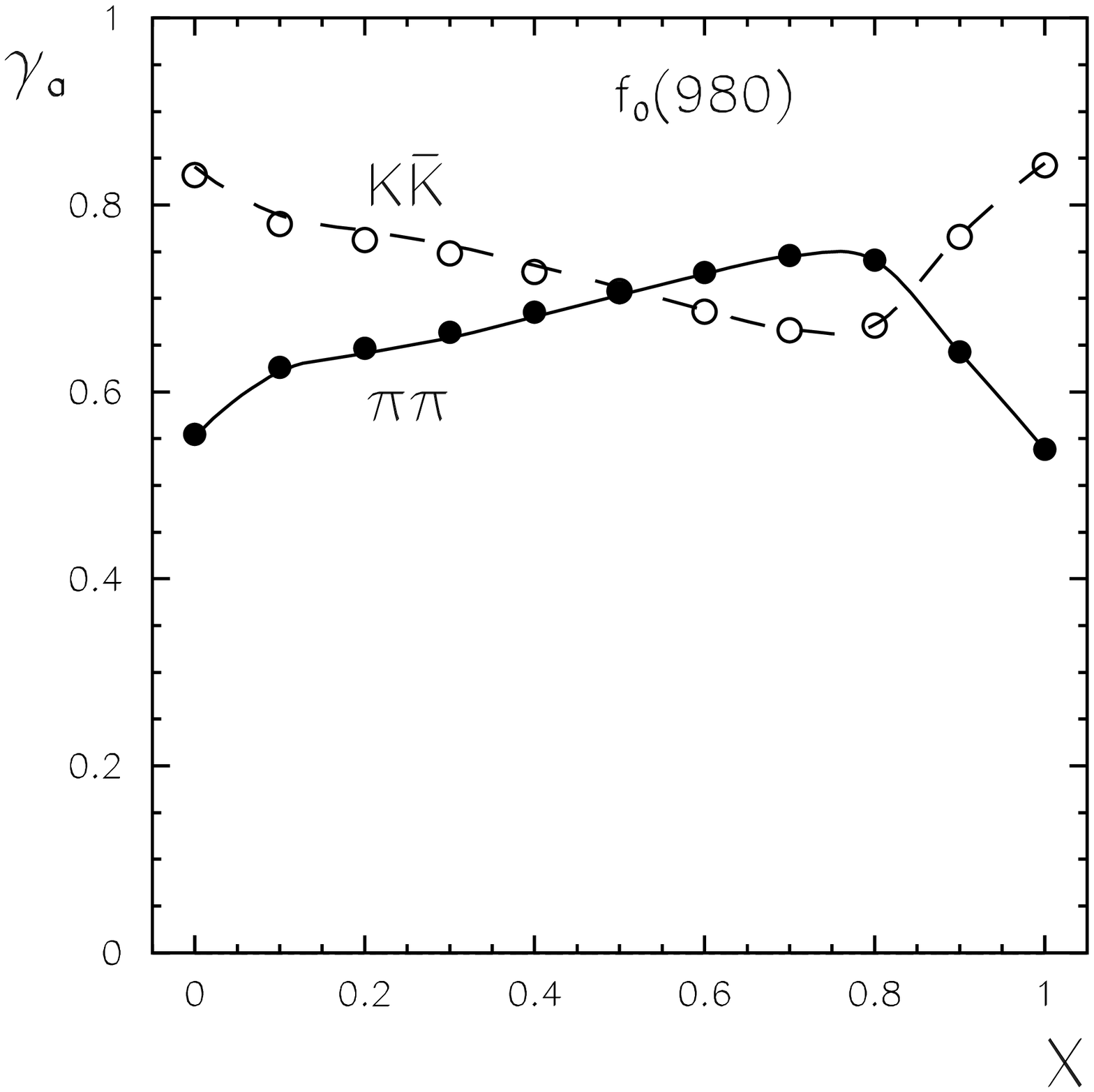,width=6cm}
            \epsfig{file=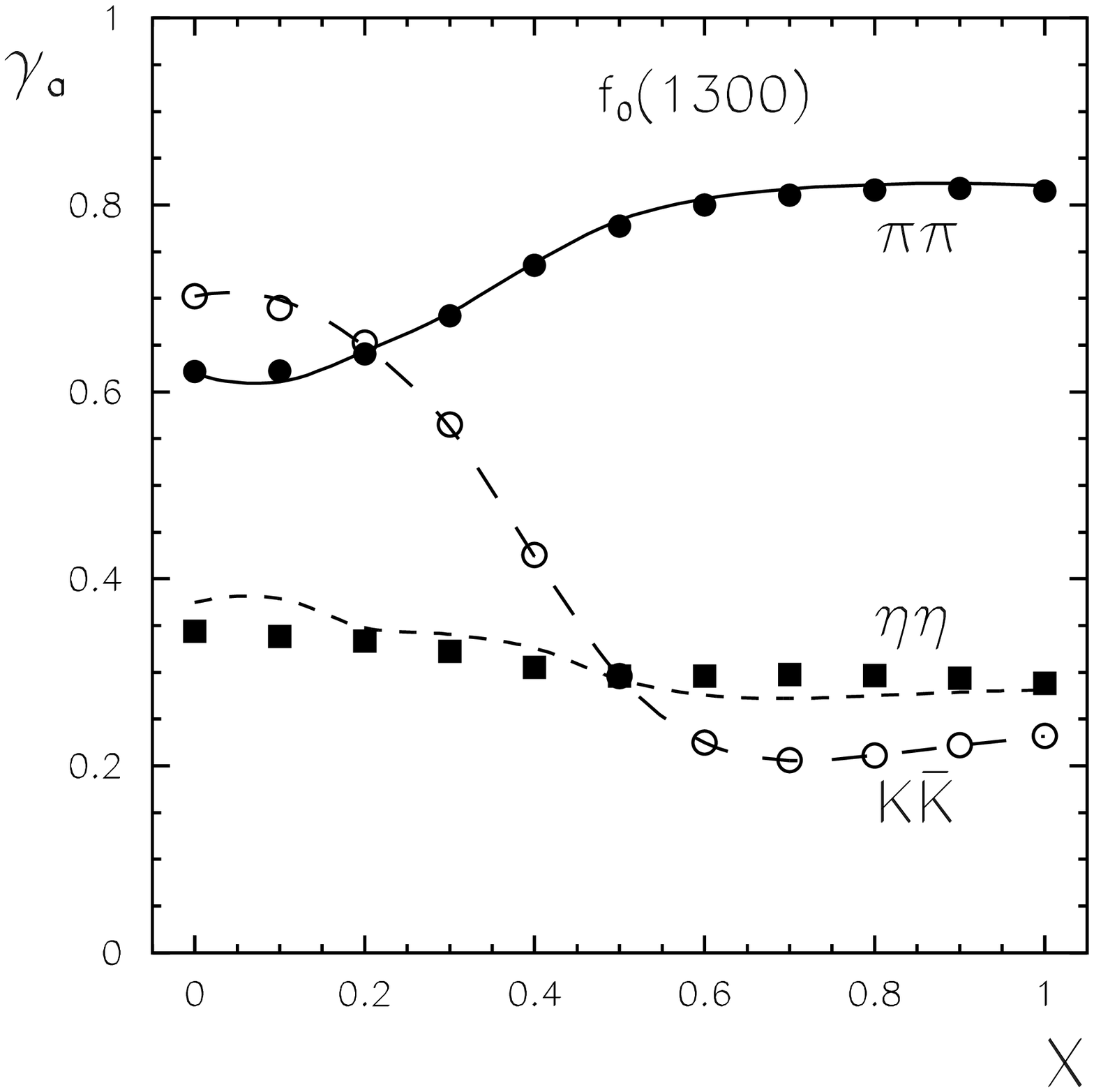,width=6cm}}
\centerline{\epsfig{file=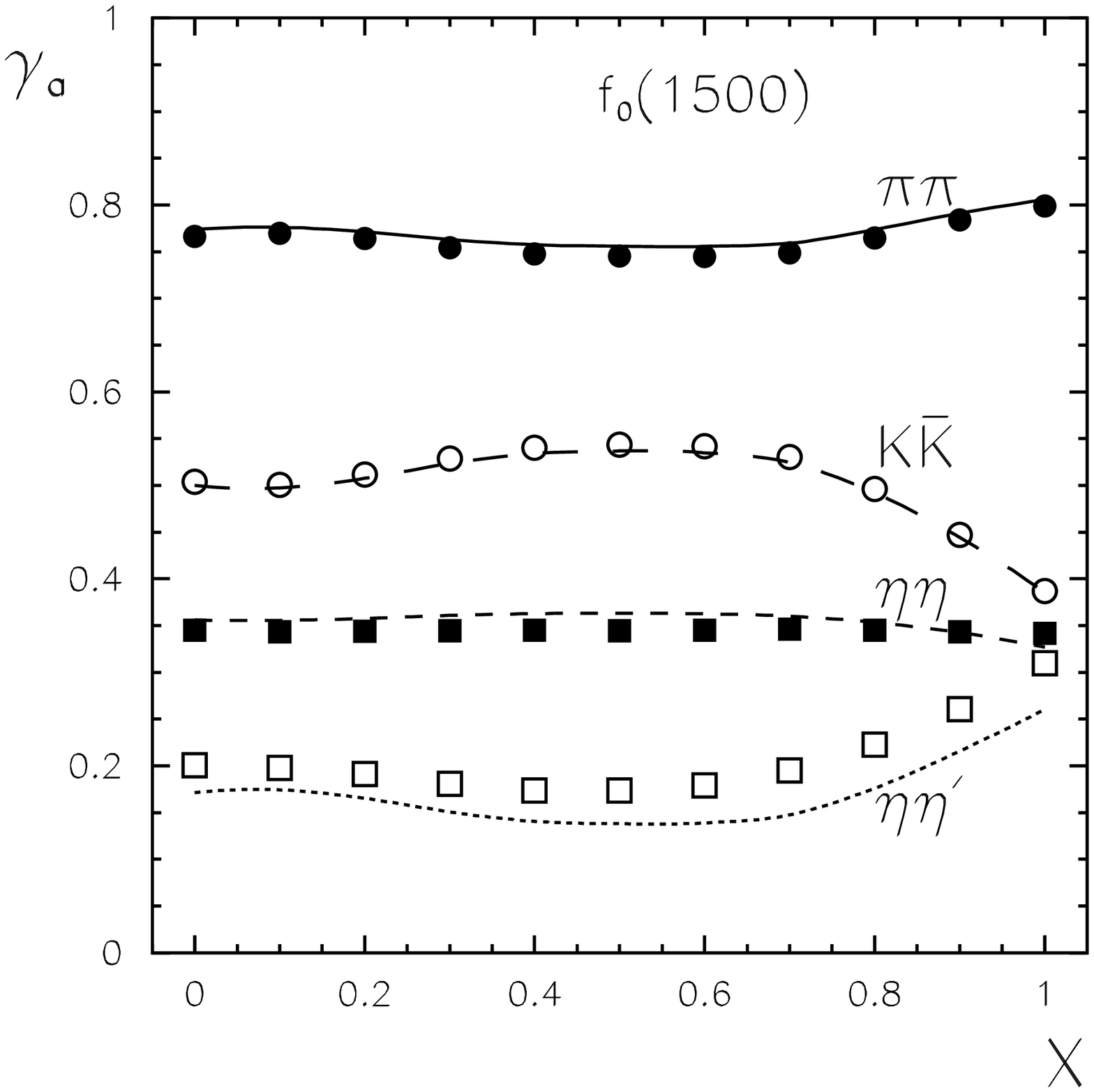,width=6cm}
            \epsfig{file=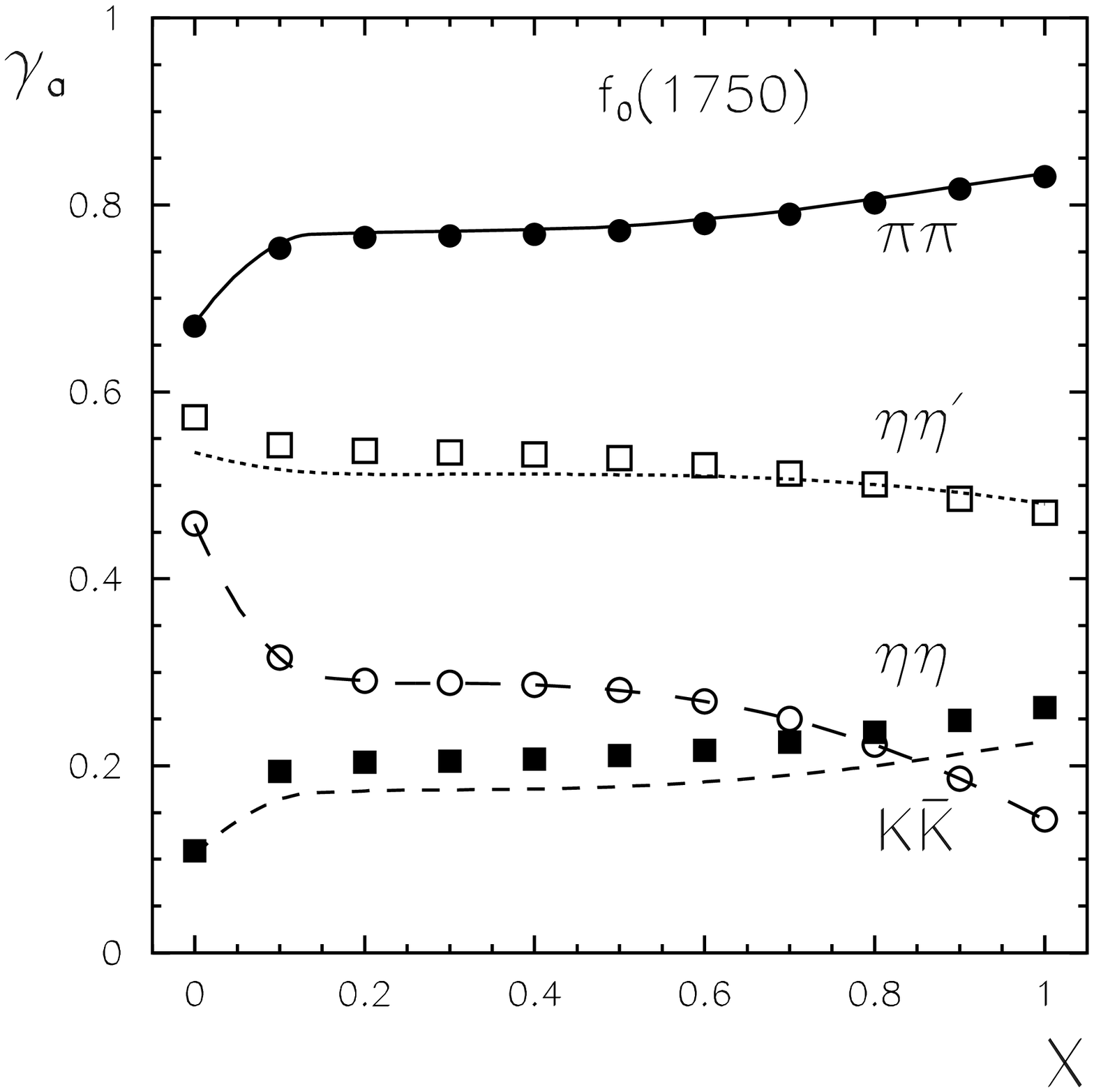,width=6cm}}

\caption{The evolution of normalized coupling constants
$\gamma_a =g_a/\sqrt{\sum_b g^2_b}$
at the onset of the decay channels for $f_0(980)$,
$f_0(1300)$, $f_0(1500)$, $f_0(1750)$.
Curves demonstrate the description of constants by formula (38).}
\end{figure}

\begin{figure}[h]
\centerline{\epsfig{file=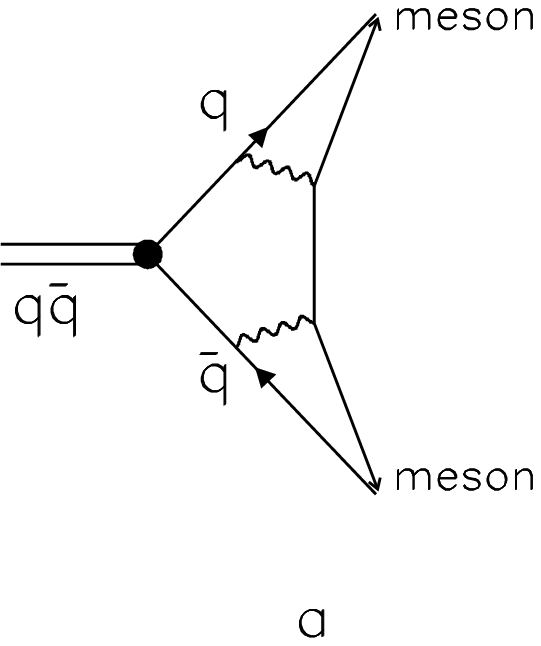,height=5cm}\hspace{0.5cm}
            \epsfig{file=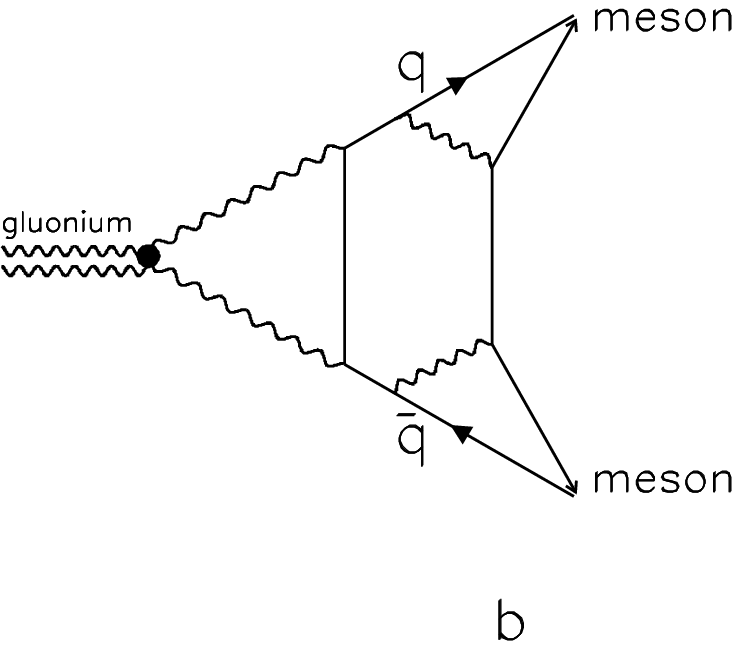,height=5cm}}
\caption{(a,b) \footnotesize Examples of planar diagrams
responsible for the decay of the  $q\bar q$-state and gluonium into
two $q\bar q$-mesons (leading terms in the $1/N$ expansion).}
\end{figure}

\begin{figure}
\centerline{\epsfig{file=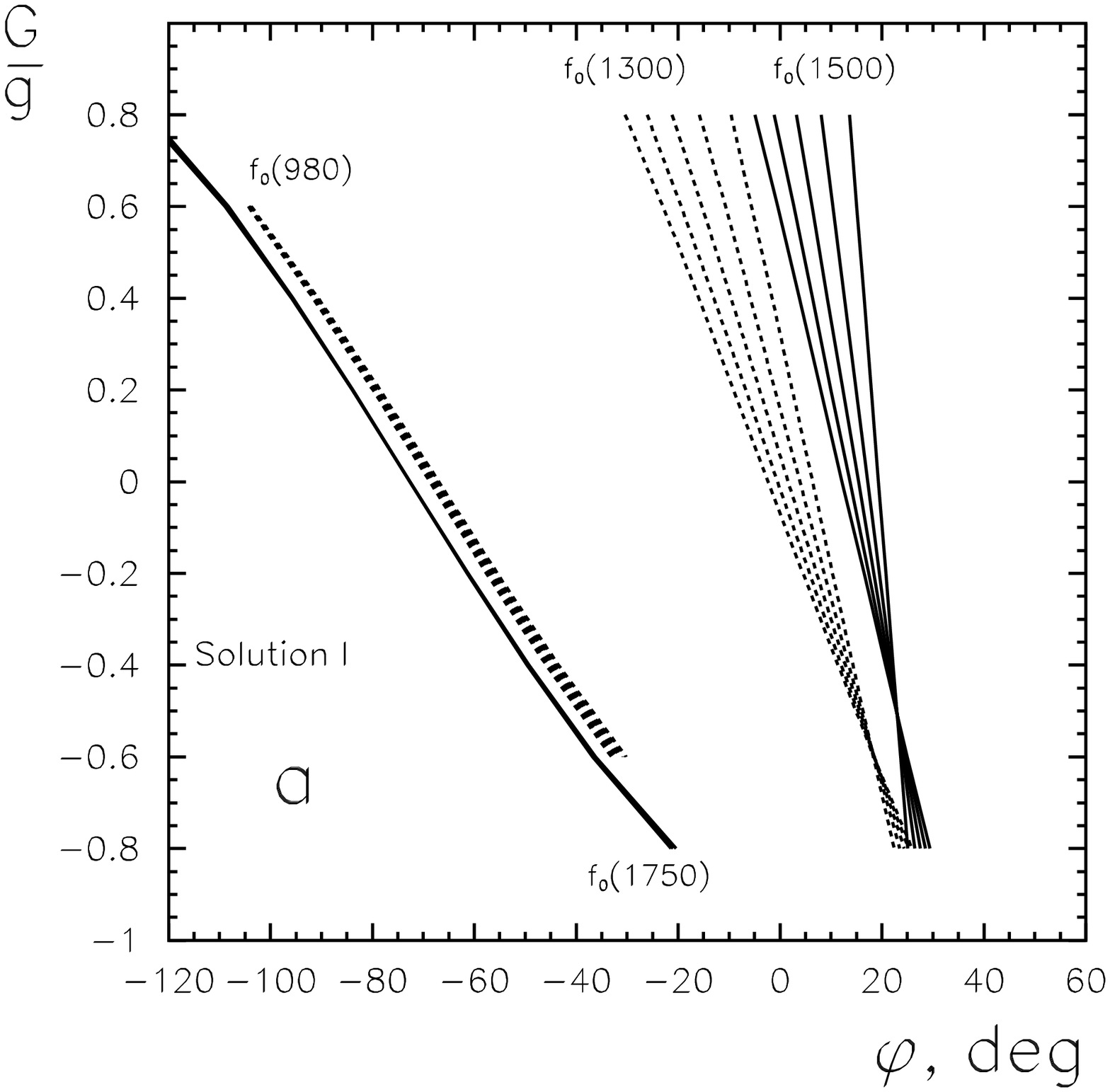,width=7cm}\hspace{0.5cm}
            \epsfig{file=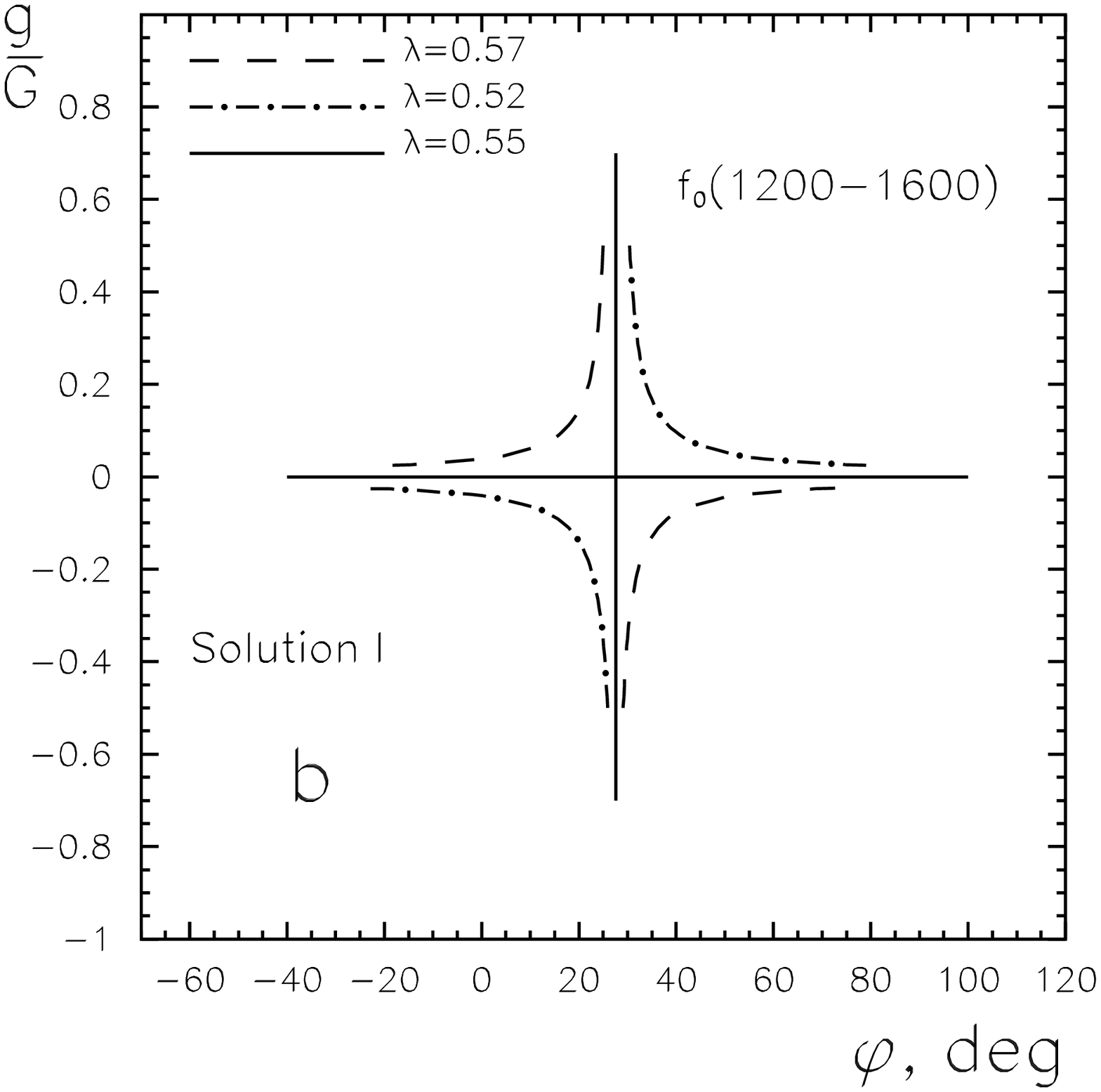,width=7cm}}
\centerline{\epsfig{file=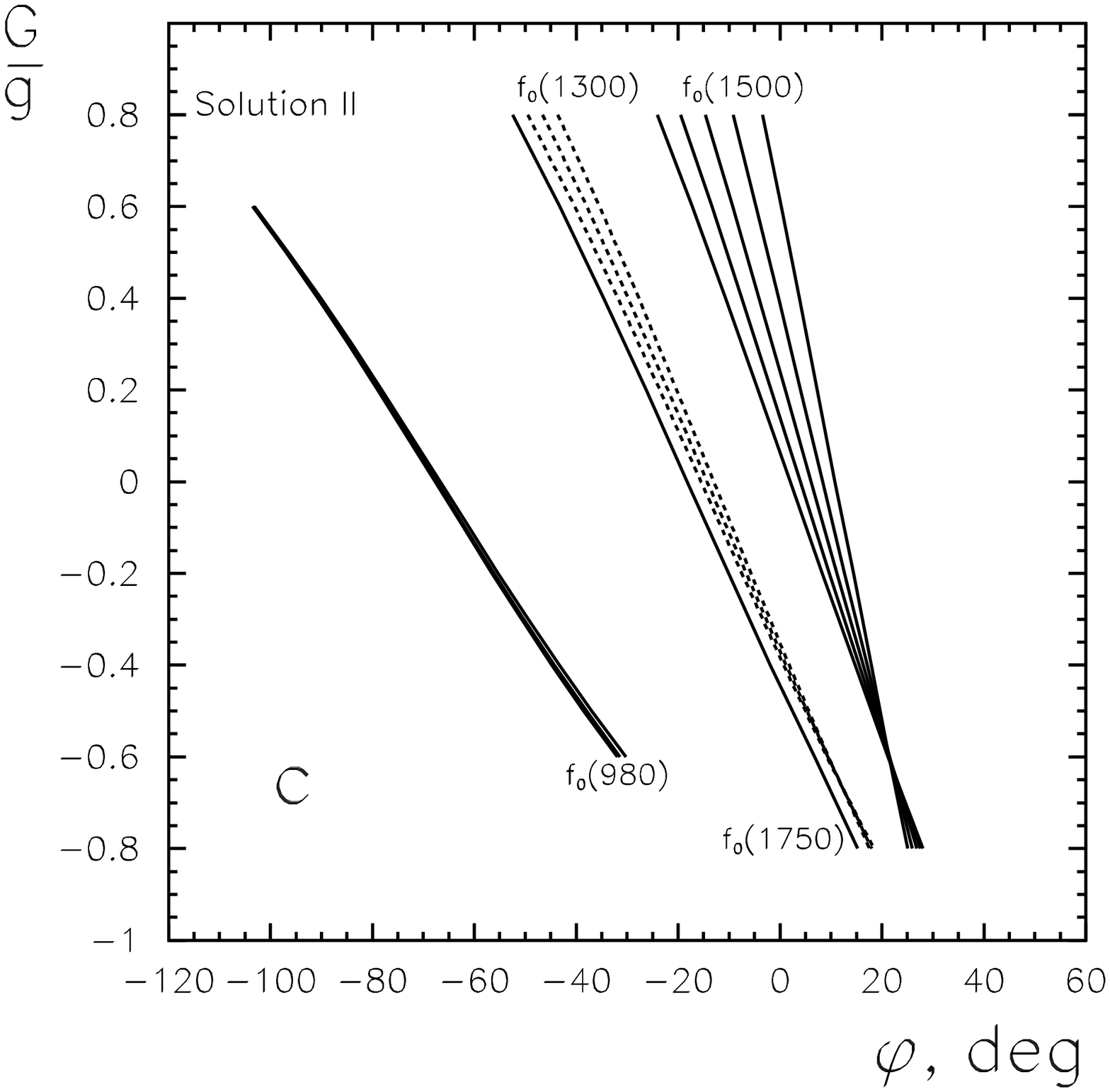,width=7cm}\hspace{0.5cm}
            \epsfig{file=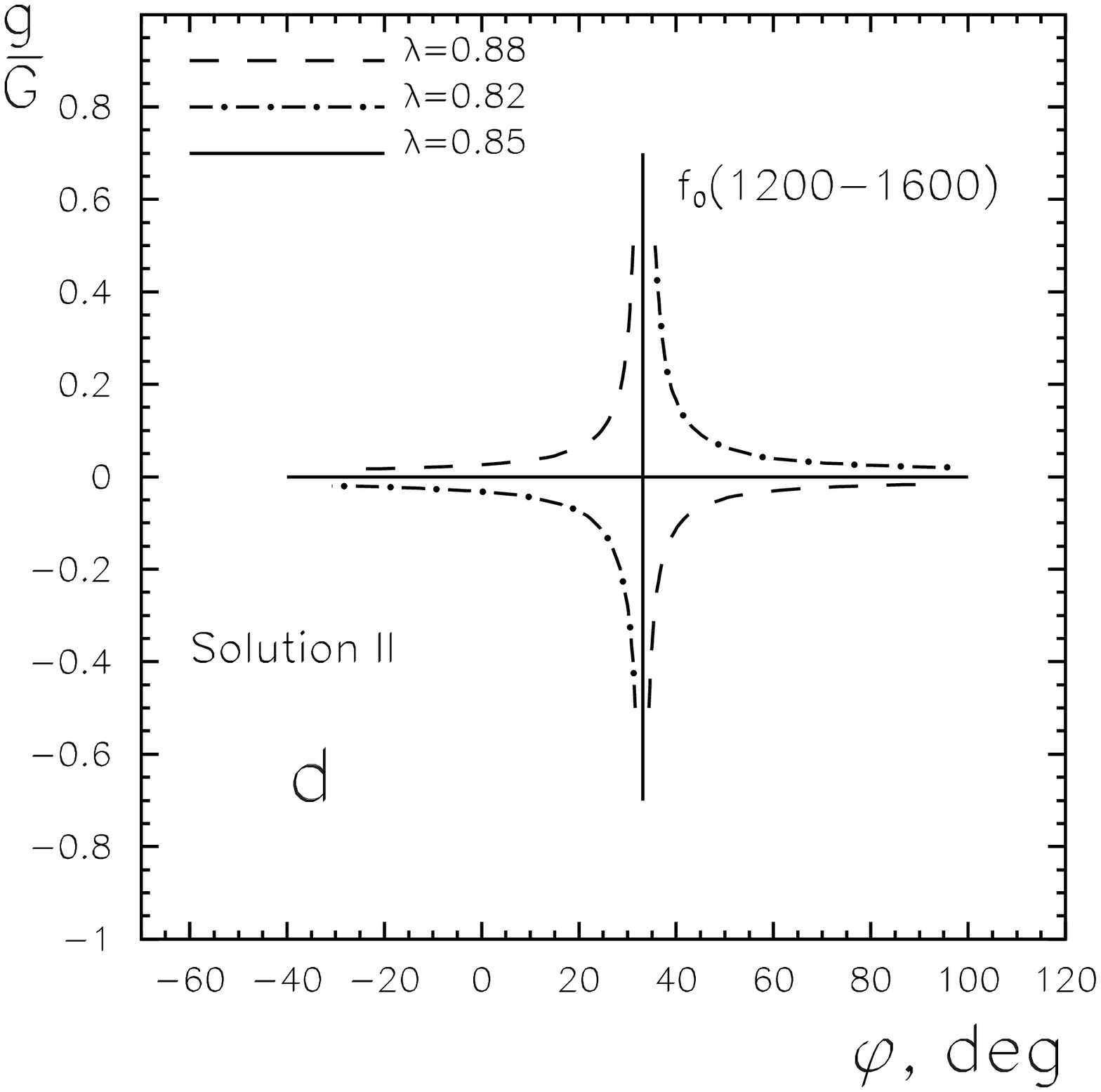,width=7cm}}
\caption{Correlation curves on the $(\varphi ,G/g)$ and $(\varphi ,g/G)$
plots for the description of the decay couplings of resonances (Table
4) in terms of quark-combinatorics relations (38).  a,c) Correlation
curves  for the $q\bar q$-originated resonances: the curves with
appropriate $\lambda$'s cover strips on the $(\varphi ,G/g)$ plane.
b,d) Correlation curves  for the glueball descendant: the curves
at appropriate
$\lambda$'s form a cross on the $(\varphi ,g/G)$ plane
with the center near $\varphi \sim 30^\circ$,  $g/G\sim 0$.}
\end{figure}

\begin{figure}[h]
\centerline{\epsfig{file=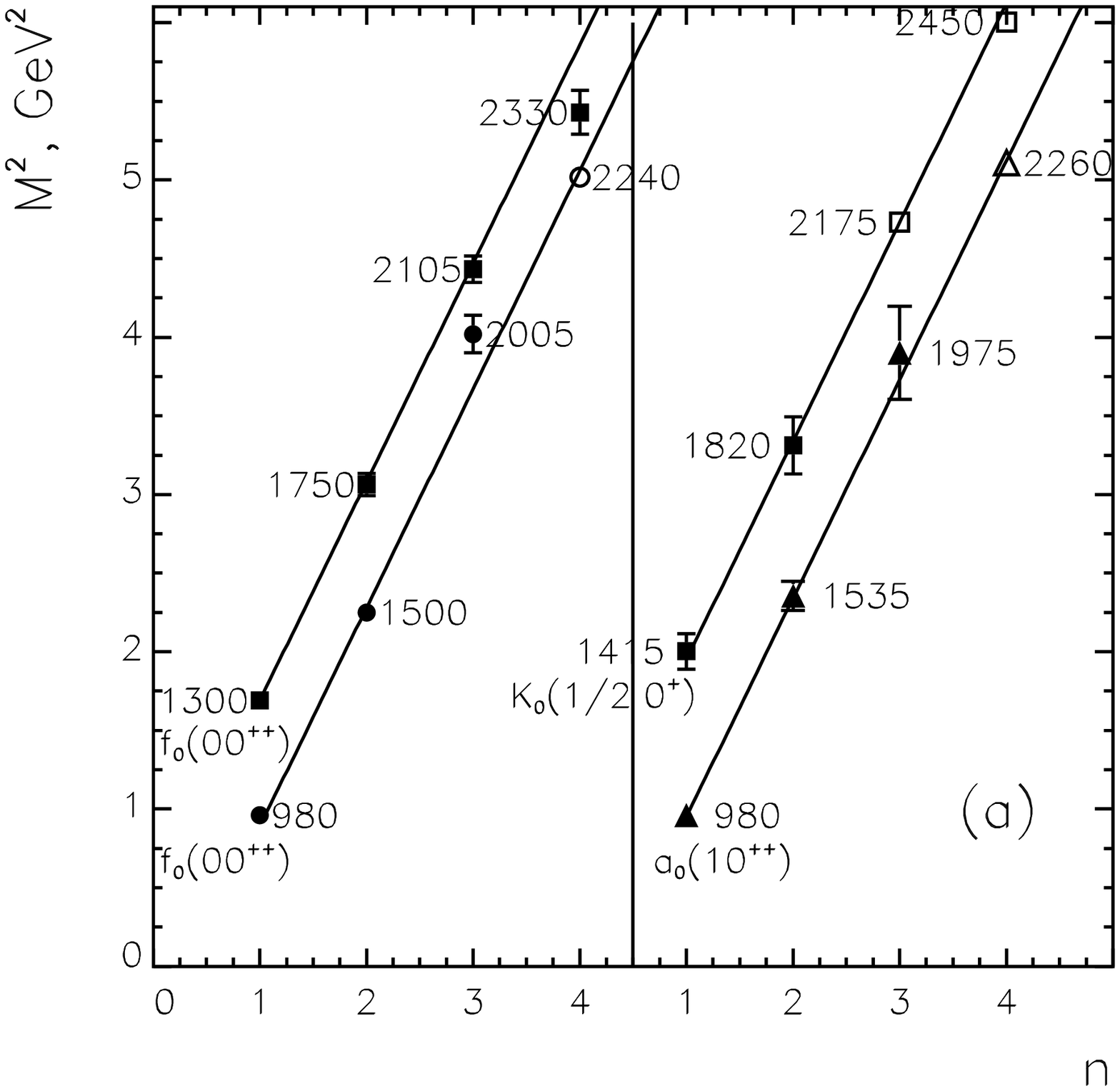,width=6.5cm}
            \epsfig{file=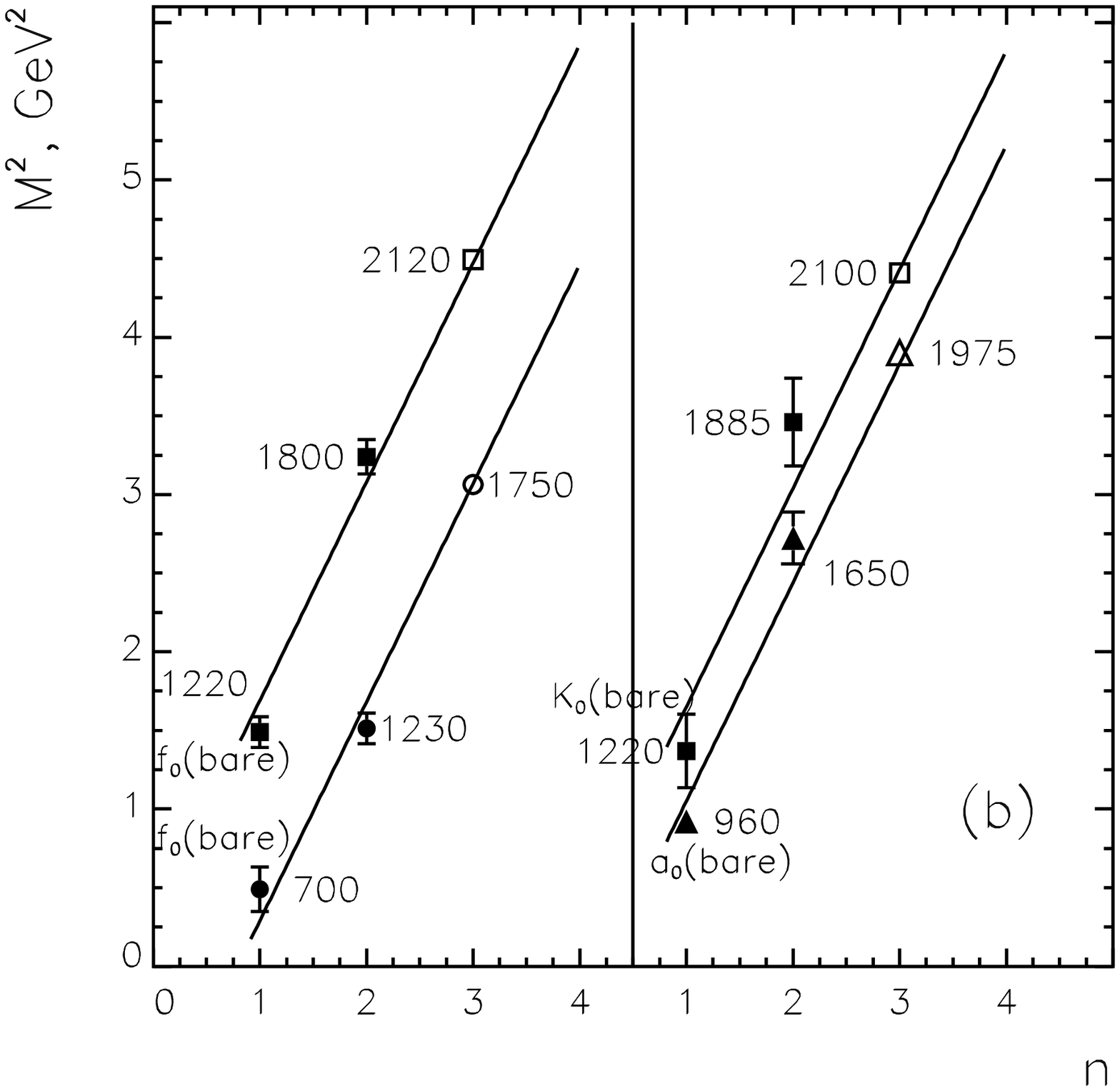,width=6.5cm}}
\caption{\footnotesize Linear trajectorieson the
$(n,M^2)$ plane for scalar resonances (a) and bare scalar states
 (b). Open circles correspond to the predicted states.}
\end{figure}
\newpage

\begin{figure}
\centerline{\epsfig{file=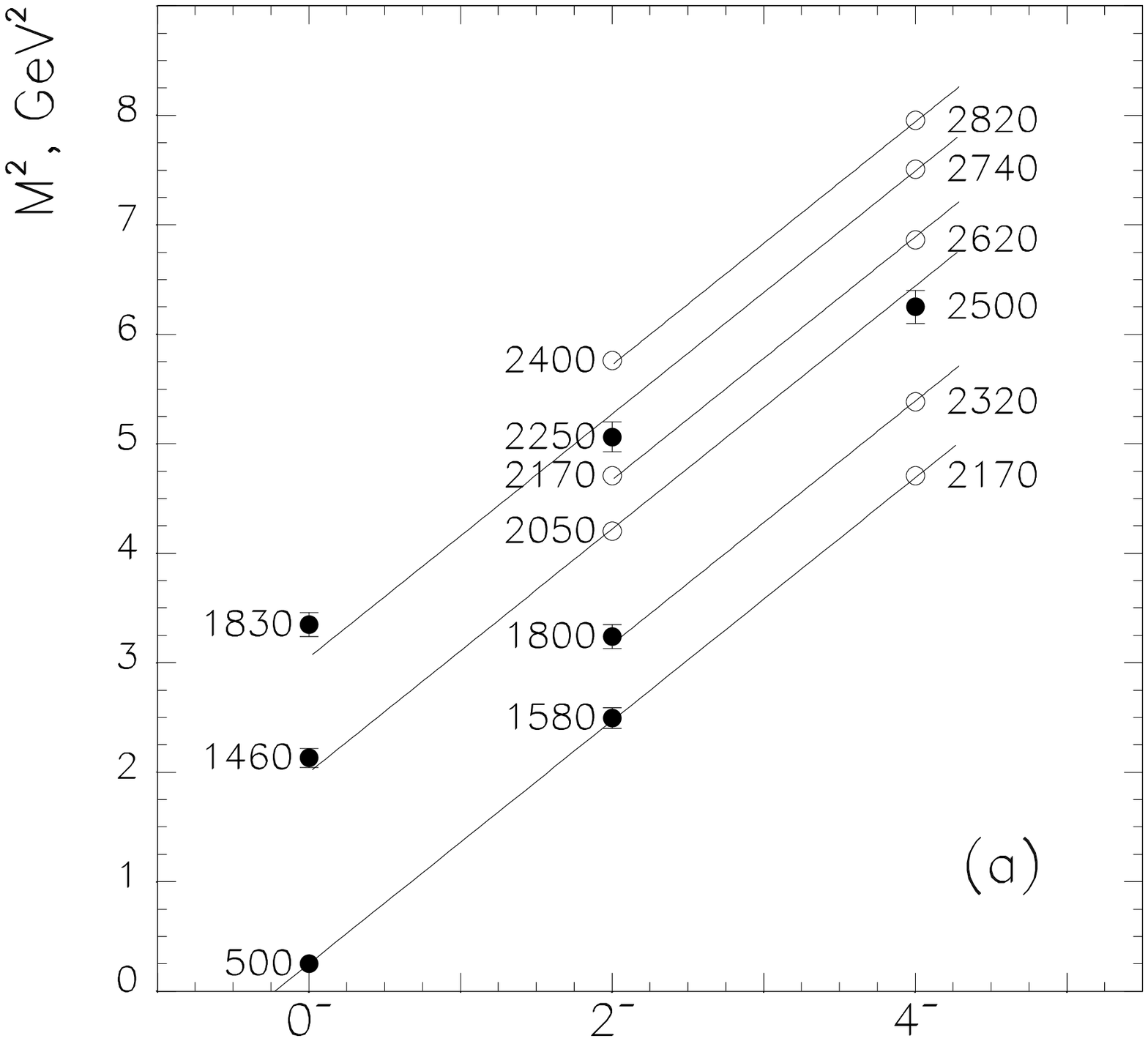,width=8cm}\hspace{-1.5cm}
            \epsfig{file=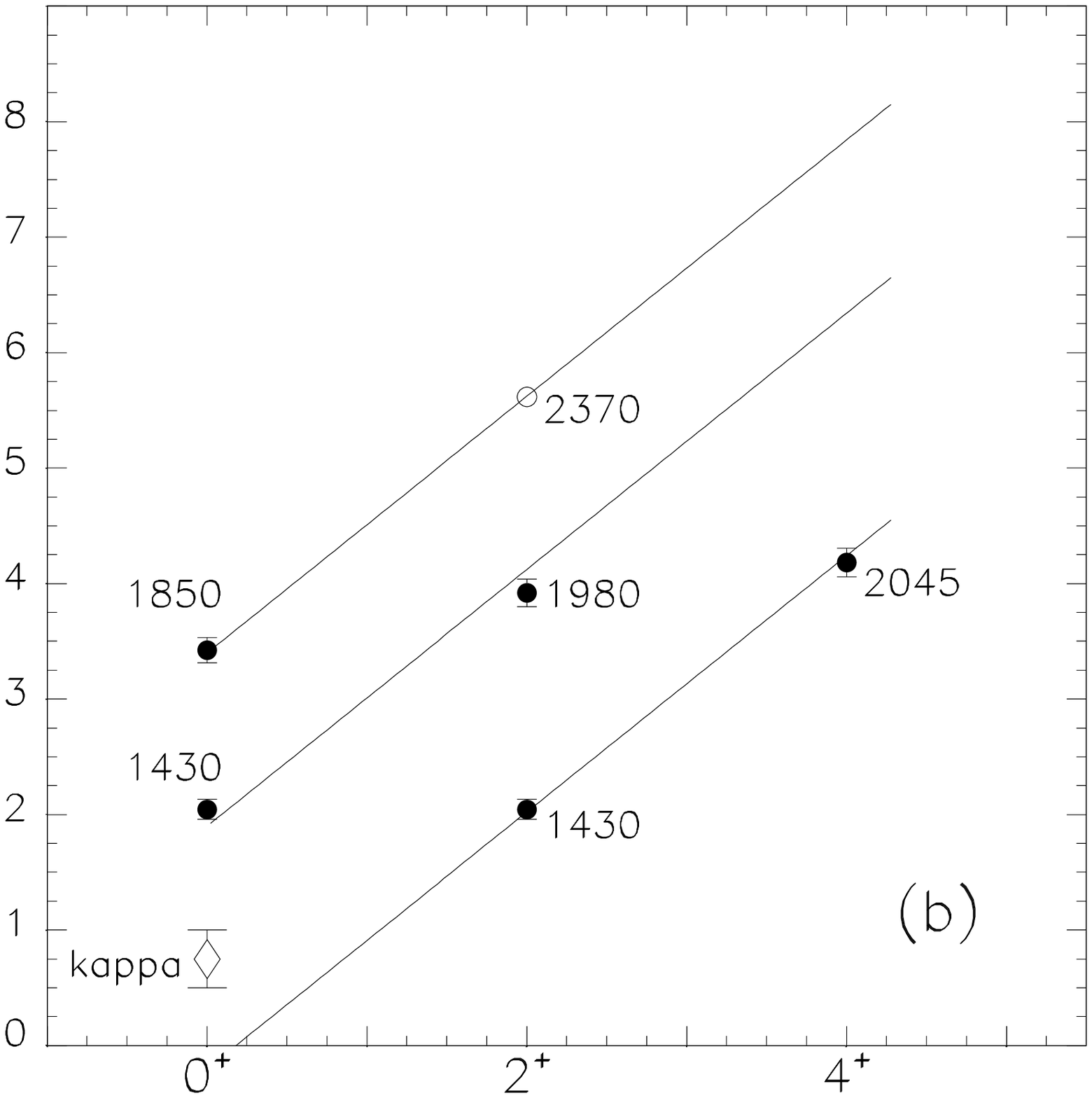,width=8cm}}
\vspace{-1.5cm}
\centerline{\epsfig{file=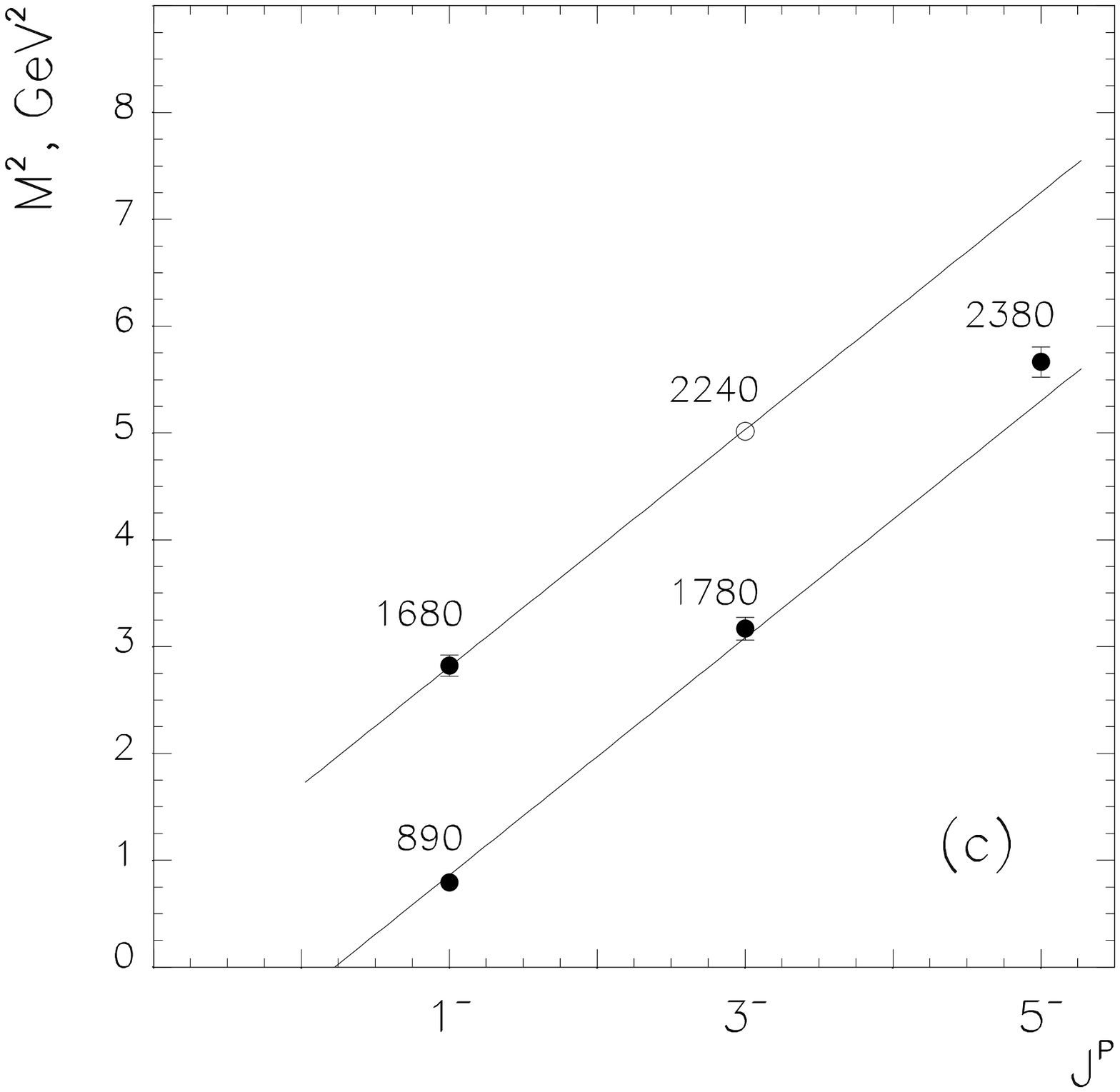,width=8cm}\hspace{-1.5cm}
            \epsfig{file=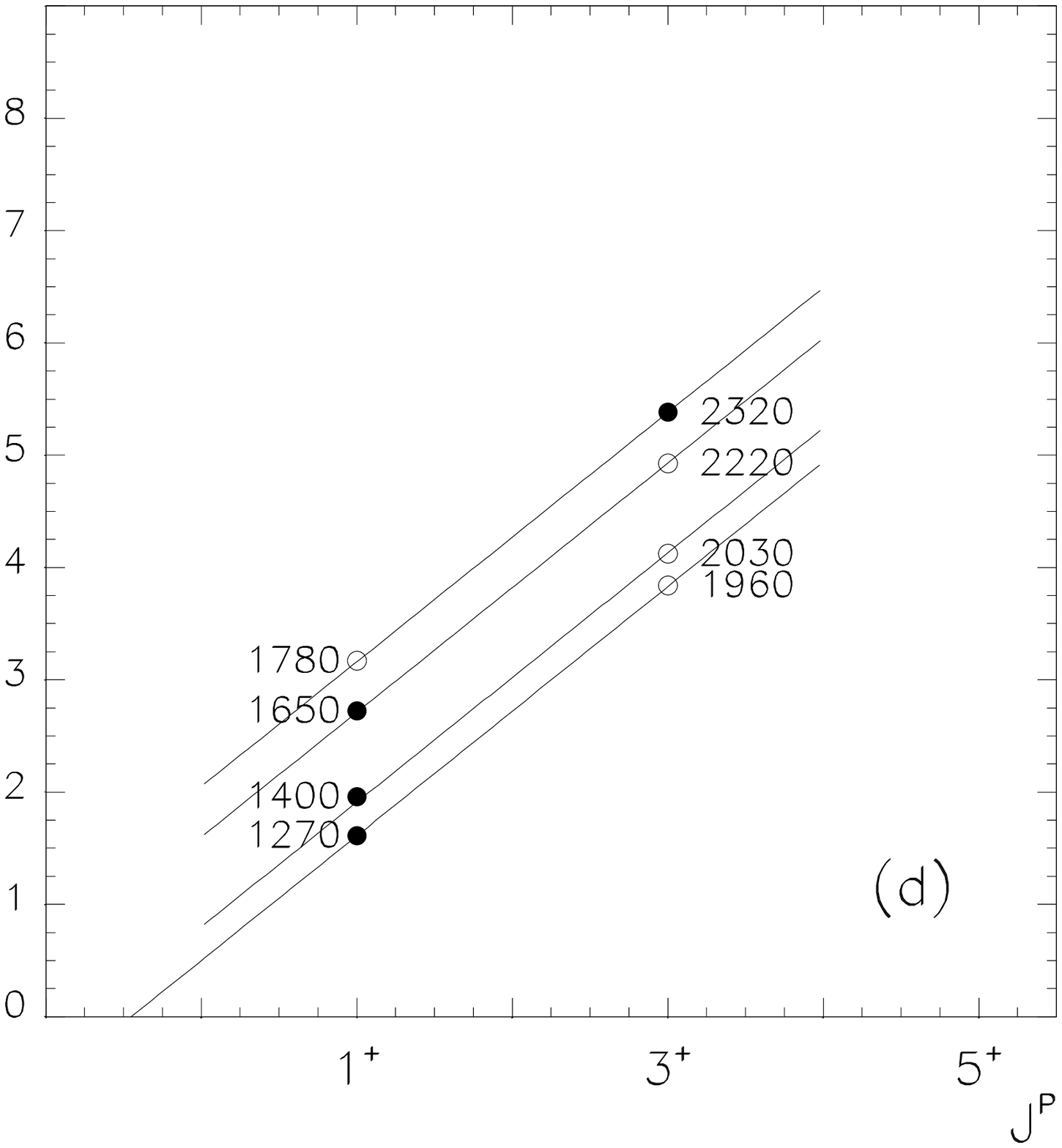,width=8cm}}
\caption{\footnotesize The $(J^P,M^2)$ planes for kaonic sector
(open circles stand for the predicted states).}
\end{figure}

\begin{figure}
\centerline{\epsfig{file=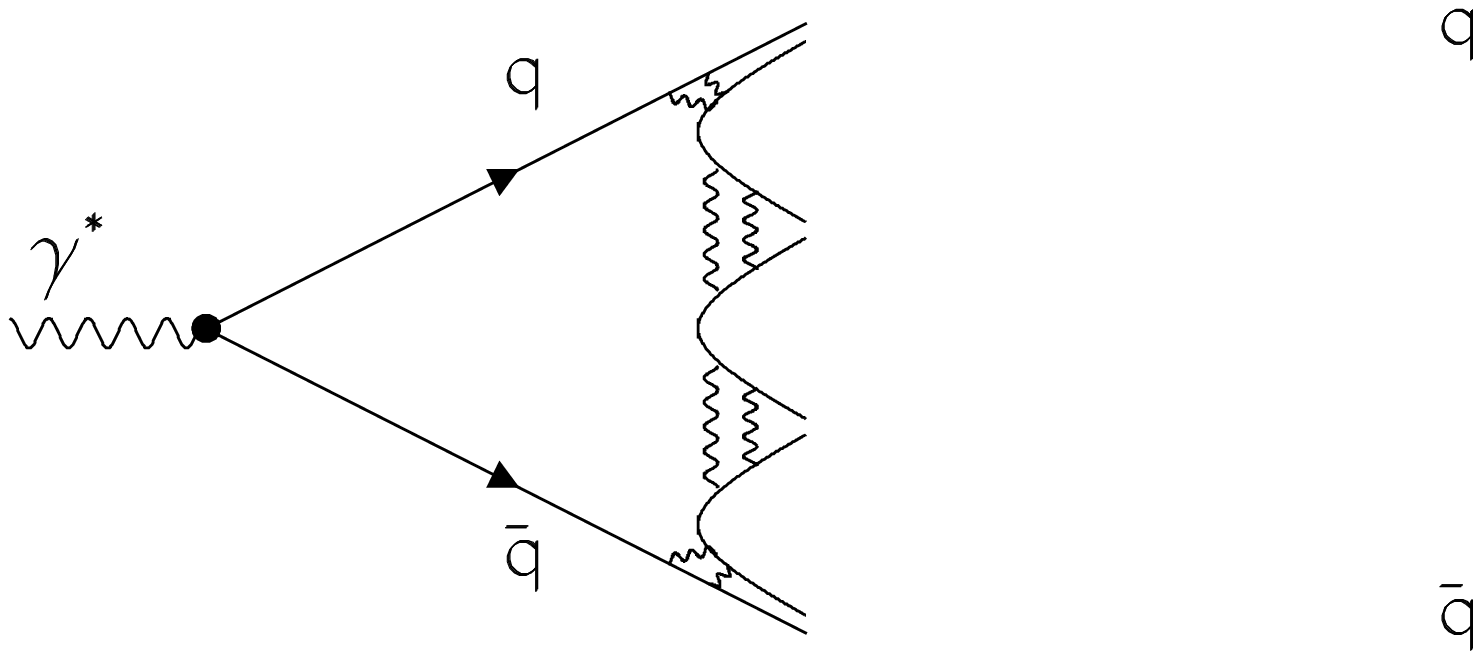,height=6cm}\hspace{-1.0cm}
            \epsfig{file=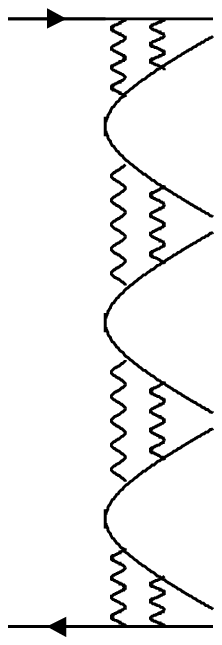,height=6cm}\hspace{-1.0cm}
            \epsfig{file=f13b.eps,height=6cm,angle=180}}
\vspace{-0.7cm}
{\large \hspace{5cm} a \hspace{6.7cm} b}
\caption{a) Quark-gluonic comb produced by breaking  a string by
quarks flowing out in  the process $e^+e^- \to \gamma^*\to q\bar q\to
mesons$.  b) Convolution of the quark-gluonic combs --- an example of
diagrams describing interaction forces in the $q\bar q$ systems at
$r\sim 2.0$ fm.}
\end{figure}

\end{document}